\documentclass[preprint,12pt,3p]{elsarticle}
\usepackage{amsfonts} % if you want blackboard bold symbols e.g. for real numbers
\usepackage{amsmath}
\usepackage{amssymb}
\usepackage{amsthm}
\usepackage{graphicx} % if you want to include jpeg or pdf pictures

\usepackage[caption=false]{subfig}
\usepackage{float} 
\usepackage[utf8]{inputenc} % use to list of figures
\graphicspath{ {figures/} }
\usepackage{array}
\usepackage{empheq} %to help box two or more equations

\usepackage{hyperref}

\journal{Physica D}

\begin{document}

\begin{frontmatter}

\title{Nonlinear Dynamics of Coupled Axion-Josephson Junction Systems}

\author[label]{Jin Yan}
\ead{j.yan@qmul.ac.uk} 
\author[label]{Christian Beck}
\ead{c.beck@qmul.ac.uk}
\address[label]{School of Mathematical Sciences, Queen Mary University of London, Mile End Road, London E1 4NS, UK}

\begin{abstract}
                                   We study the classical dynamics of an axion field (the signal)
                 that is coupling into a Josephson junction (the detector) by means of a capacitive coupling of arbitrary size. Depending on the size of the coupling constant and the initial conditions, we find a rich phase space structure of this nonlinear problem. We present
                  general analytic solutions of the equations of motion in the limit of small amplitudes of the angle variables, and discuss both the case of no dissipation and the case of dissipation in the system. The effect of a magnetic field is investigated as well, leading to topological phase transitions in the phase space structure.
\end{abstract} 


\end{frontmatter}

\section{Introduction}

Axions are exotic particles that are predicted in many extensions of the standard model of
elementary particle physics \cite{ax1,ax2,ax3,ax4,ax5,dark1,dark2,dark3,beck1,beck2,beck3,beck4}.
The QCD axion is one of the main candidates for dark matter in the universe. The equations
of motion of axions are very similar to those of Josephson junctions. This analogy has
been discussed in detail in \cite{beck1,beck2,beck3,beck4} and possible detection schemes for axions
or axion-like particles 
have been proposed based on this analogy, using Josephson junctions \cite{tinkham} as detectors, and assuming
the possibility of a synchronisation between the axion and Josephson phase angle.
%Indeed, if axions are the main ingredient of dark matter in our Galaxy, then a huge number
%of axions is constantly flowing through the Earth, similarly as neutrinos do.
%However, nothing is known about the size of the coupling constant of axions within
%a given Josephson junction environment, assuming the Josephson junction
%is used as a detector\cite{beck2,beck3,beck4}, so one has to study models and see what the predictions are
%for a given size of the coupling constant. In principle, the axion coupling
%to tunneling Cooper pairs in a Josephson junction can be stronger
%than the axion coupling in vacuum, for reasons discussed in \cite{beck2,beck3}. In this way, by comparison with
% experiments, one can then restrict the parameter space
%of axion couplings that are a priori possible in a Josephson environment.

In this paper we are interested
in the general behaviour of the classical nonlinear field equations that decribe this problem. We will study systematically the nonlinear
dynamics aspects of this coupled initial value problem, i.e.  a classical axion field that is
coupled into a Josephson junction in a capacitive way, meaning the interaction strength
is proportional to $c (\ddot{\varphi} - \ddot{\theta})$, where $\varphi$ is the Josephson phase difference
and $\theta$ the axion misalignment angle, $c$ is the coupling constant. We keep the size of the coupling constant $c$ as an arbitrary parameter and
will study both very small, intermediate, and large values of $c$. Mathematically, the problem is equivalent to two coupled
Josephson junctions, one with phase angle $\varphi$ and the other one with phase angle $\theta$.
Surprisingly, a very rich and complex phase space structure arises as a function of the coupling, the frequency ratio and
the initial conditions, which we will describe in detail in the following sections. For previous work related to coupled Josephson junction-like systems, see e.g., \cite{coupJJ1, coupJJ2, coupJJ3, coupJJ4}. 

For small
elongations of the angle variables and velocities, the problem becomes linear and describes
two coupled harmonic oscillators. Of course this linear case is exactly solvable, and we will compare
in our paper carefully the nonlinear effects due to the cosine potentials of the axion and junction
as compared to the case of coupled harmonic oscillators with just a quadratic potential. We will deal with our coupled
system in a general mathematical way, allowing in principle for
arbitrary parameter combinations and analysing the classical phase space structure of this nontrivially coupled
nonlinear system. It should be clear that this system has many different applications in physics: it describes
not only the possible coupling of axions into a Josephson environment, but also the coupling
of two classical $q$-bits \cite{coup1,coup2}, which are coupled with a capacity, of relevance for nanotechnological applications.
One can also think of two coupled pendulas that are coupled in a way that is proportional to
the acceleration difference between the two pendulas, i.e. a classical mechanics problem with constraints
in a constant gravitational field. Thus many generic physical
interpretations are possible for the dynamical system that we systematically study in the following.

This paper is organized as follows:

In section 2 we study the phase space structure as a function of the coupling constant $c$, for suitable
initial conditions that are physically motivated. Particular
emphasis is put on coupling constants that are of the order $c \sim 10^{-3}$, which
is a typical physically realized coupling strength in coupled $q$-bits \cite{coup2}. In section 3 we focus on the dependence with respect to the ratio of Josephson junction frequency and axion frequency, emphasizing the sensitivity on this parameter, including resonance effects. In section 4 we study the dependence on initial conditions,
in particular how the phase diagram changes when the initial $\dot{\varphi} (0)$ of the measuring Josephson
junction is varied (in experiments, this can be easily achieved by varying the applied bias voltage).
In section 5 we present analytic solutions of the general initial value problem in the limit case of small amplitudes
and small angular velocities, where the problem reduces to two coupled (in a capacitive way) harmonic oscillators. In this section we will deal with arbitrary
coupling strengths, without and with dissipation. 
The effect of an external magnetic field is discussed in section 6. Finally, our concluding remarks
are given in section 7.

\section{Nontrivial phase space structure as a function of the coupling strength $c$}

We start by introducing the classical equations of motion of axions coupled into a Josephson junction environment.
As worked out in \cite{beck1}, the coupled system of equations  is given by
\begin{subequations}
\begin{align}
\ddot{\varphi} + a_{1} \dot{\varphi} + b_{1} \sin{\varphi} &= c(\ddot{\theta} - \ddot{\varphi})\\
\ddot{\theta} + a_{2} \dot{\theta} + b_{2} \sin{\theta} &= c(\ddot{\varphi} - \ddot{\theta})
\end{align}
\label{1}
\end{subequations}
where $\varphi (t)$ and $\theta (t)$ are the phase angle variables for the Josephson junction and axion, respectively. The parameters in the above equations, namely, the dissipation coefficients $(a_{1}, a_{2})$ , the frequency parameters $(b_{1}, b_{2}) $, and the coupling constant $c $ depend on details of the physical model considered. For the measuring Josephson junction, $b_1=\omega^2$ corresponds to the plasma frequency $\omega$ of the junction, and $a_1$ is given by $a_1=1/RC$, where $R$ is the shunt resistance and $C$ the capacity. For dark matter axions, the corresponding $(a_2,b_2)$ parameters as well as the coupling strength $c$ are unknown, although some conjectures have been formulated \cite{beck1,beck2,beck3,beck4} and some new experiments search in the relevant parameter region\cite{ex1,ex2,ex3}. $b_2$ is given by the square of the axion mass. In the early universe one has $a_2=3H$ where $H$ is the Hubble constant. At current times, $a_2=0$ in very good approximation. To the best of our knowledge, a detailed and systematic investigation of the mathematical properties of the system \eqref{1} as a function of the parameters is lacking, but this is higly relevant for future experimental axion searches that use Josephson junctions or coupled arrays of Josephson junctions as possible novel types of axion detectors
\cite{beck2,beck3,beck4}. Hence, in the following we explore the mathematical properties of the coupled system in more detail.

For simplicity, consider first the case with no dissipation ($a_{1} = a_{2} = 0$). Due to its nonlinearity, the system exhibits complex behavior as a function of the parameters, which in the following is investigated numerically by writing \eqref{1} as a system of four first-order differential equations and using the fourth-order Runge-Kutta method. Apart from the three parameters $b_{1}, b_{2}$ and $c$, the four initial conditions $(\varphi , \dot{\varphi}, \theta , \dot{\theta})_{t = 0}$ are to be specified. 

To study how the size of coupling constant $c$ affects the behaviour of the system we fix $b_{1} = b_{2} = 1$ and vary $c$ from $10^{-6}$ to $1$. The initial conditions are chosen to be $(\varphi , \dot{\varphi}, \theta , \dot{\theta})_{t = 0} = (0, 2, 0, 0)$ such that, in the absence of the axion, the Josephson junction would behave like a pendulum in a gravitational field just reaching the highest unstable point with zero momentum. The angular velocity of the axion is assumed to be small, and put equal to zero in our simulations in the following.

\begin{figure} [H]
\centering
\subfloat[\footnotesize $c = 2\times 10^{-6}$]{
\includegraphics[width = 0.33\textwidth]{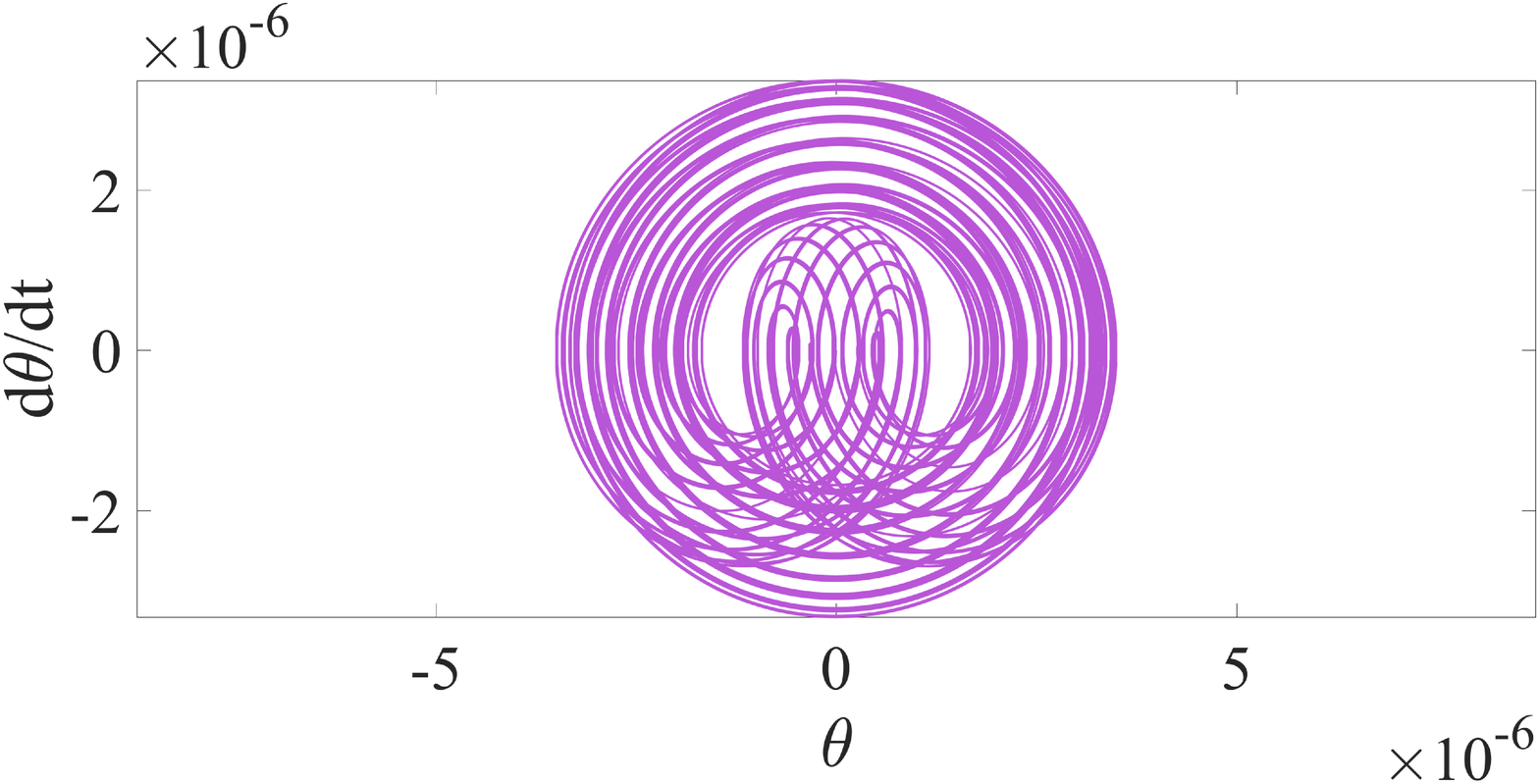} %---c = 2e-6
\label{c2e-6}
}
\subfloat[\footnotesize $c = 3.85\times 10^{-6}$]{
\includegraphics[width = 0.33\textwidth]{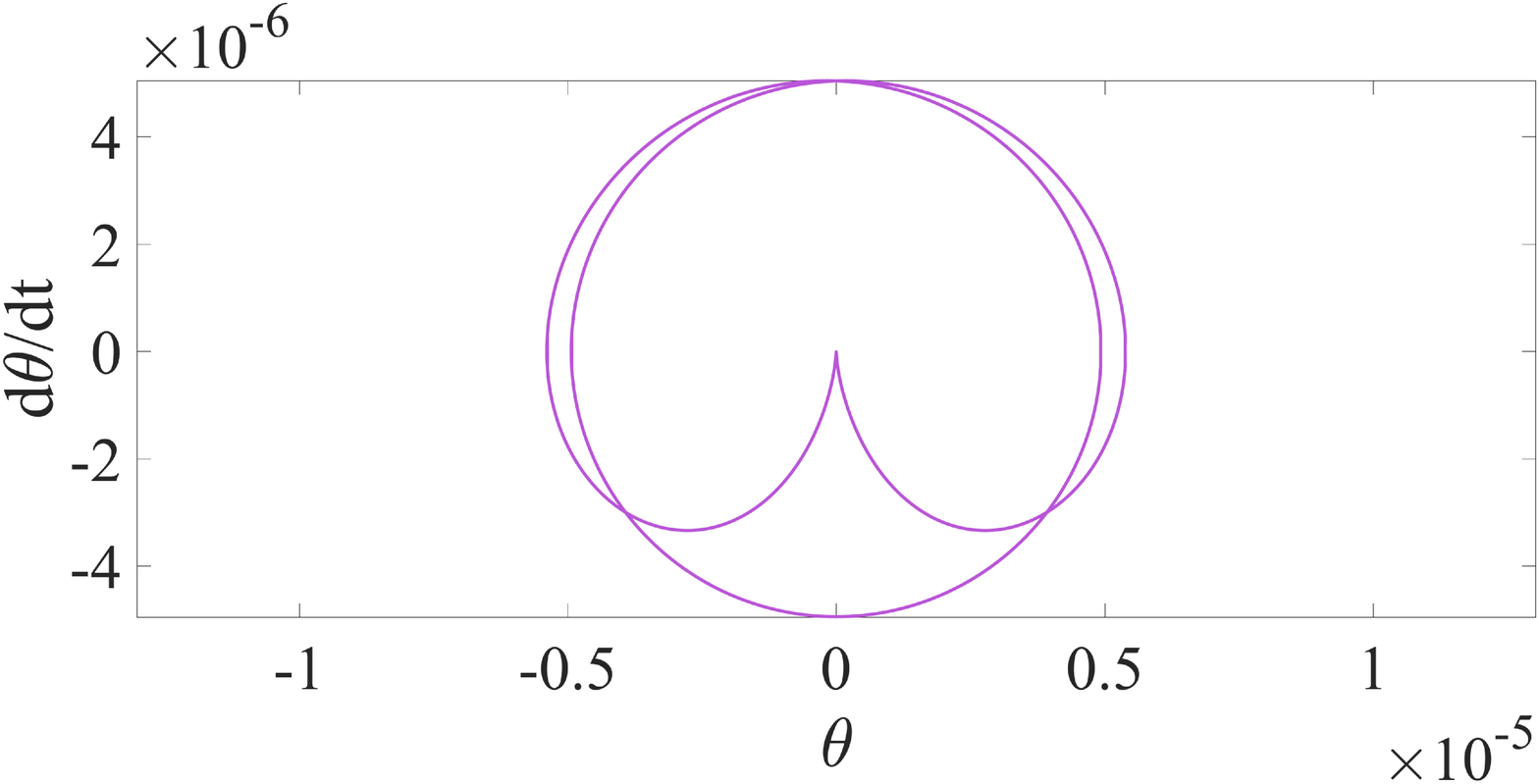} %---c = 3.85e-6
\label{c3pt85e-6}
}
\subfloat[\footnotesize $c = 6\times 10^{-6}$]{
\includegraphics[width = 0.33\textwidth]{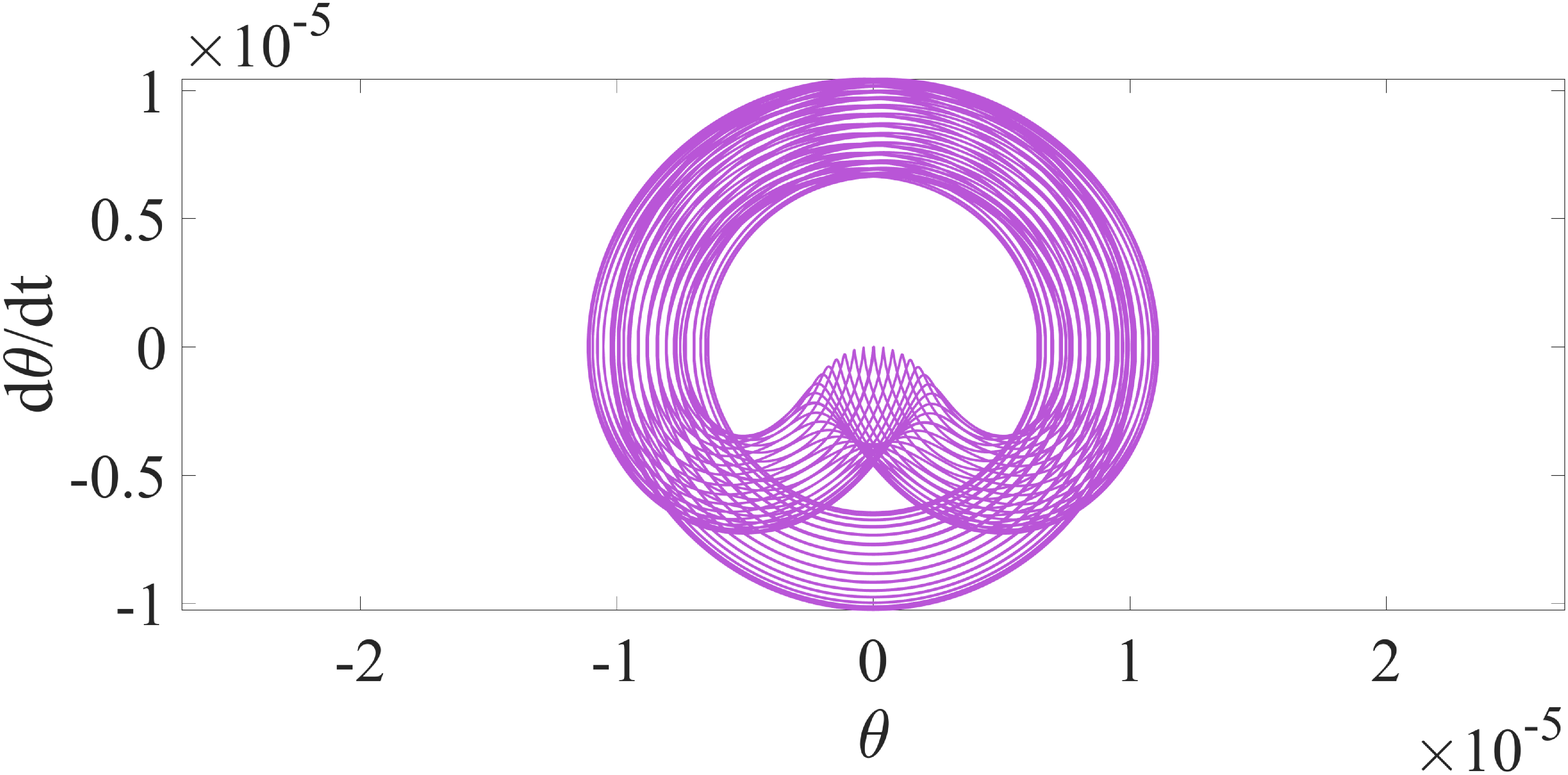} %---c = 6e-6
\label{c6e-6}
}
\qquad
\subfloat[\footnotesize $c = 1\times 10^{-3}$]{
\includegraphics[width = 0.33\textwidth]{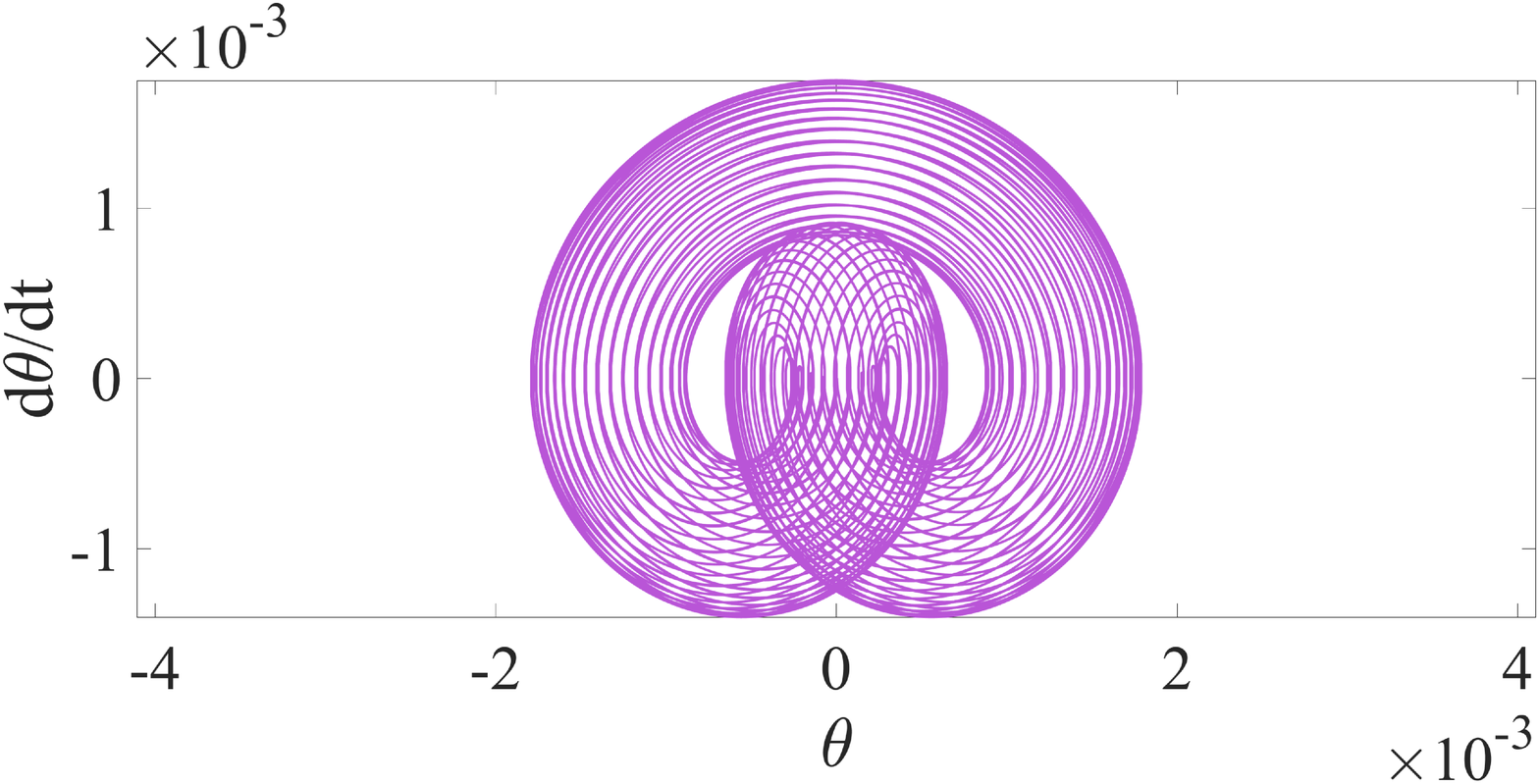} %------c = 1e-3
\label{c1e-3}
}
\subfloat[\footnotesize $c = 2.0545\times 10^{-3}$]{
\includegraphics[width = 0.33\textwidth]{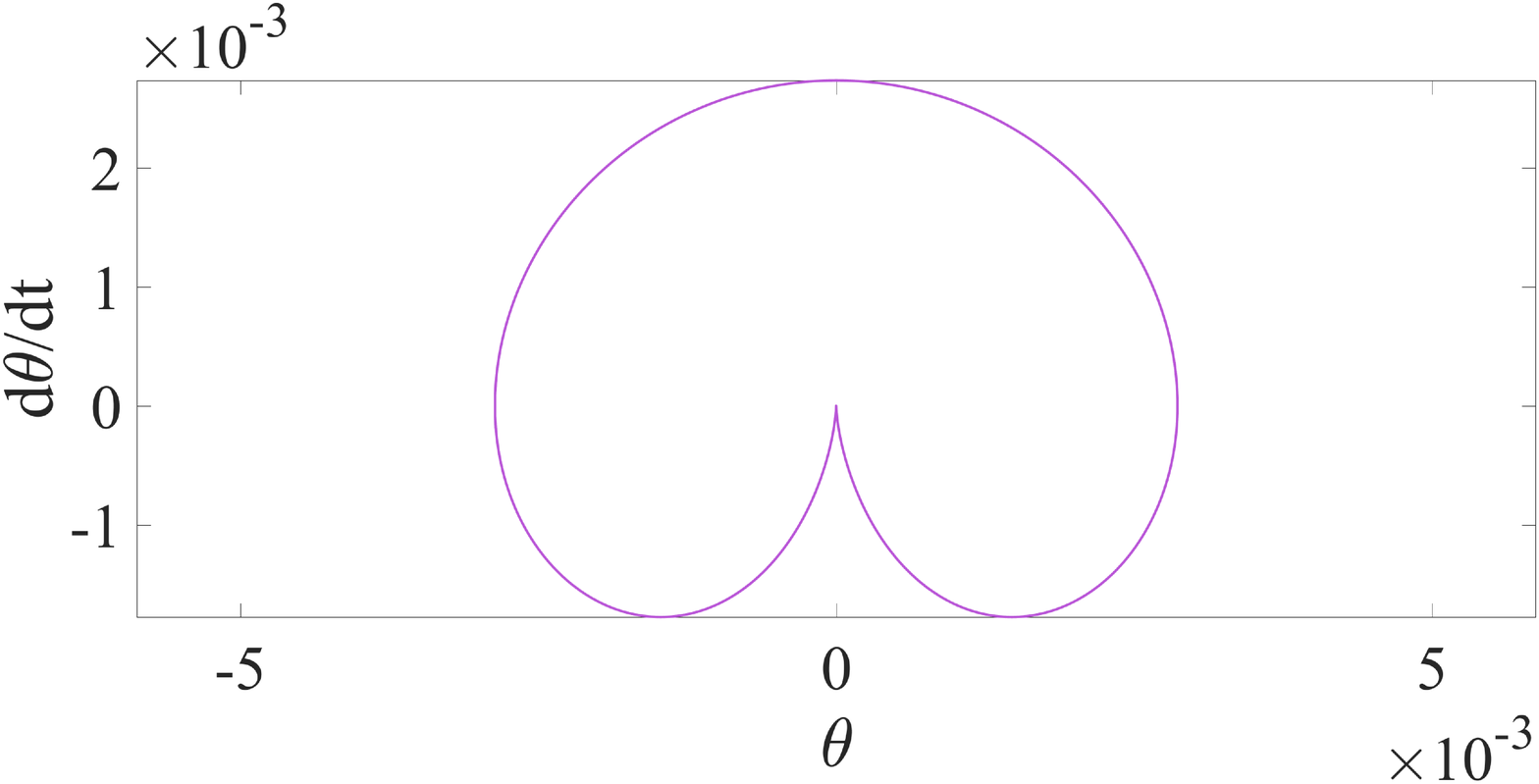} %------c = 2.0545e-3 (crit)
\label{ccrit}
}
\subfloat[\footnotesize $c = 4\times 10^{-3}$]{
\includegraphics[width = 0.33\textwidth]{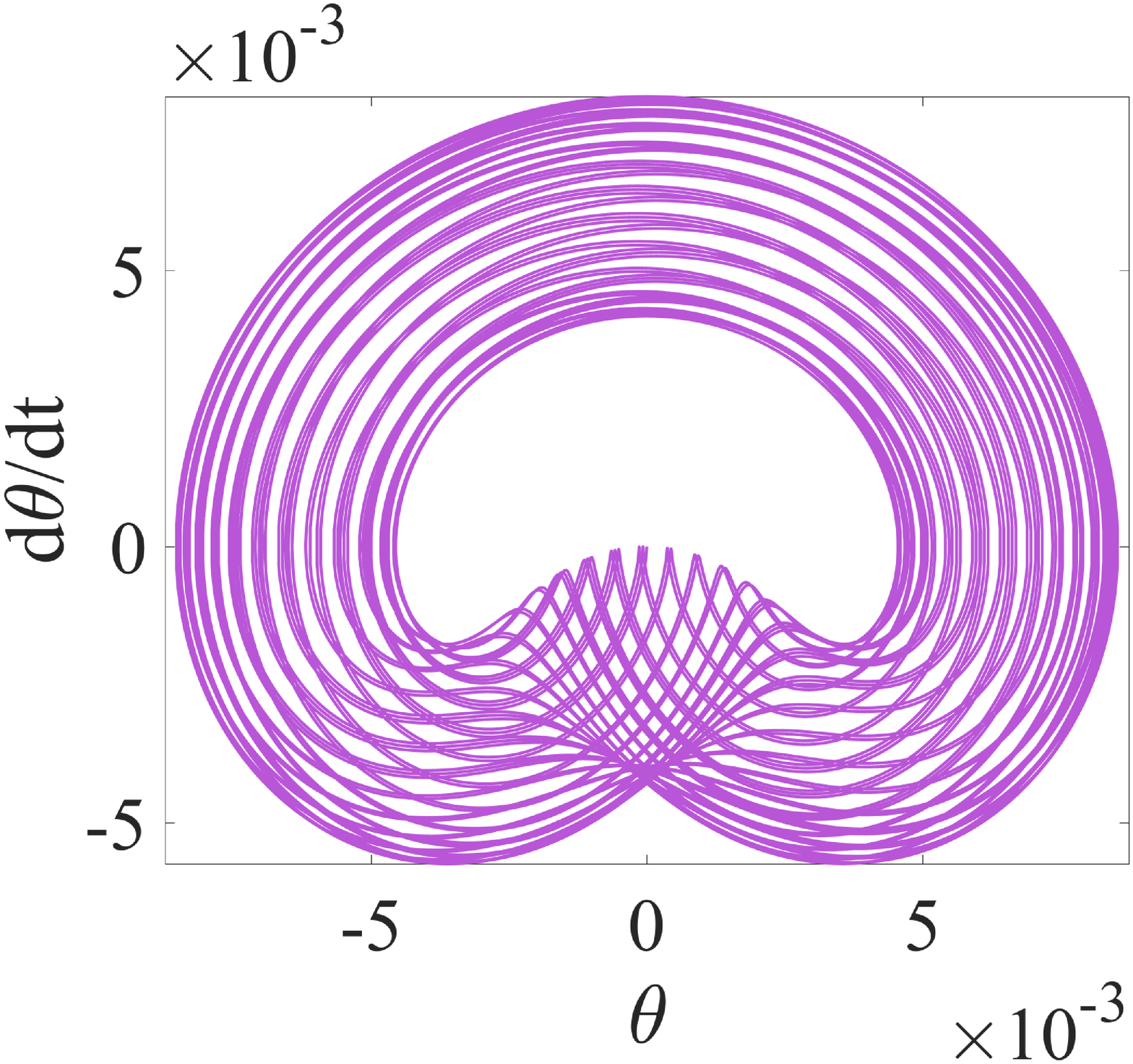} %------c = 4e-3
\label{c4e-3}
}
\qquad
\subfloat[\footnotesize $c = 1\times 10^{-3}$]{
\includegraphics[width = 0.33\textwidth]{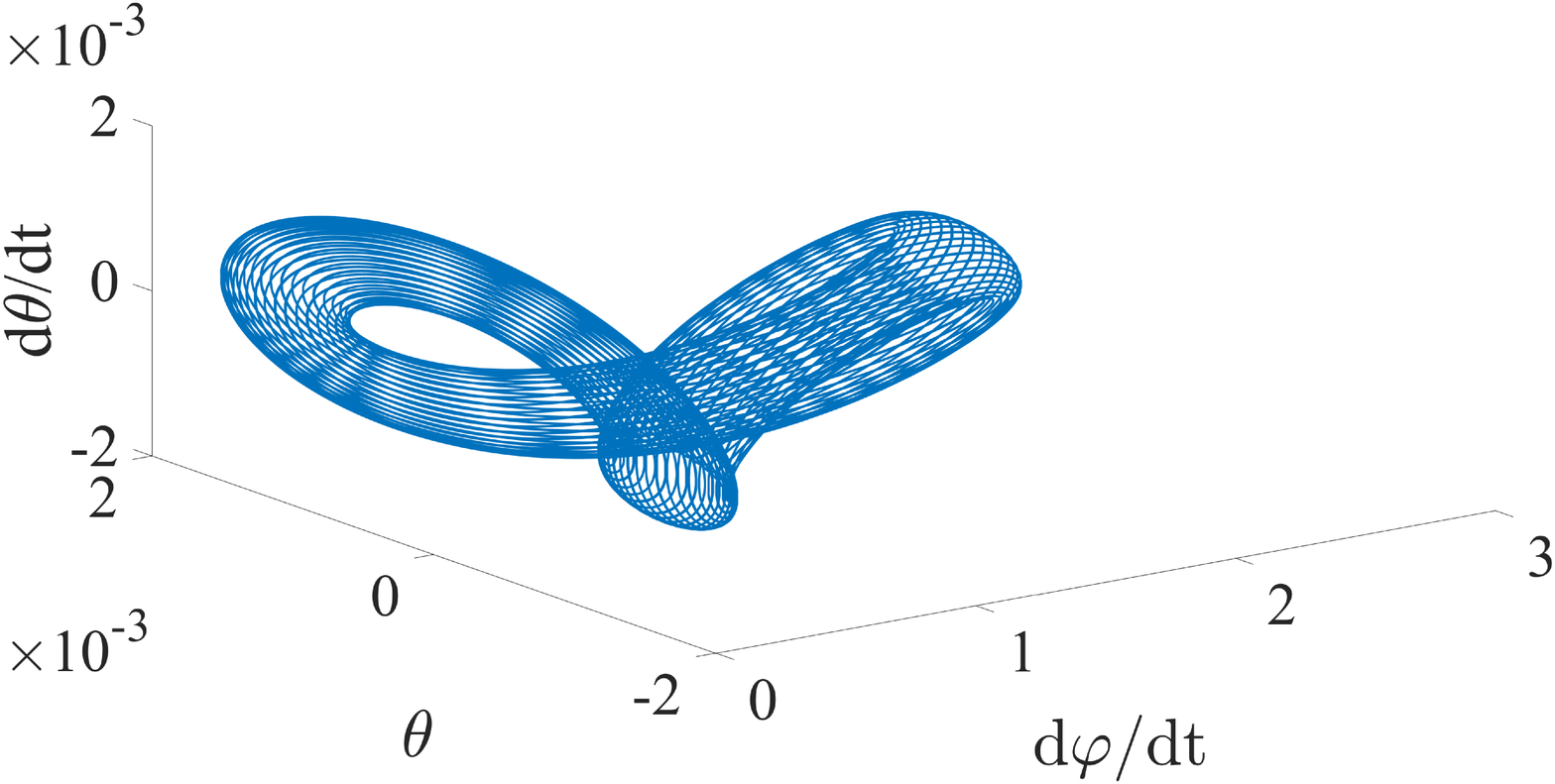} %---------3D 1e-3
}
\subfloat[\footnotesize $c = 2.0545\times 10^{-3}$]{
\includegraphics[width = 0.33\textwidth]{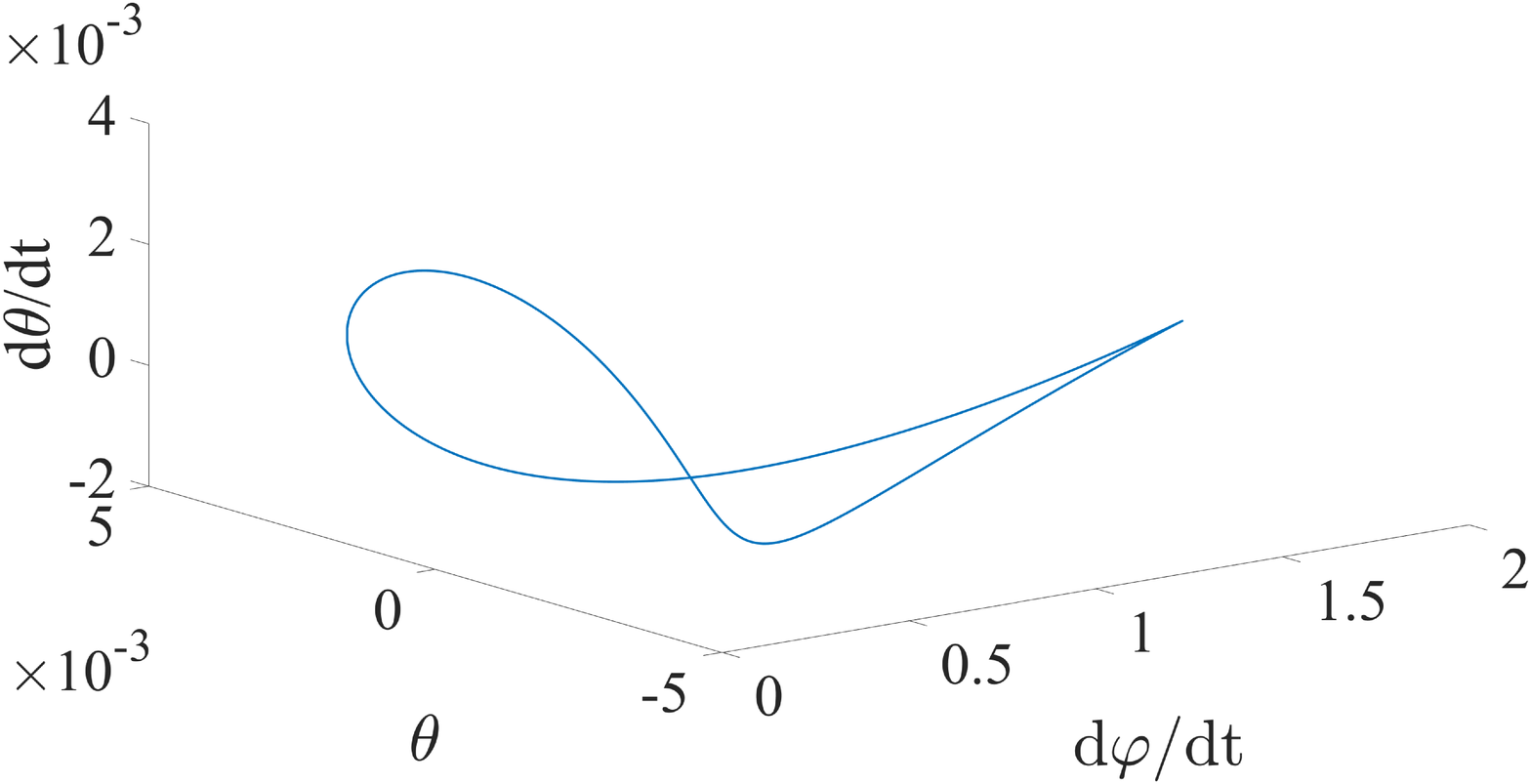} %---------3D crit
\label{3Dccrit}
}
\subfloat[\footnotesize $c = 4\times 10^{-3}$]{
\includegraphics[width = 0.33\textwidth]{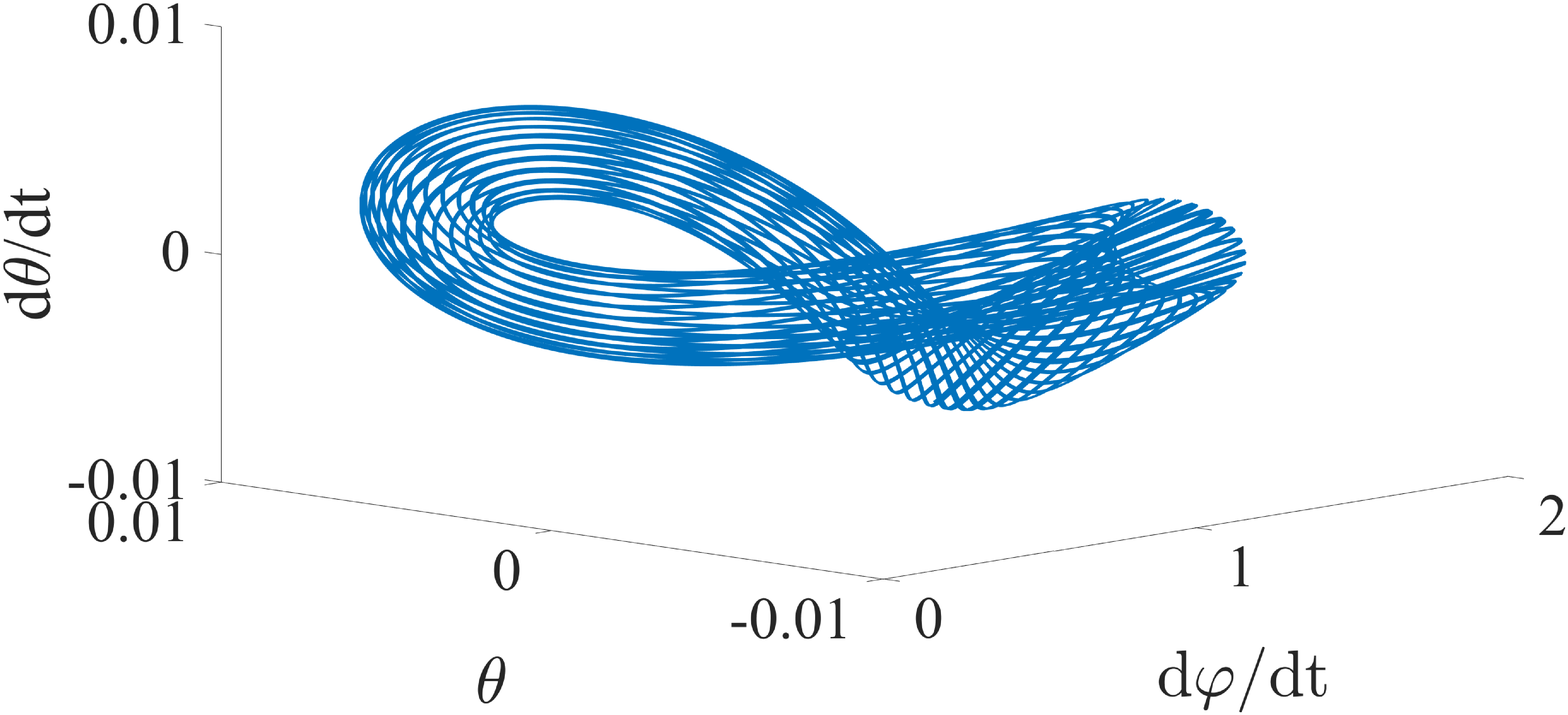} %---------3D 4e-3
}
\caption{Eversion processes of the axion trajectory for weak couplings: upper two rows: axion phase portraits in the $(\theta, \dot{\theta})$-plane; last row: the corresponding $(\dot{\varphi}, \theta, \dot{\theta})$-subspace ($t_{max} = 500$)} %with increasing scales
\label{c-weak}
\end{figure}

\begin{figure} [H]
\centering
\subfloat[\footnotesize $c = 0.743$]{
\includegraphics[width = 0.33\textwidth]{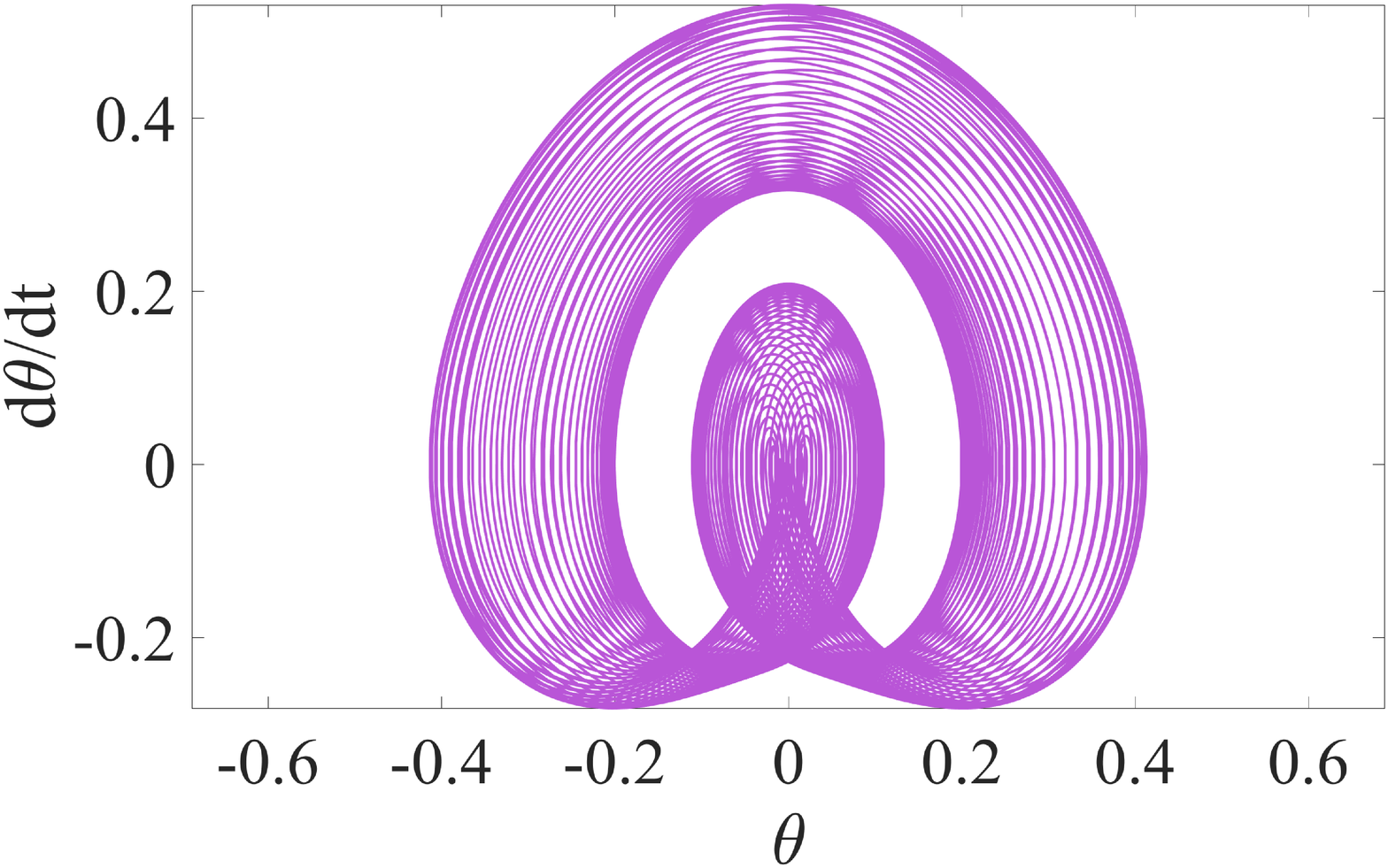} %---c = 0.743
}
\subfloat[\footnotesize $c = 0.75$]{
\includegraphics[width = 0.33\textwidth]{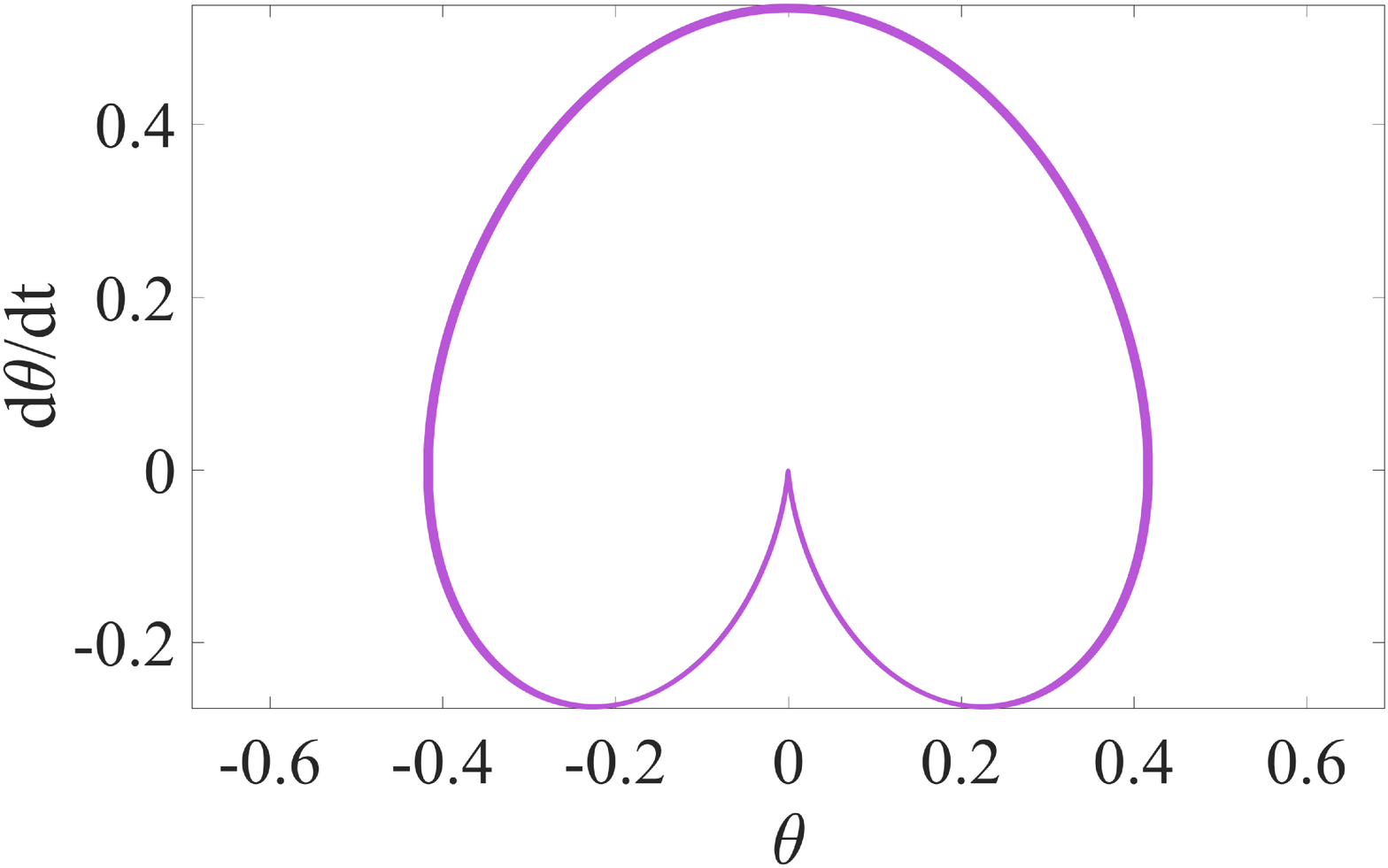} %---c = 0.75
\label{c0pt75}
}
\subfloat[\footnotesize $c = 0.8$]{
\includegraphics[width = 0.33\textwidth]{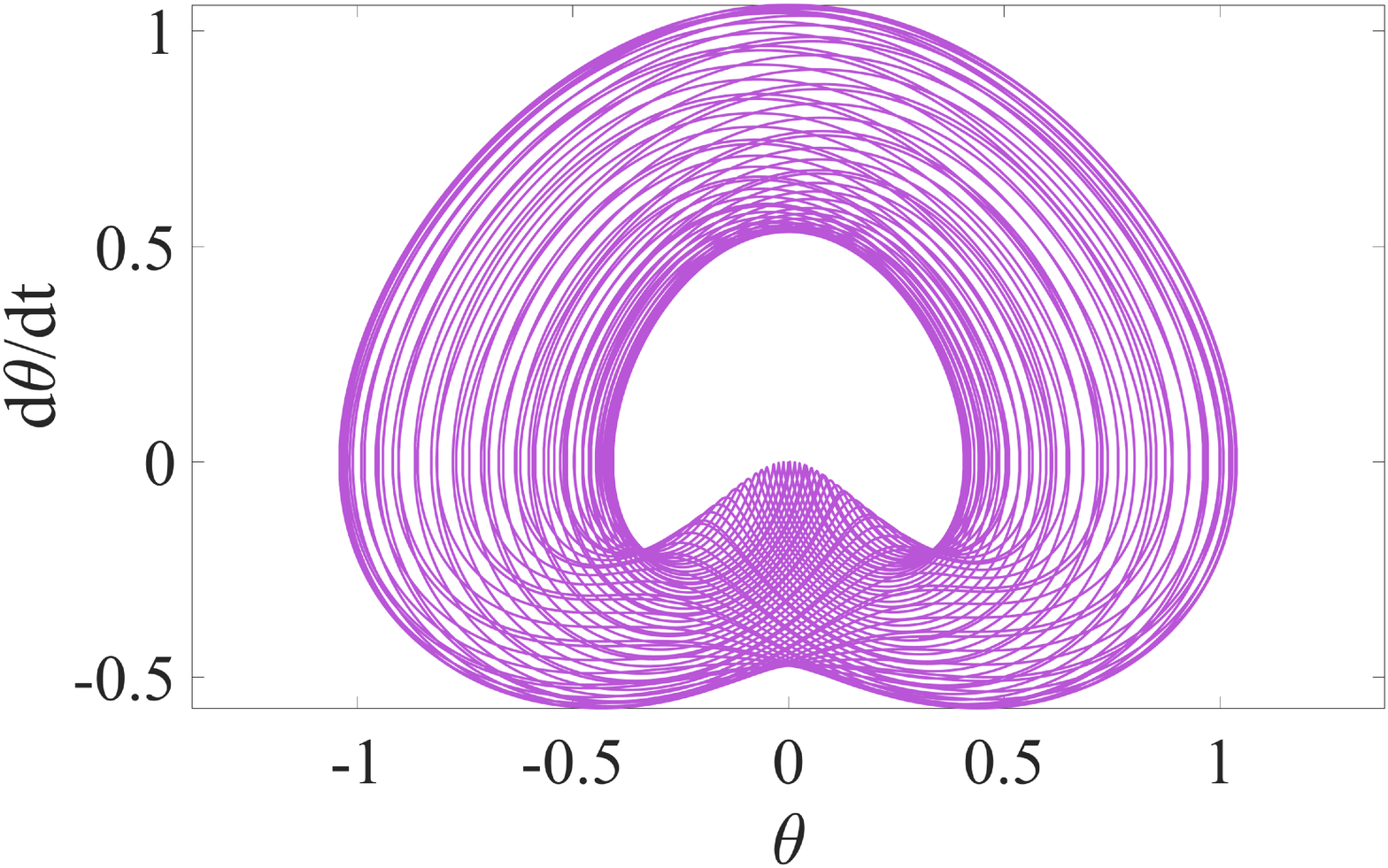} %---c = 0.8
}
\qquad
\subfloat[\footnotesize $c = 0.743$]{
\includegraphics[width = 0.33\textwidth]{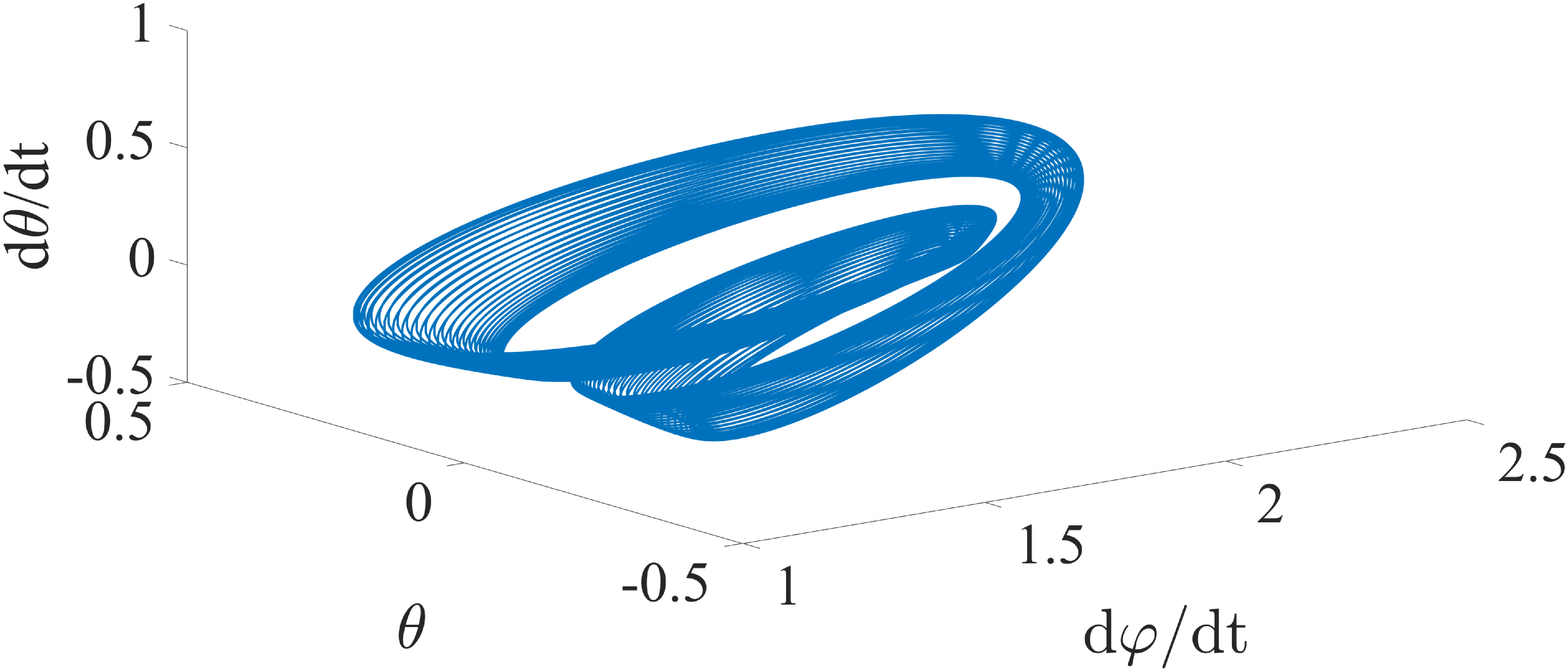} %---------3D 0.743
}
\subfloat[\footnotesize $c = 0.75$]{
\includegraphics[width = 0.33\textwidth]{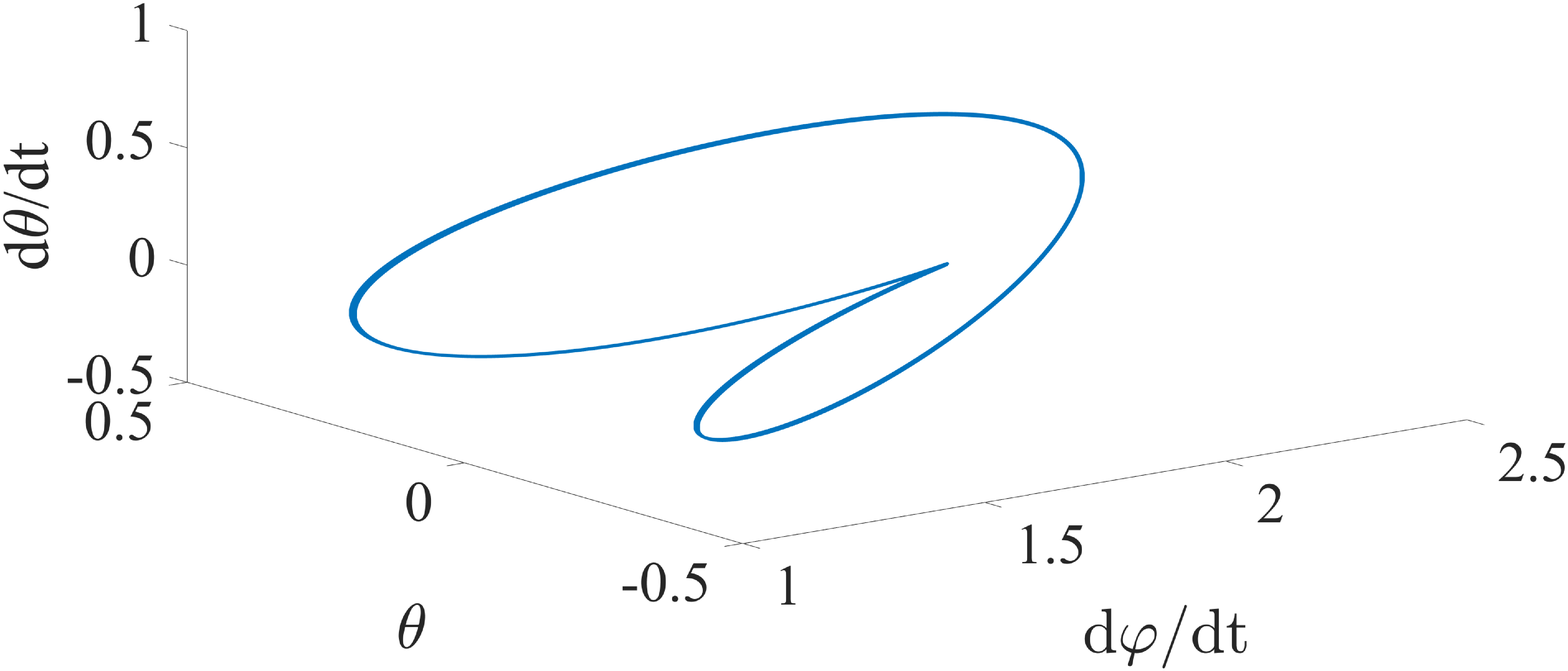} %---------3D 0.75
}
\subfloat[\footnotesize $c = 0.8$]{
\includegraphics[width = 0.33\textwidth]{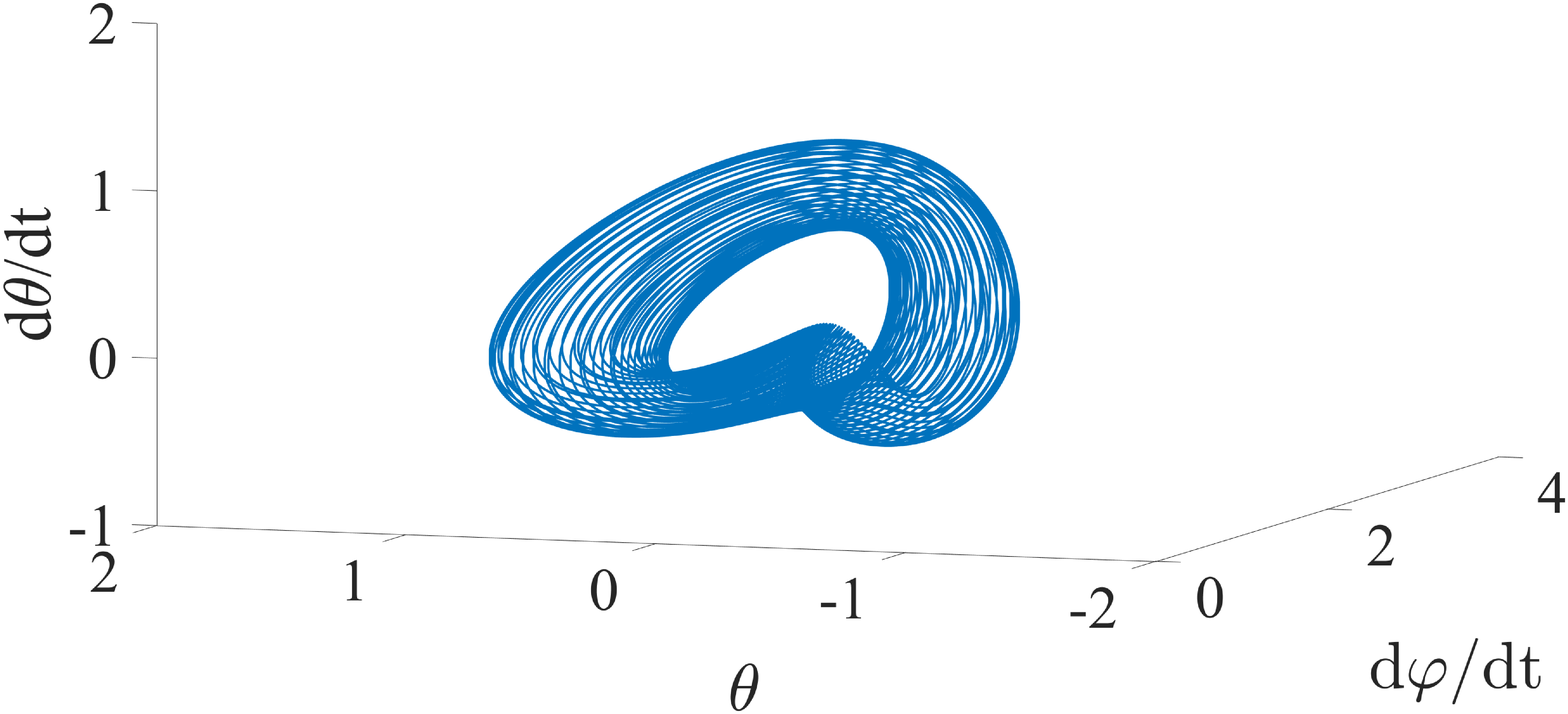} %---------3D 0.8
}
\caption{Same as Fig.1 but for medium couplings: upper row: axion phase portraits in the $(\theta , \dot{\theta})$-plane; lower row: the corresponding $(\dot{\varphi}, \theta, \dot{\theta})$-subspace ($t_{max} = 500$)} %with increasing scales
\label{c0pt75s}
\end{figure}

%Intuitively, a small coupling $c$ would only perturb the motion from the standard RSJ (resistively shunted Josephson junction) model, and this small coupling would excite the axion to have oscillation with small amplitudes. As the coupling $c$ increases, its amplitudes is expected to be large, and the Josephson junction would be influenced by their interaction. 
%
Figure \ref{c-weak} shows some phase space trajectories of the axion for small values of the coupling parameter $c$. The phase space portraits for the axion undergo a bifurcation-like process which we may call an {\em eversion process}\footnote{The word {\em eversion} means `turning inside out', which is borrowed from differential topology. If we regard the trajectory as a curve lying on the surface of some three-dimensional manifold, then by varying the parameter the inside of the surface is turned outside smoothly and continuously.}. 
%see this movie of such an explicit eversion: https://www.youtube.com/watch?v=wO61D9x6lNY (Outside In)
This happens, for example, when we vary $c$
from $2\times 10^{-6}$ to $6\times 10^{-6}$ (see figs.\ref{c2e-6}, \ref{c3pt85e-6} and \ref{c6e-6}). A similar process also happens when $c$ varies
in the region $c = 1\times 10^{-3}\sim 4\times 10^{-3}$ (figs.\ref{c1e-3}, \ref{ccrit} and \ref{c4e-3}): a `pretzel-shaped' trajectory is deformed into a simple `cardioid' and then becomes an `inside-out pretzel'. As the phase $\varphi$ of the Josephson junction is unbounded (i.e., $\varphi$ is monotonically increasing with time), the whole four-dimensional phase trajectory is projected onto the $(\dot{\varphi}, \theta, \dot{\theta})$-subspace to illustrate the relation among the three bounded variables, see the last row in fig.\ref{c-weak}. The cusp appearing in fig.\ref{3Dccrit} (or \ref{ccrit}) corresponds to the largest angular velocity of the Josephson junction, at which $\max \dot{\varphi} = \dot{\varphi}(0) = 2$. The eversion occurs again at $c = 0.74 \sim 0.84$ (fig.\ref{c0pt75s}), but the cusp in fig.\ref{c0pt75} does not correspond to $\max \dot{\varphi}$: the Josephson junction gains energy from the axion via a medium coupling so that the largest angular velocity exceeds the initial one. 

\section{Dependence on the frequency ratio}

As already mentioned, future detectors for galactic axions passing through the Earth may be based on Josephson junctions or
arrays of Josephson junctions \cite{beck2,beck3,beck4, ex2}
and in such a setting one expects resonance effects if the axion mass coincides with the plasma frequency of the junctions.
Recall that the two parameters $b_{1}$ and $b_{2}$ are given by the square of the plasma frequency of the Josephson junction and the mass of the axion, respectively, and as outlined in \cite{beck1} these parameters have similar order of magnitude.
 At a resonance point one has $b_1/b_2=1$, but we now want to explore what happens if the ratio is just close to 1.

We are interested in the dependence on the frequency ratio $b_{1}/b_{2}$ for the non-dissipative ($a_{1} = a_{2} = 0$) case. 
In the following numerical experiment we fix the coupling $c = 2.0545\times 10^{-3}$ (a `cardioid' for $b_{1} = b_{2} = 1$) and, since the initial conditions are chosen to be $(\varphi , \dot{\varphi}, \theta , \dot{\theta})_{t = 0} = (0, 2, 0, 0)$, we expect that the parameter $b_{1}$ has a stronger influence on the system than $b_{2}$ does. In addition, $b_{1}$ characterises the frequency of the Josephson junction, which can be easily adjusted in experiments; so we set $b_{2} = 1$ and let $b_{1}$ slightly deviate from $b_{2}$, which experimentally corresponds to searching for an
axion mass resonance in a given vicinity of the plasma frequency.

\begin{figure} [H]
\centering
\subfloat[\footnotesize $b_{1}/b_{2} = 1/0.9979$]{
\includegraphics[width = 0.33\textwidth]{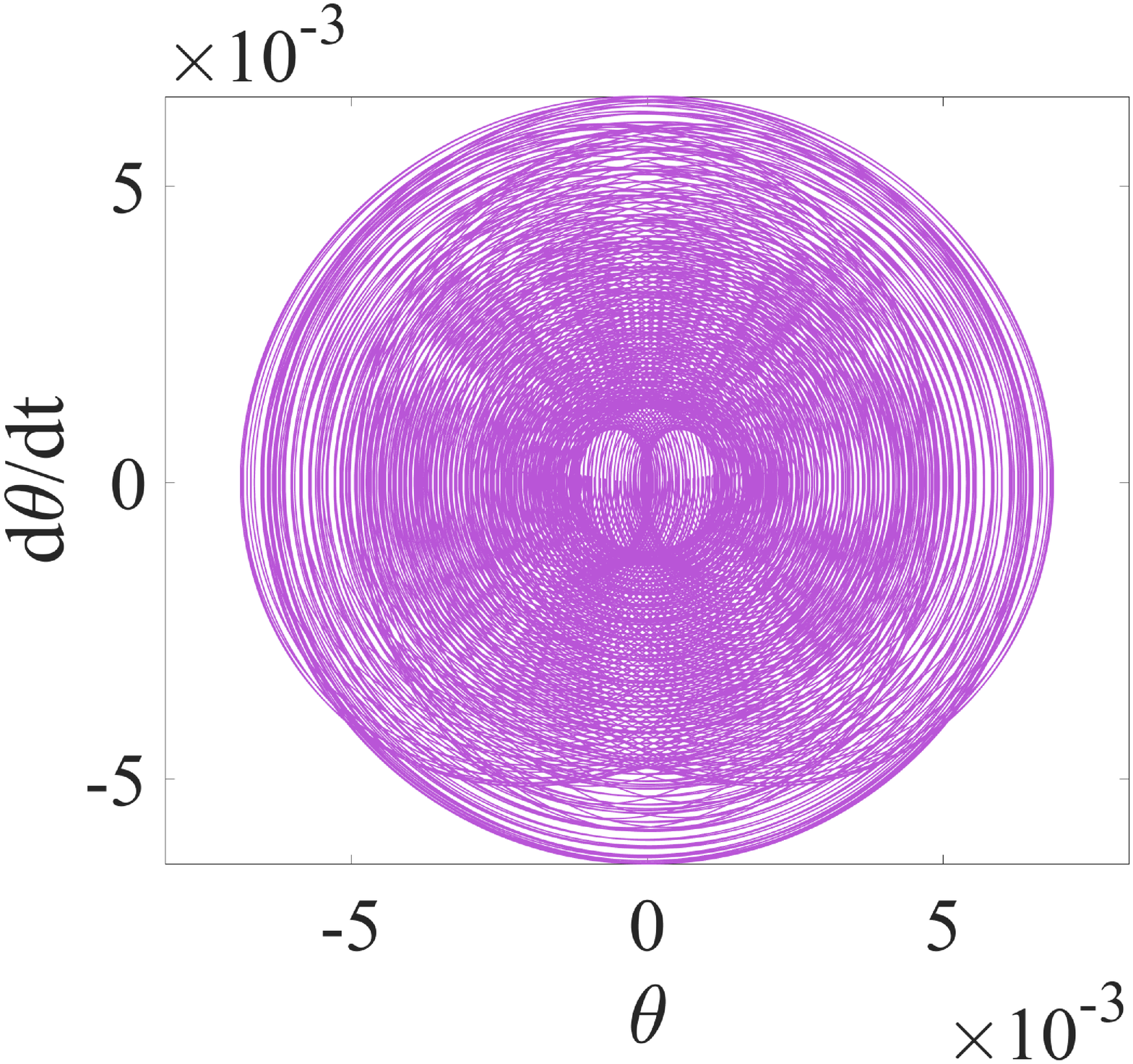} %---b1 = 1/0.9979
}
\subfloat[\footnotesize $b_{1}/b_{2} = 1/0.9978$]{
\includegraphics[width = 0.33\textwidth]{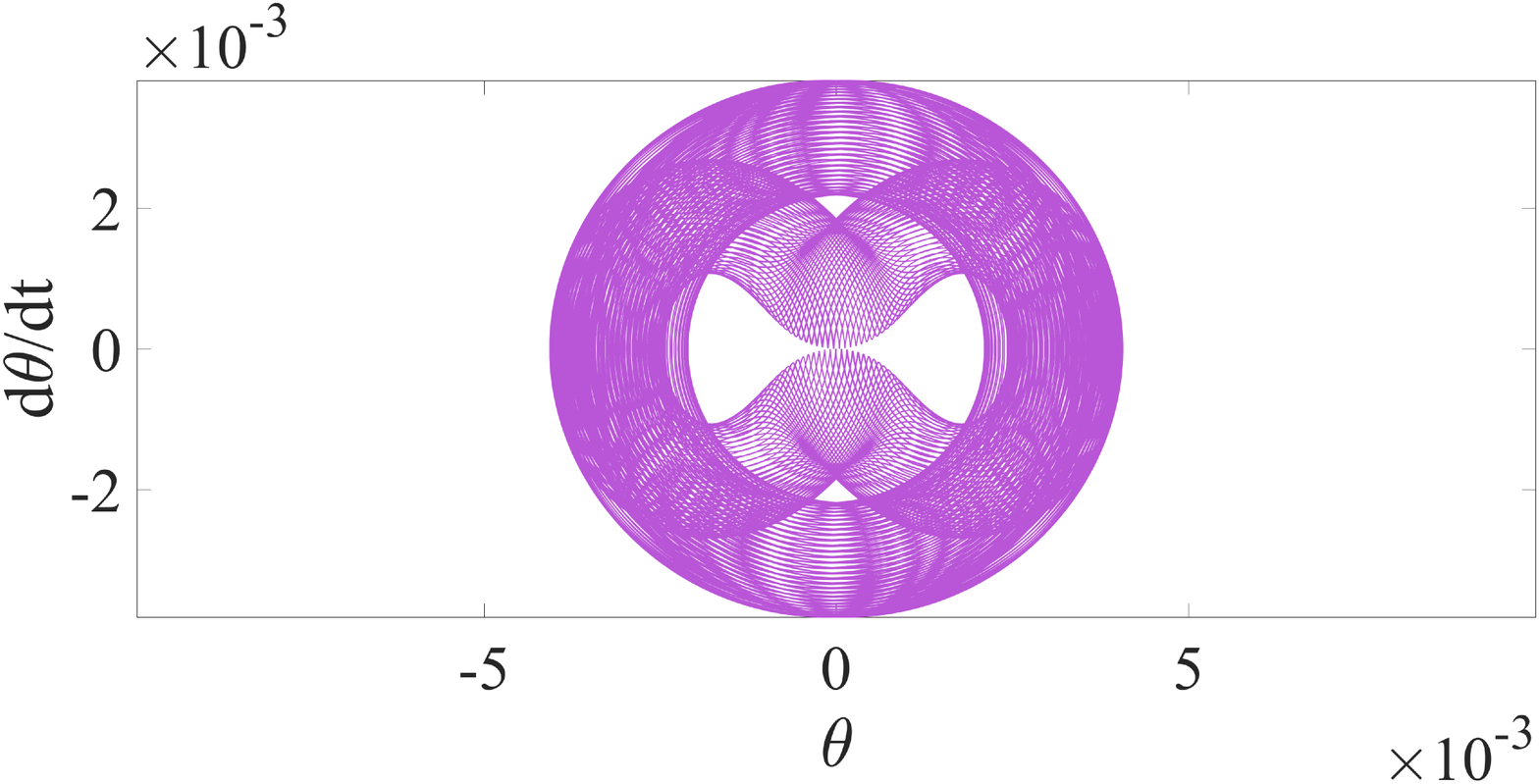} %---b1 = 1/0.9978
}
\subfloat[\footnotesize $b_{1}/b_{2} = 1/0.9975$]{
\includegraphics[width = 0.33\textwidth]{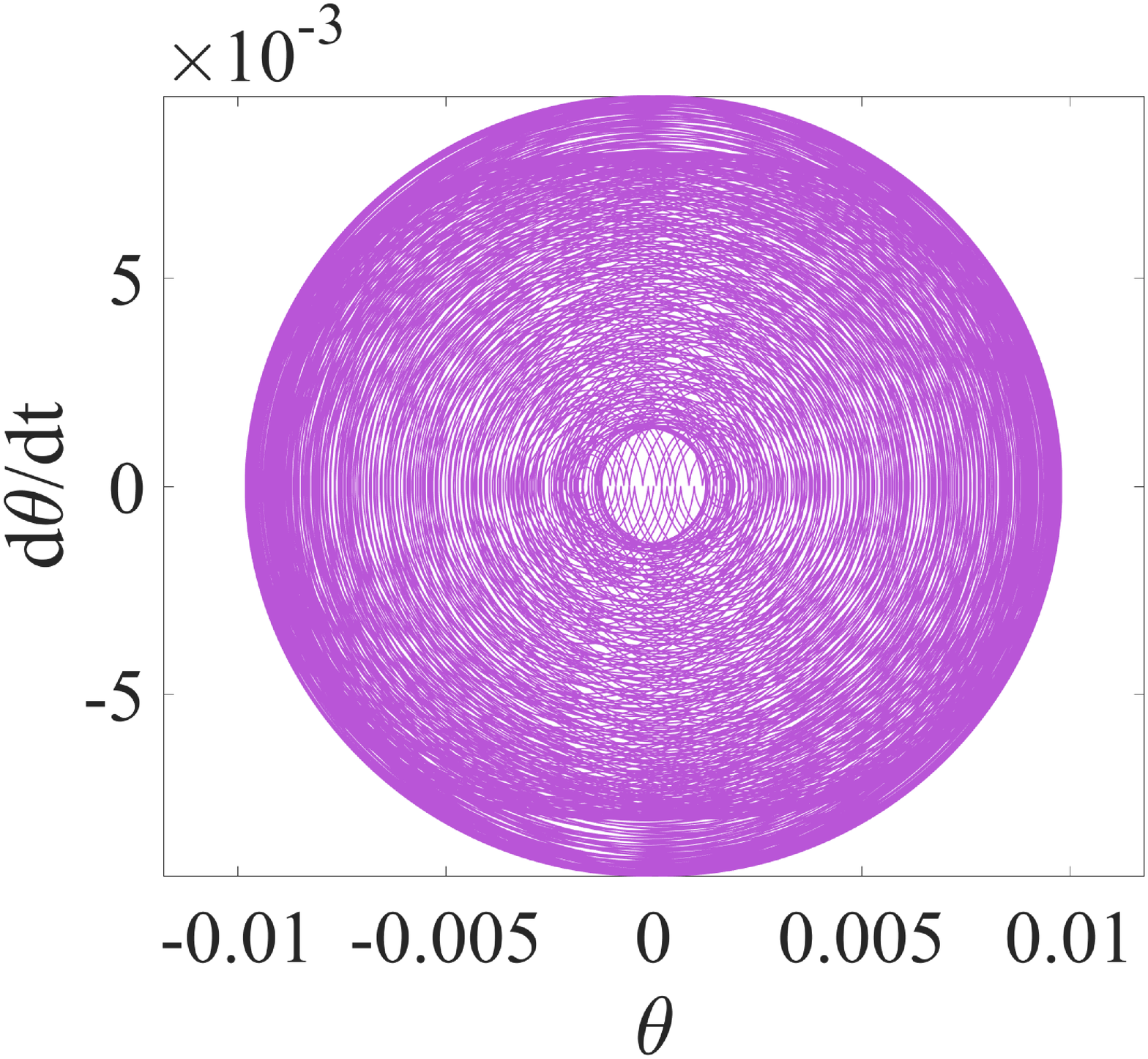} %---b1 = 1/0.9975
}
\qquad
\subfloat[\footnotesize $1/0.9979$]
{
\includegraphics[width = 0.33\textwidth]{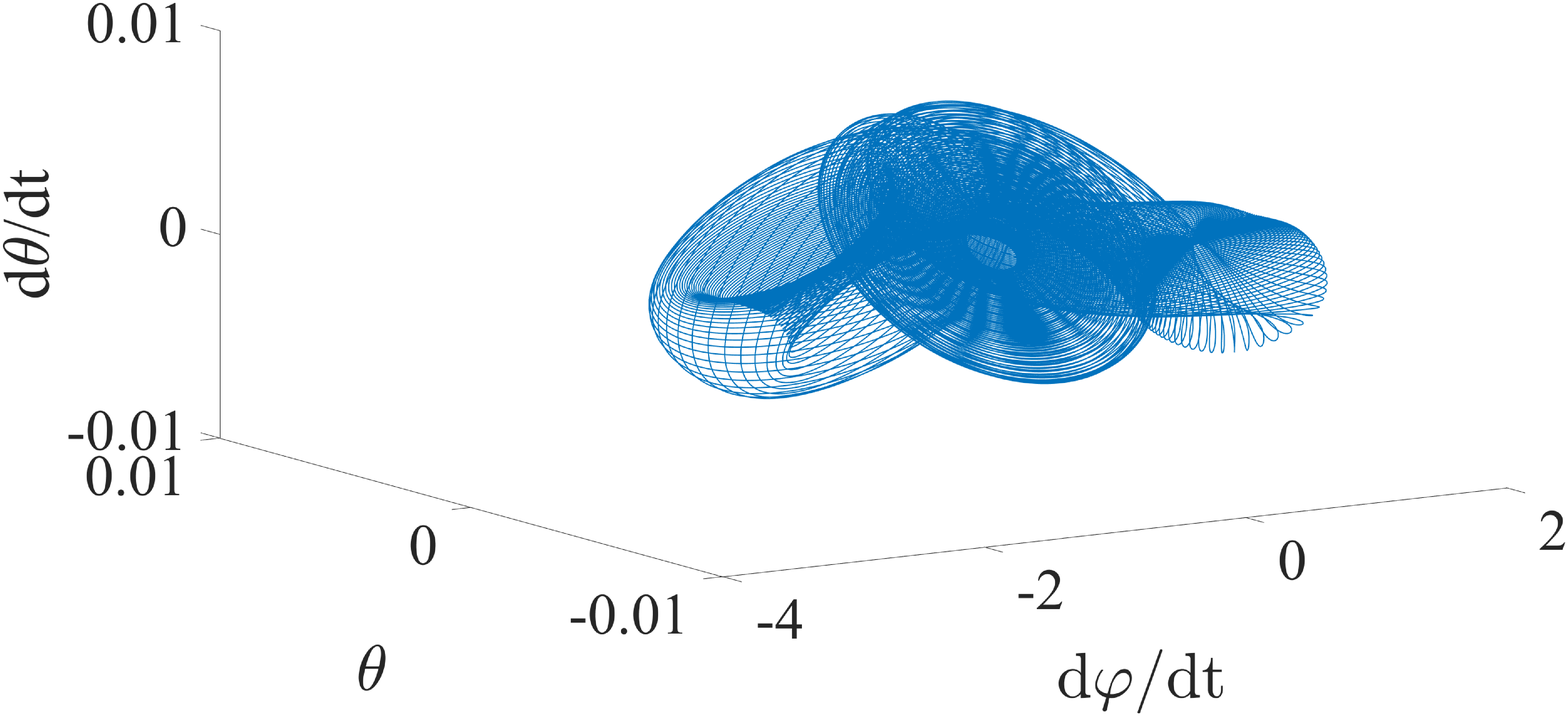} %---3D blue (*3)
}
\subfloat[\footnotesize $1/0.9978$]
{
\includegraphics[width = 0.33\textwidth]{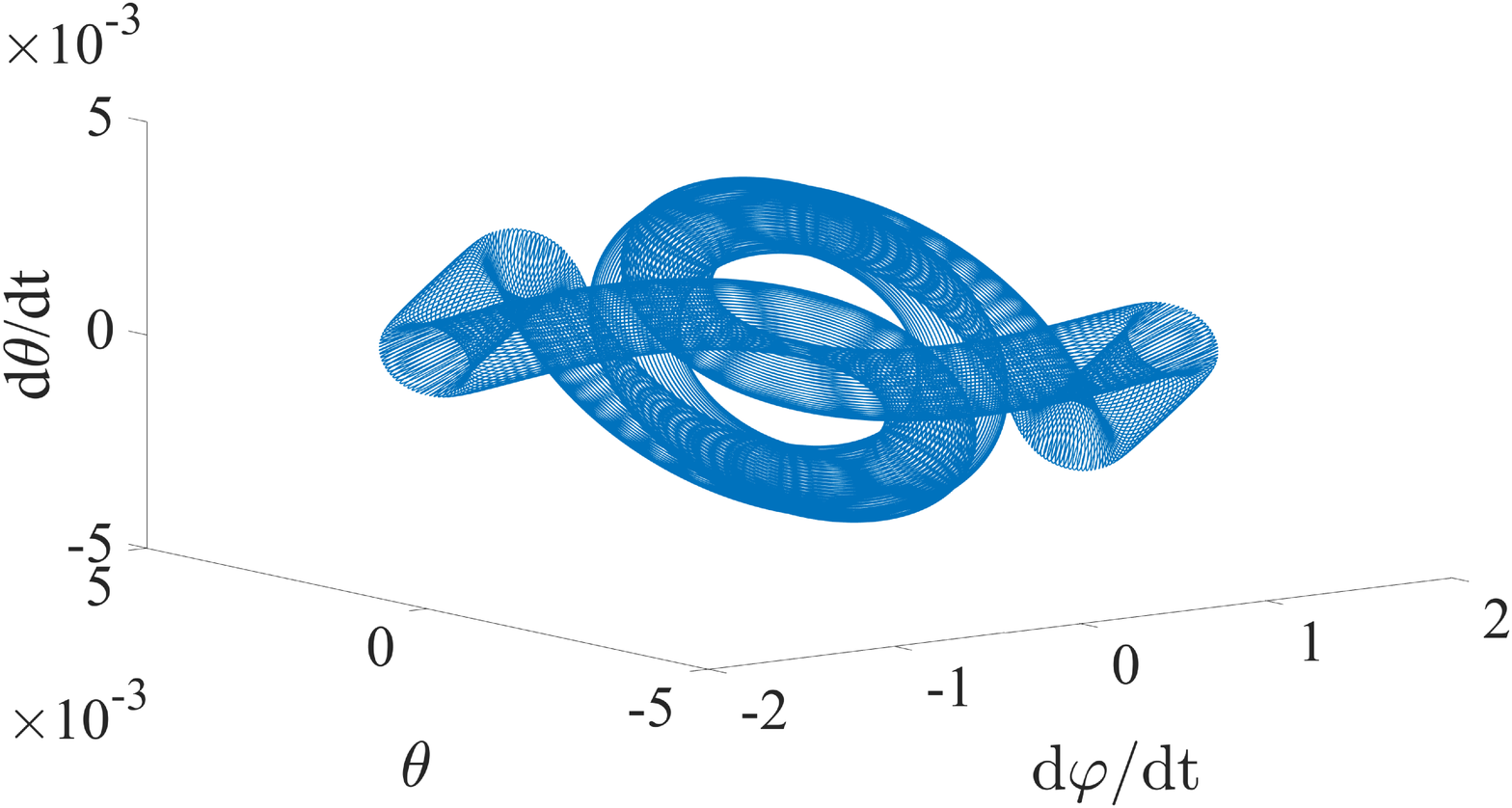} 
}
\subfloat[\footnotesize $1/0.9975$]
{
\includegraphics[width = 0.33\textwidth]{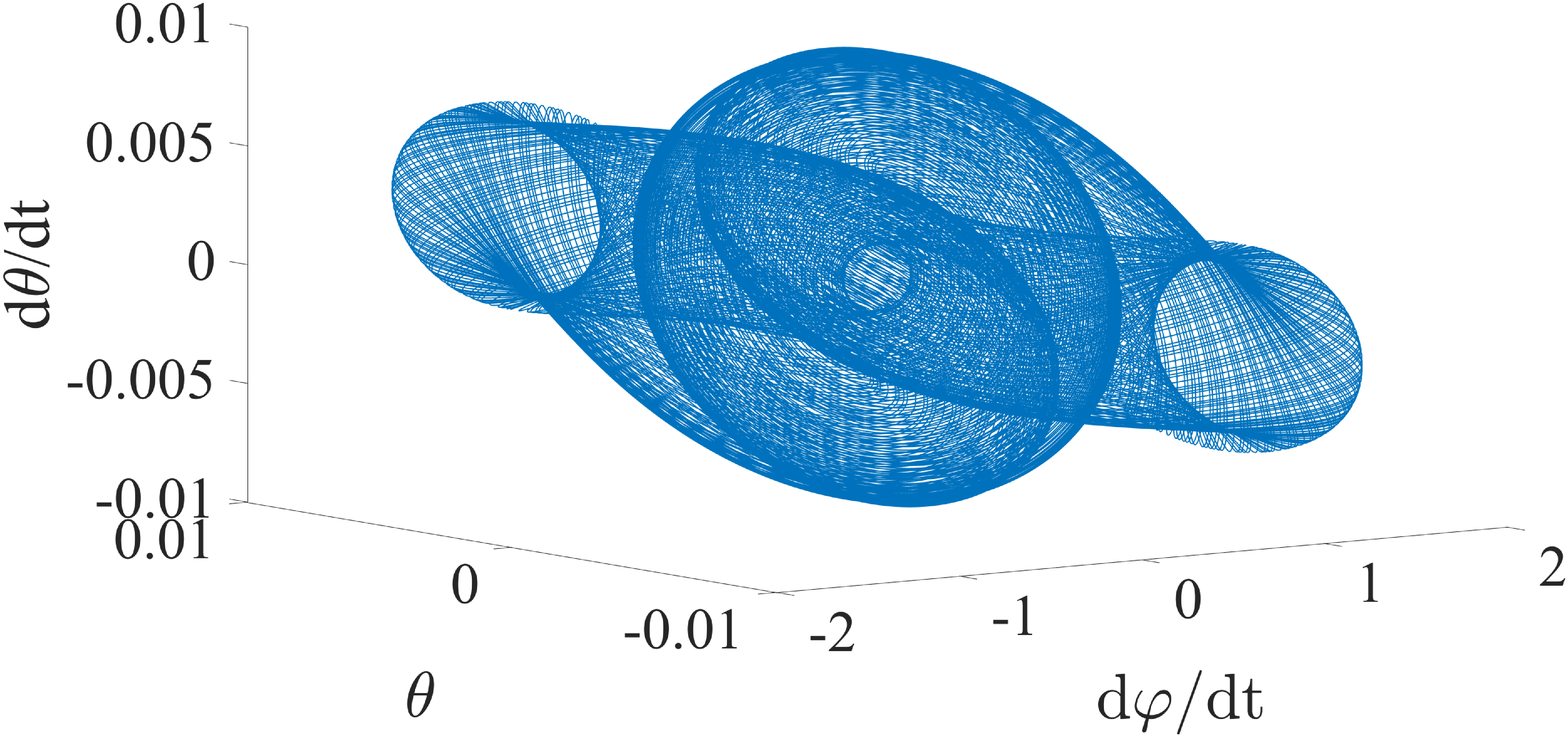} 
}
\qquad
\subfloat[\footnotesize $1/0.9979$]
{
\includegraphics[width = 0.33\textwidth]{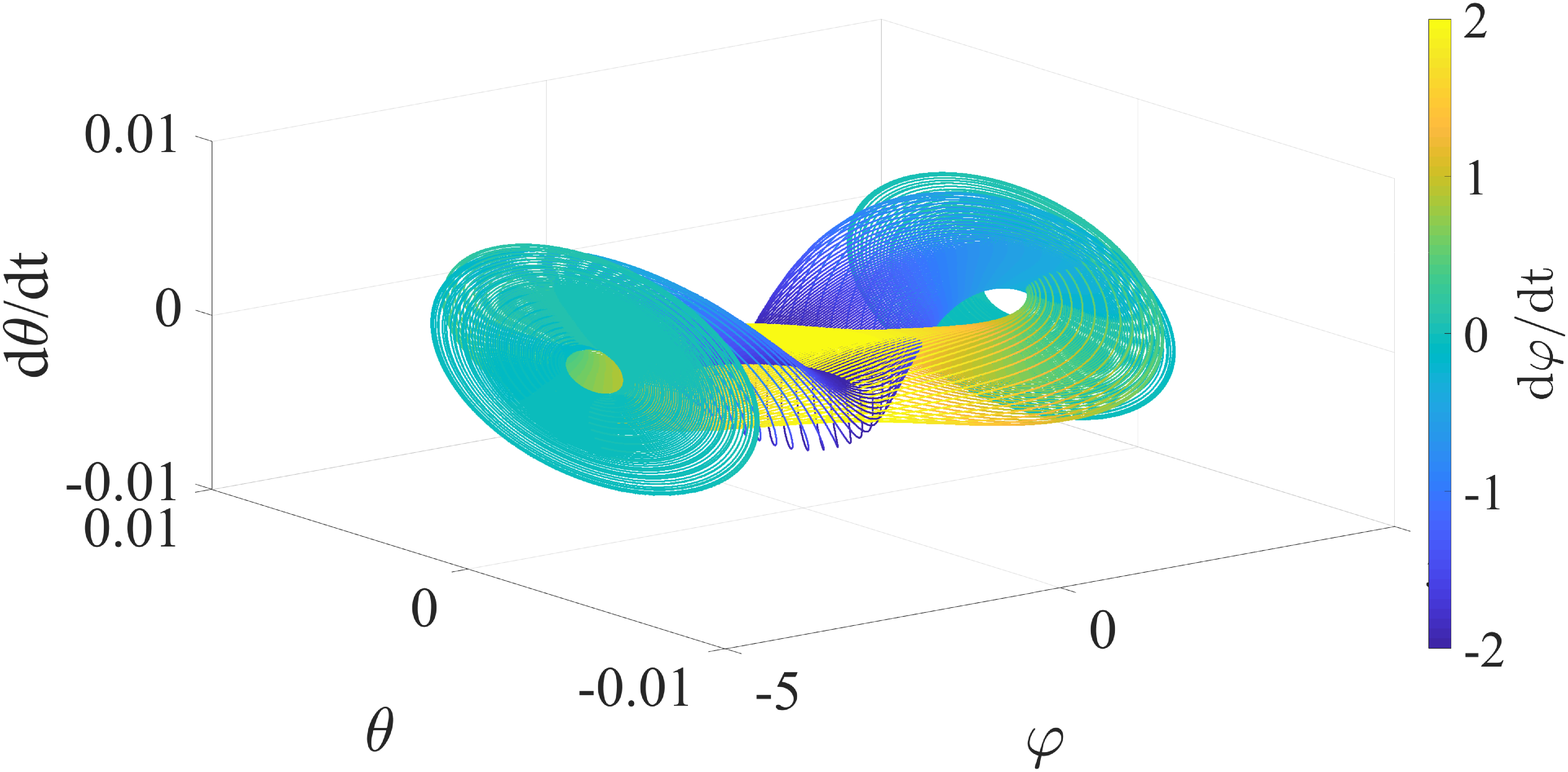} 
}
\subfloat[\footnotesize $1/0.9978$]
{
\includegraphics[width = 0.33\textwidth]{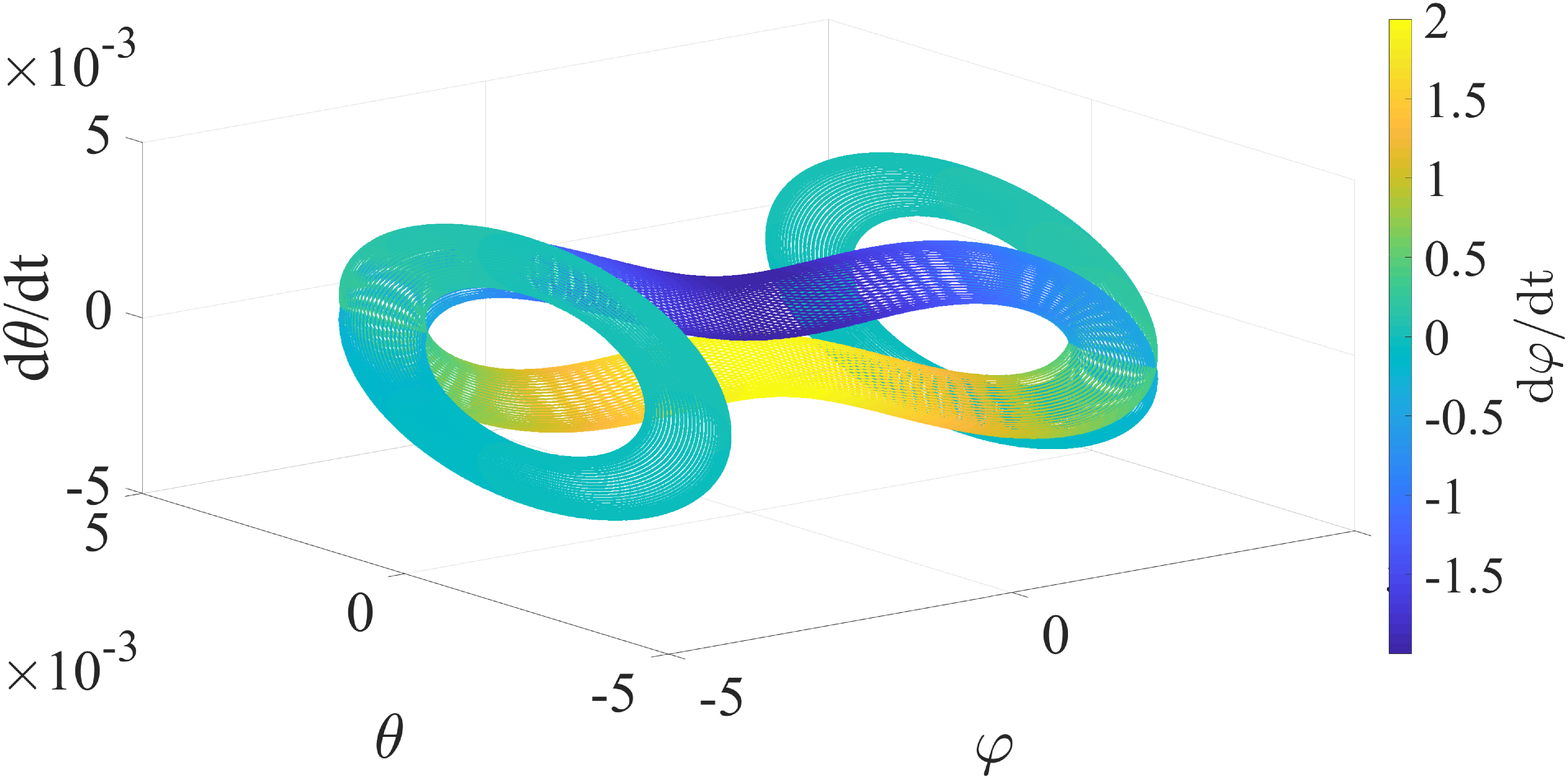} 
}
\subfloat[\footnotesize $1/0.9975$]
{
\includegraphics[width = 0.33\textwidth]{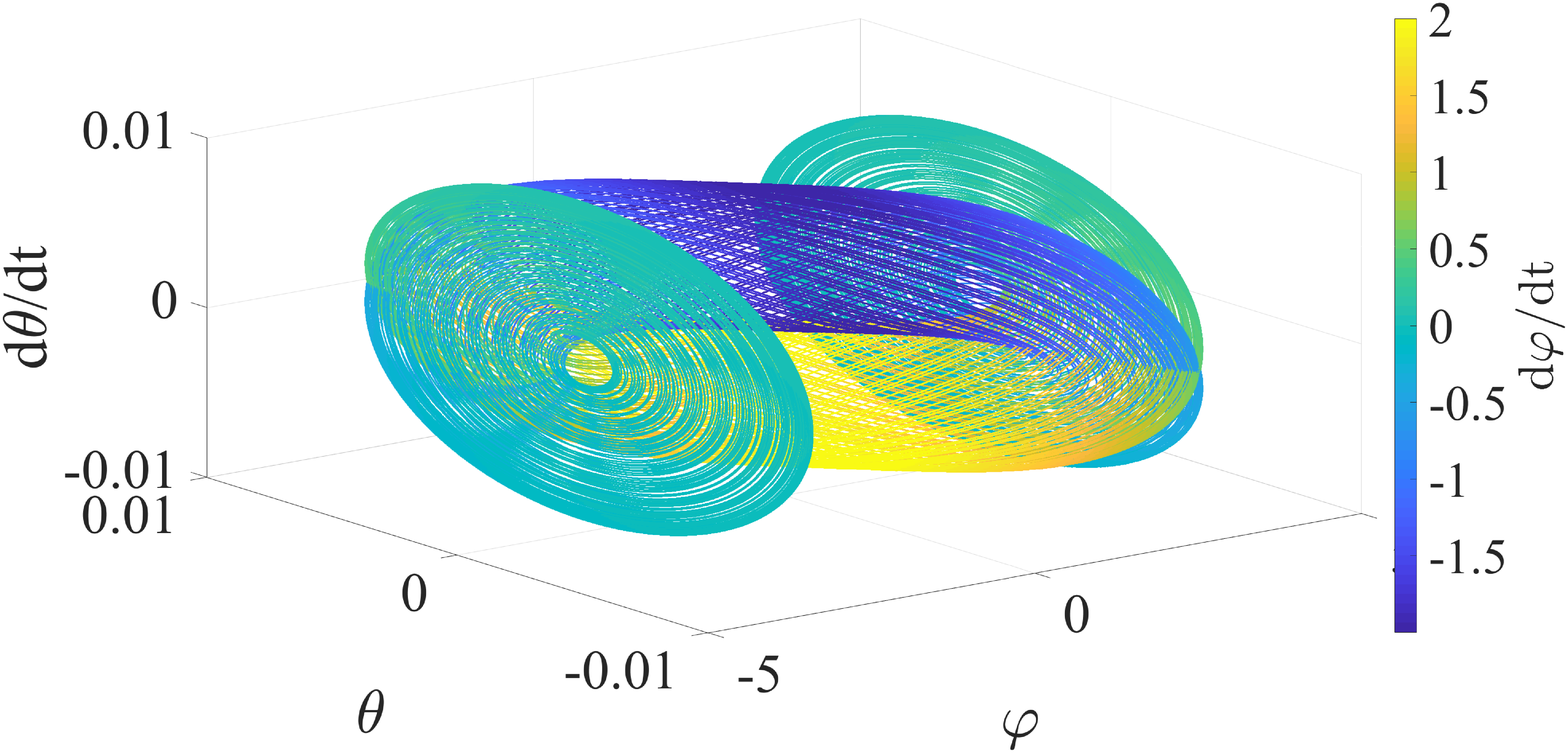} 
}
\qquad
\subfloat[\footnotesize $1/0.9979$: projection]
{
\includegraphics[width = 0.33\textwidth]{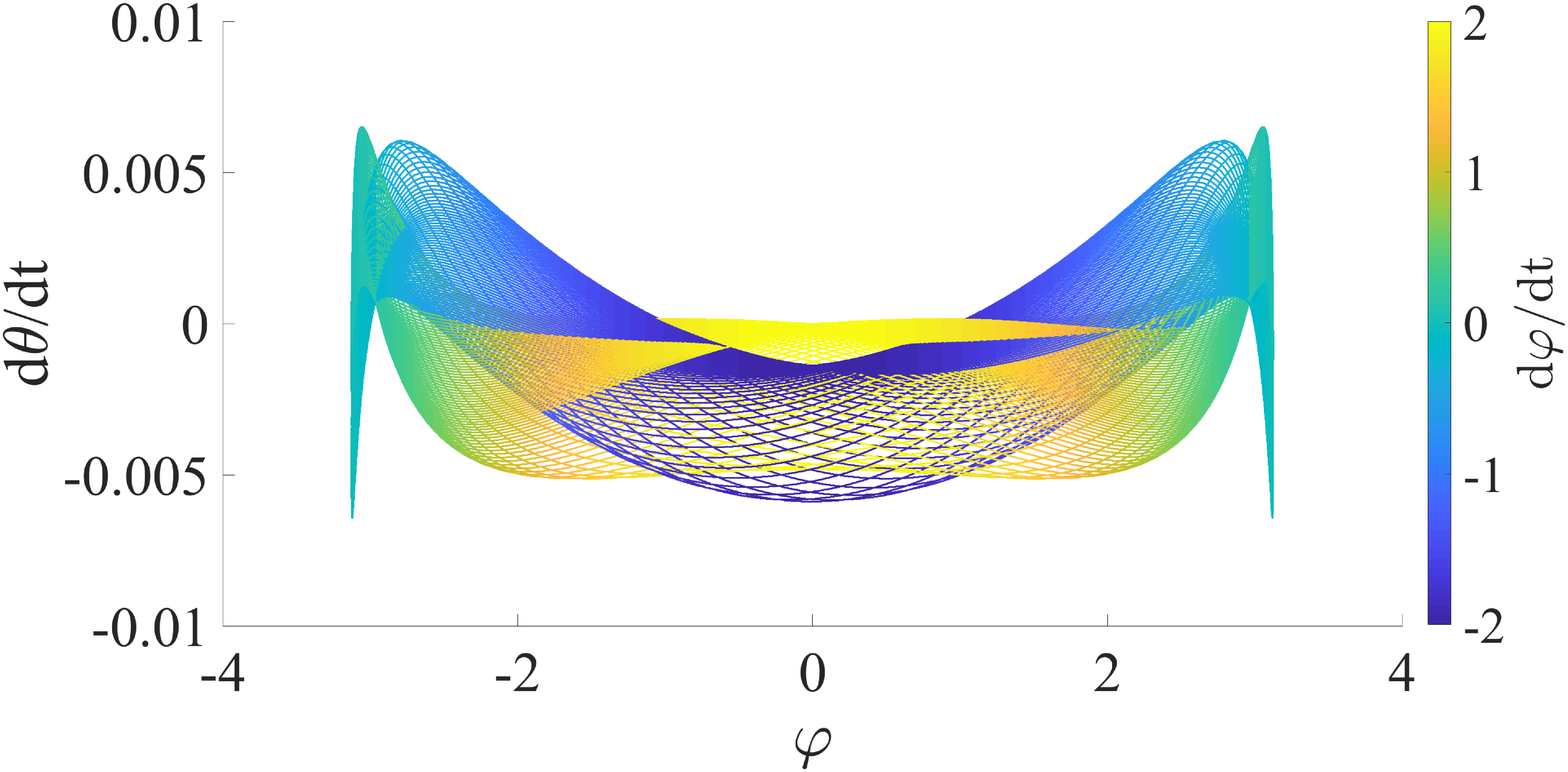} 
}
\subfloat[\footnotesize $1/0.9978$: projection]
{
\includegraphics[width = 0.33\textwidth]{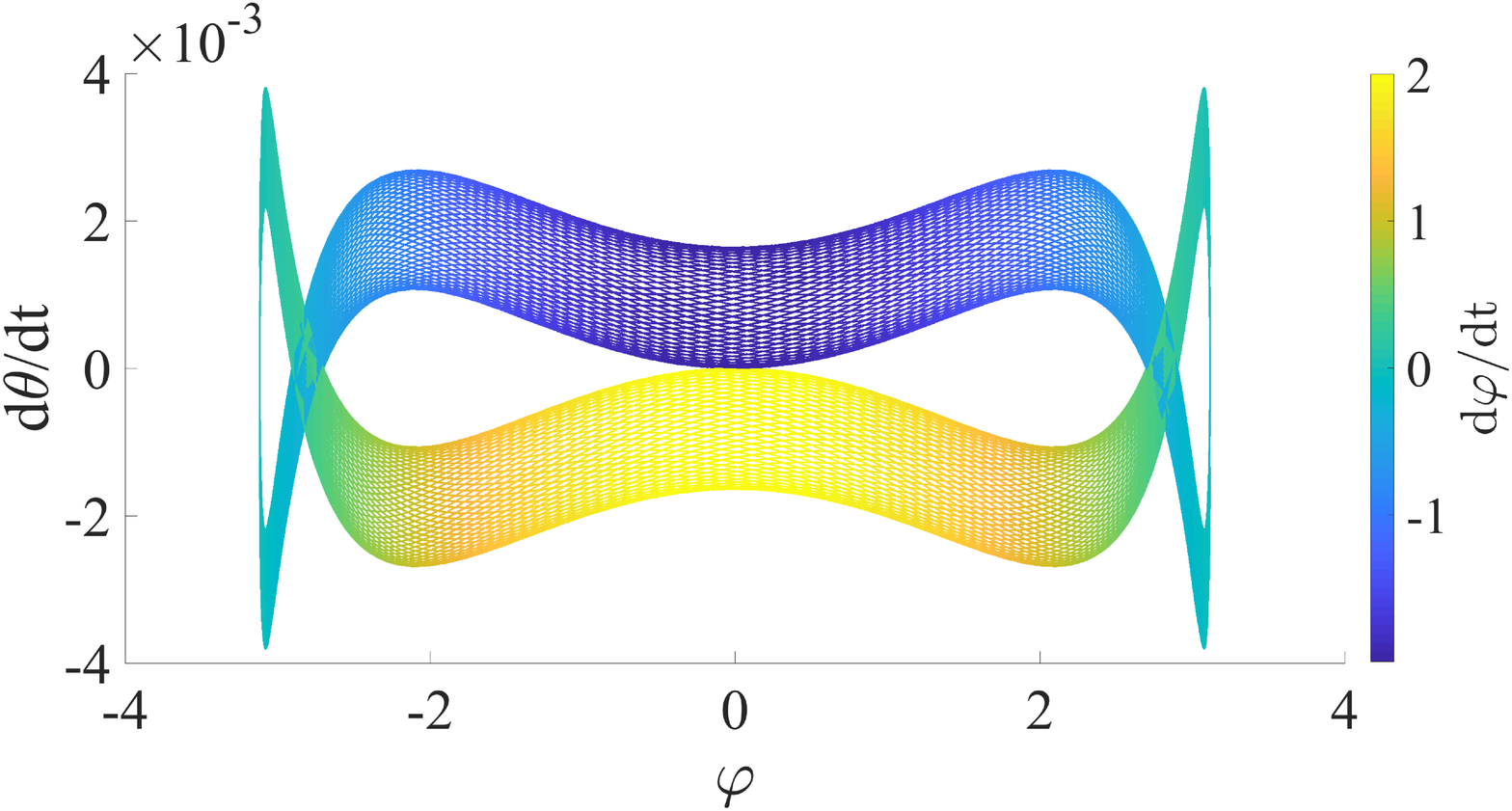} 
}
\subfloat[\footnotesize $1/0.9975$: projection]
{
\includegraphics[width = 0.33\textwidth]{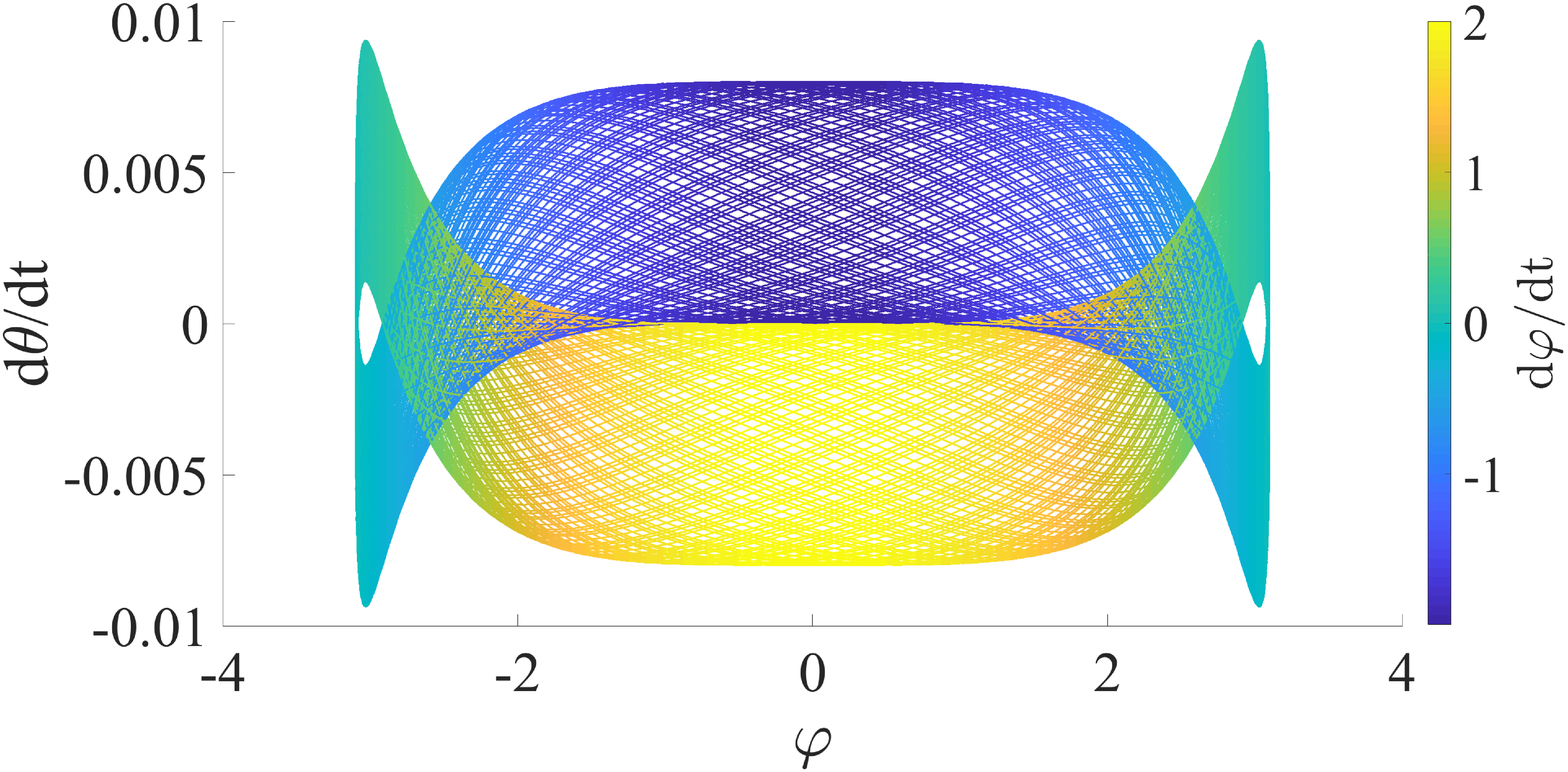} 
}
\caption{Frequency ratio dependence: First row: the axion phase portraits in the $(\theta, \dot{\theta})$-plane; second row: the $(\dot{\varphi}, \theta, \dot{\theta})$-subspace; third row: the $(\varphi, \theta, \dot{\theta})$-subspace with color coding $\dot{\varphi} \in [-2, 2]$; last row: the projection onto $(\varphi, \dot{\theta})$ with color coding $\dot{\varphi} \in [-2, 2]$ ($t_{max} = 2500$). One notices an extremely sensitive
dependence on the ratio $b_1/b_2$.} 
\label{vary-freq-ratio}
\end{figure}

Figure \ref{vary-freq-ratio} shows some phase space trajectories for different values of the frequency ratio. The first row shows the phase portraits of the axion for $b_{1} = 1/0.9979$, $1/0.9978$ and $1/0.9975$, respectively. As in this case all the four variables are bounded, we plot in the second and the third rows the projections onto the $(\dot{\varphi}, \theta, \dot{\theta})$-subspace and the $(\varphi, \theta, \dot{\theta})$-subspace, which respectively show the axion dynamics in relation to the angular velocity and the angle of the Josephson junction. In addition, the fourth variable, $\dot{\varphi}$, is indicated by color in the last two rows of the figure.

Notice that the two cyan `wheels' in the $(\varphi, \theta, \dot{\theta})$-subspace indicate that the angular velocity of the Josephson junction, $\dot{\varphi}$, is nearly zero, which corresponds to a large angle $\varphi$ (unstable top stationary point of pendulum), but conditioned on this small velocity the axion's motion is nontrivial; the energy has been transformed from the Josephson junction to the axion so that it could oscillate with a relatively large amplitude. The dark blue and light yellow bands connecting these two `wheels' are equivalent due to orientation symmetry of the rotation in the Josephson junction's phase ($\dot{\varphi} = -v$ for dark blue and $\dot{\varphi} = v$ for light yellow). As the ratio $b_{1}/b_{2}$ changes, the two bands can cross each other, indicating that whenever $\varphi \rightarrow 0$ with $\dot{\varphi} < 0$ the axion angular velocity is always negative ($\dot{\theta} < 0$), which is the opposite to the case that the two bands do not cross - their oscillations always have opposite orientations: $\dot{\theta} > 0$ when $\dot{\varphi} < 0$. This might indicate a method in the experiment to determine the frequency ratio - by looking at the direction of the  motion. 

It is interesting to notice the extreme sensitivity with respect to the parameter $b_{1}$: in the case of $b_{1} = b_{2} = 1$ the phase variable of the Josephson junction increases monotonically, indicating that it undergoes circular motion; while for $b_{1}$ slightly larger than $1$ (say, $1/0.9979 \leq b_{1}\leq 1/0.9$) it oscillates between two angles, just like a  pendulum cannot reach the highest point if the energy is not large enough to overcome the gravitational potential at the top. The phase trajectory in this case is just a simple `eye-shaped' closed curve. The three-dimensional subspace projections of the whole four-dimensional phase portrait change nontrivially, but the deformation still resembles an eversion process, as seen in the previous section. 

For the case that $b_{1}$ largely deviates from $b_{2} = 1$ (for example $b_{1} \sim 0.1 $ or $b_1\sim 2$) the topology of the phase trajectory does not change significantly under parameter changes.

\section{Dependence on the initial angular velocity}

\subsection{No dissipation}
There are only two degrees of freedom in the choices of initial conditions since we can always set the initial phases, $\varphi (0)$ and $\theta (0)$, to be zero by a simple coordinate transformation. In the absence of dissipation for both the Josephson junction and the axion ($a_{1} = a_{2} = 0$), 
for our numerical experiment we fix the weak coupling as $c = 2.0545\times 10^{-3}$, set $b_{1} = b_{2} = 1$ and vary the initial angular velocity of the Josephson junction ($\dot{\varphi}(0)$) from $1$ to $10$, while keeping the axion's initial angular velocity small, say, $\dot{\theta}(0) = 0$.
The initial angular velocity of the Josephson junction can be easily manipulated in experiments, by applying a constant voltage
across the junction. 

\begin{figure} [H]
\centering
\subfloat[\footnotesize $\dot{\varphi}(0) = 1.93$]{
\includegraphics[width = 0.33\textwidth]{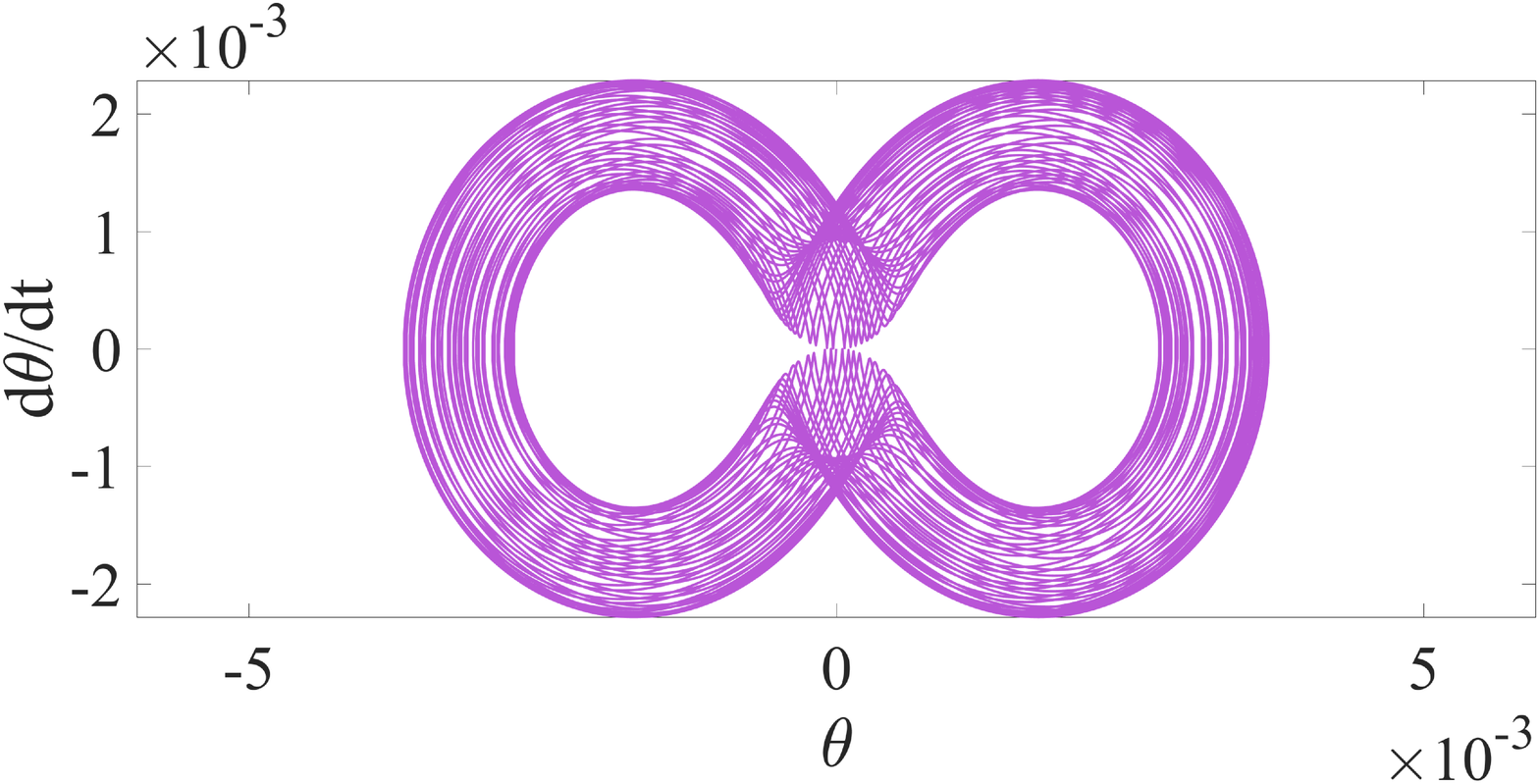} %---v1 = 1.93
}
\subfloat[\footnotesize $\dot{\varphi}(0) = 1.95$]{
\includegraphics[width = 0.33\textwidth]{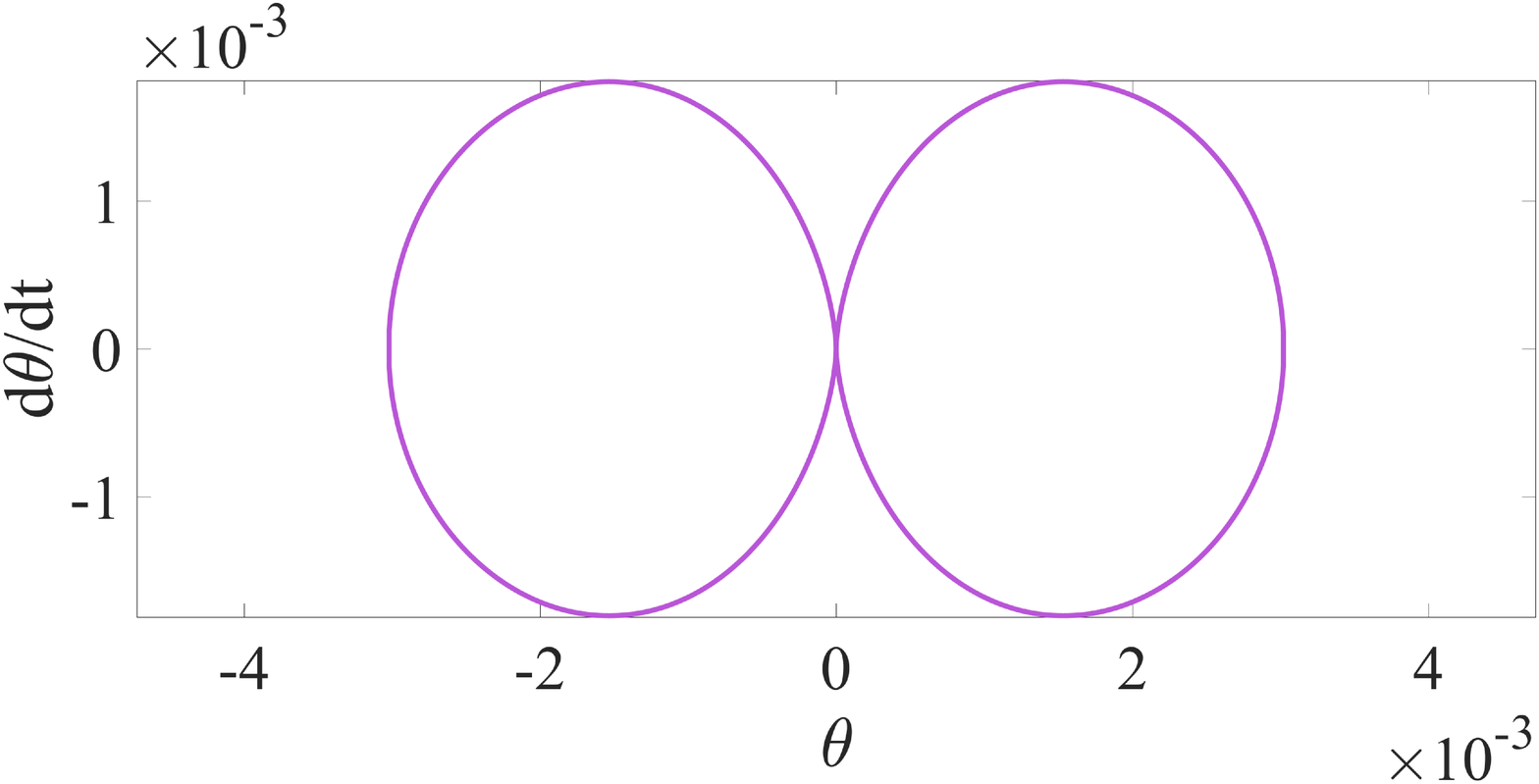} %---v1 = 1.95
}
\subfloat[\footnotesize $\dot{\varphi}(0) = 1.97$]{
\includegraphics[width = 0.33\textwidth]{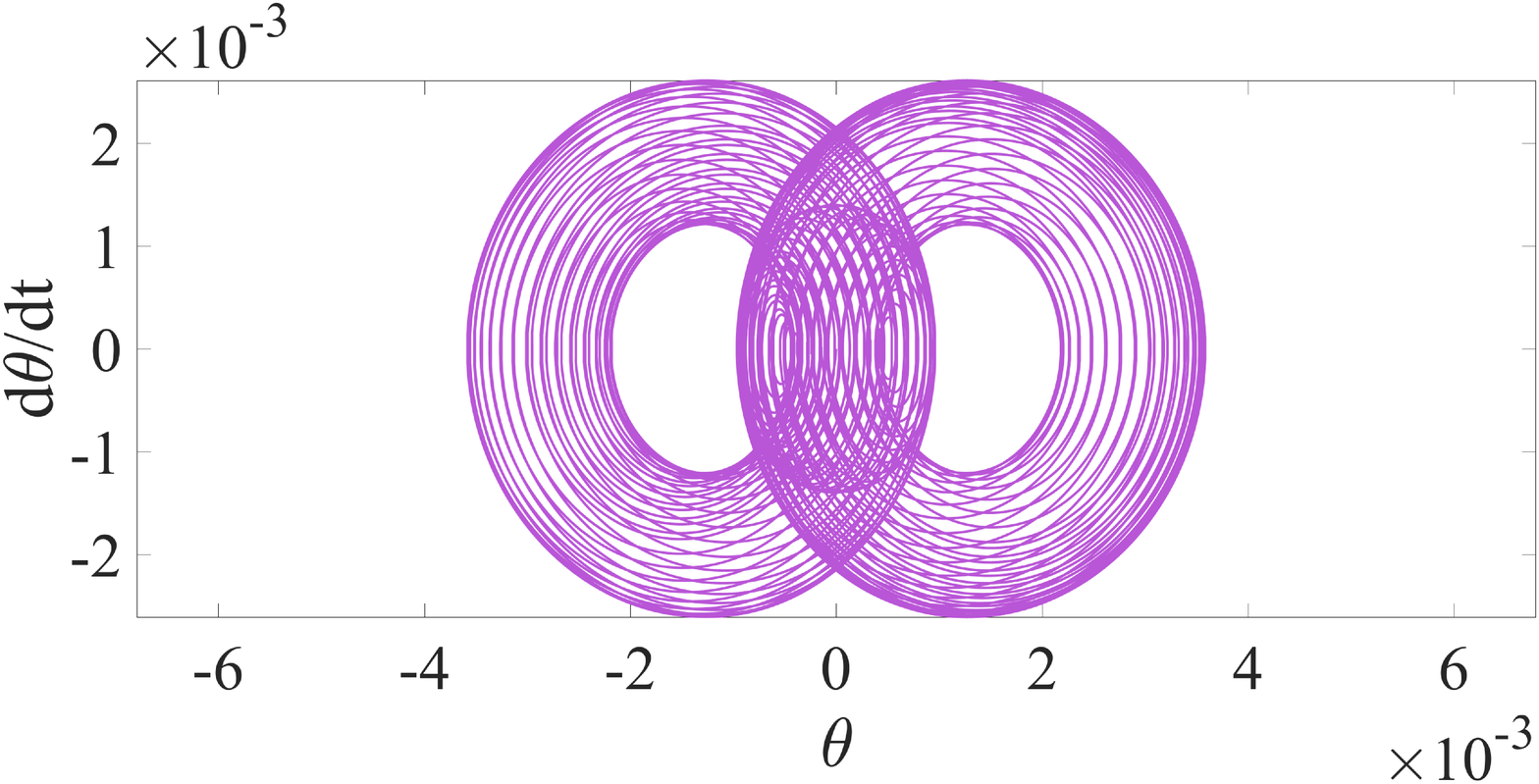} %---v1 = 1.97
}
\qquad
\subfloat[\footnotesize $\dot{\varphi}(0) = 1.93$]{
\includegraphics[width = 0.33\textwidth]{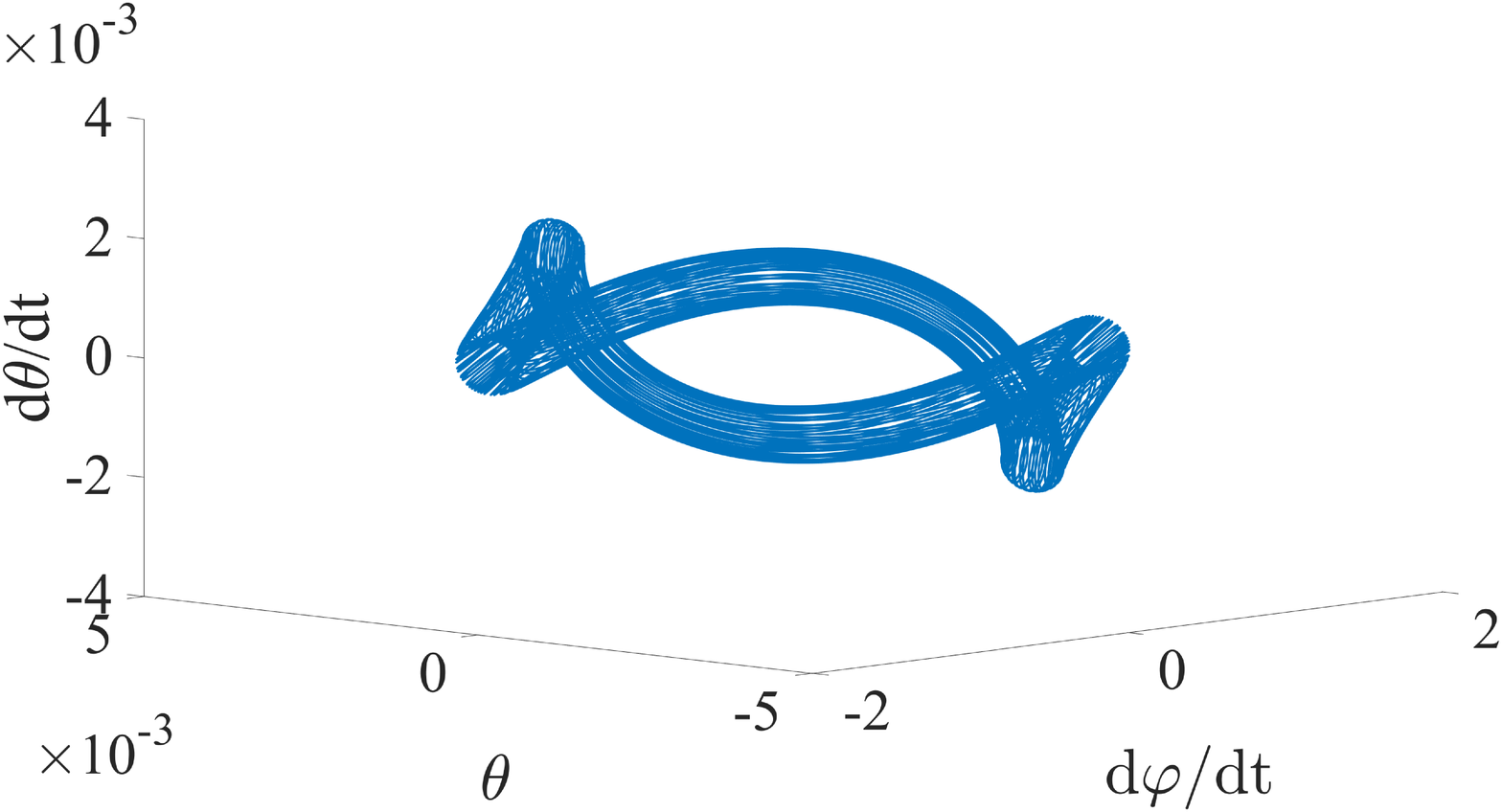} %---------3D 1.93
}
\subfloat[\footnotesize $\dot{\varphi}(0) = 1.95$]{
\includegraphics[width = 0.33\textwidth]{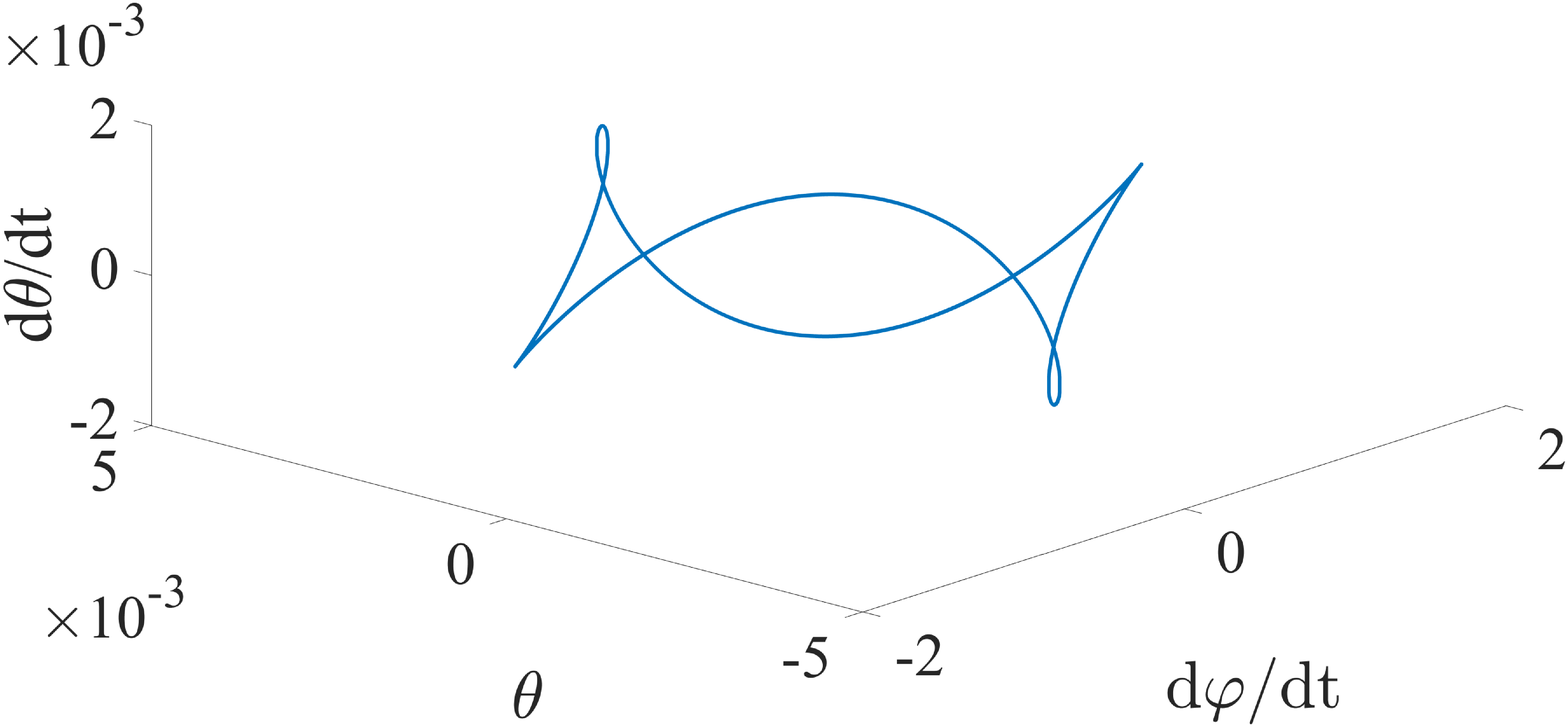} %---------3D 1.95 (crit2)
}
\subfloat[\footnotesize $\dot{\varphi}(0) = 1.97$]{
\includegraphics[width = 0.33\textwidth]{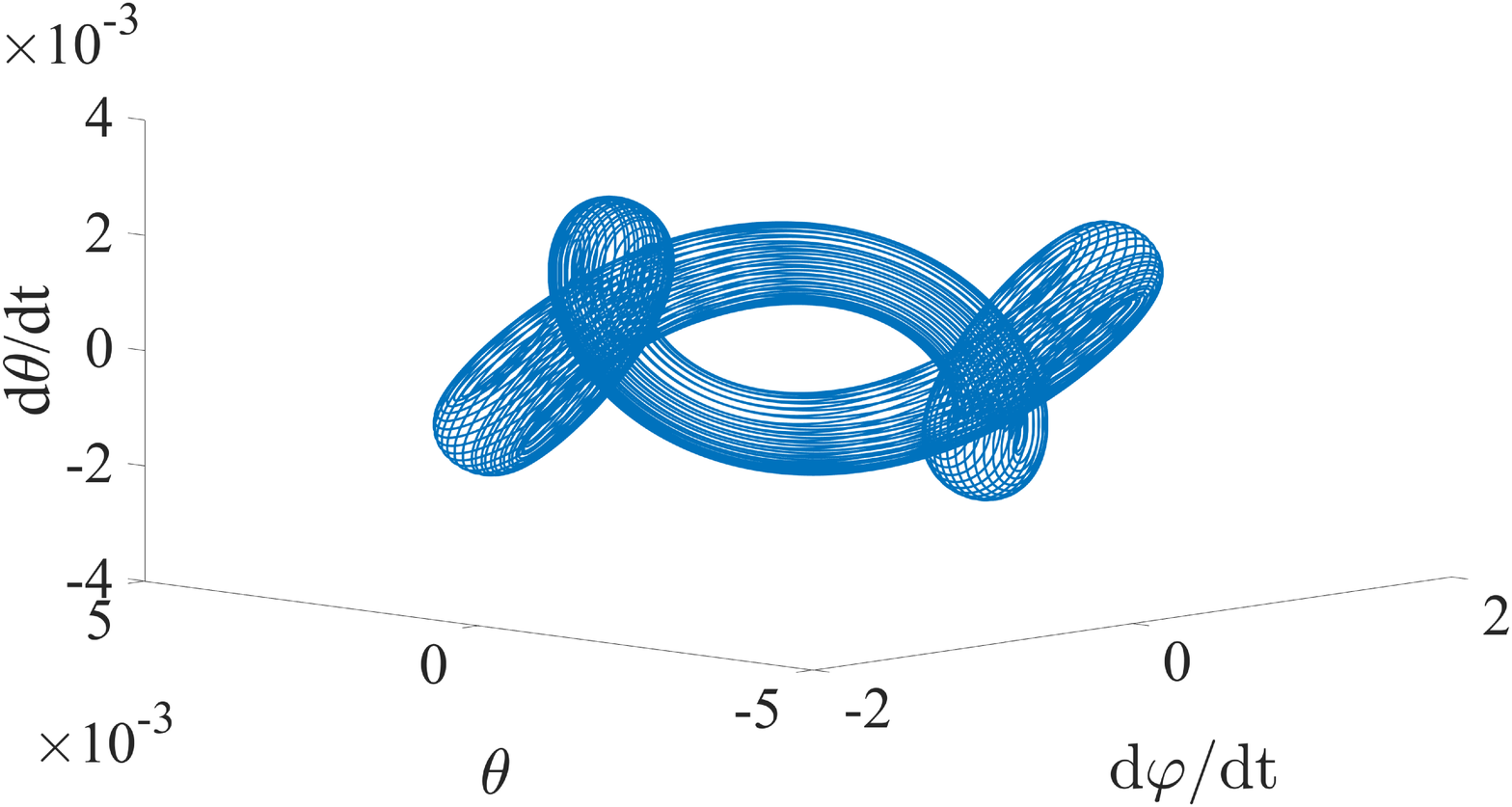} %---------3D 1.97
}
\qquad
\subfloat[\footnotesize $\dot{\varphi}(0) = 1.999$]{
\includegraphics[width = 0.33\textwidth]{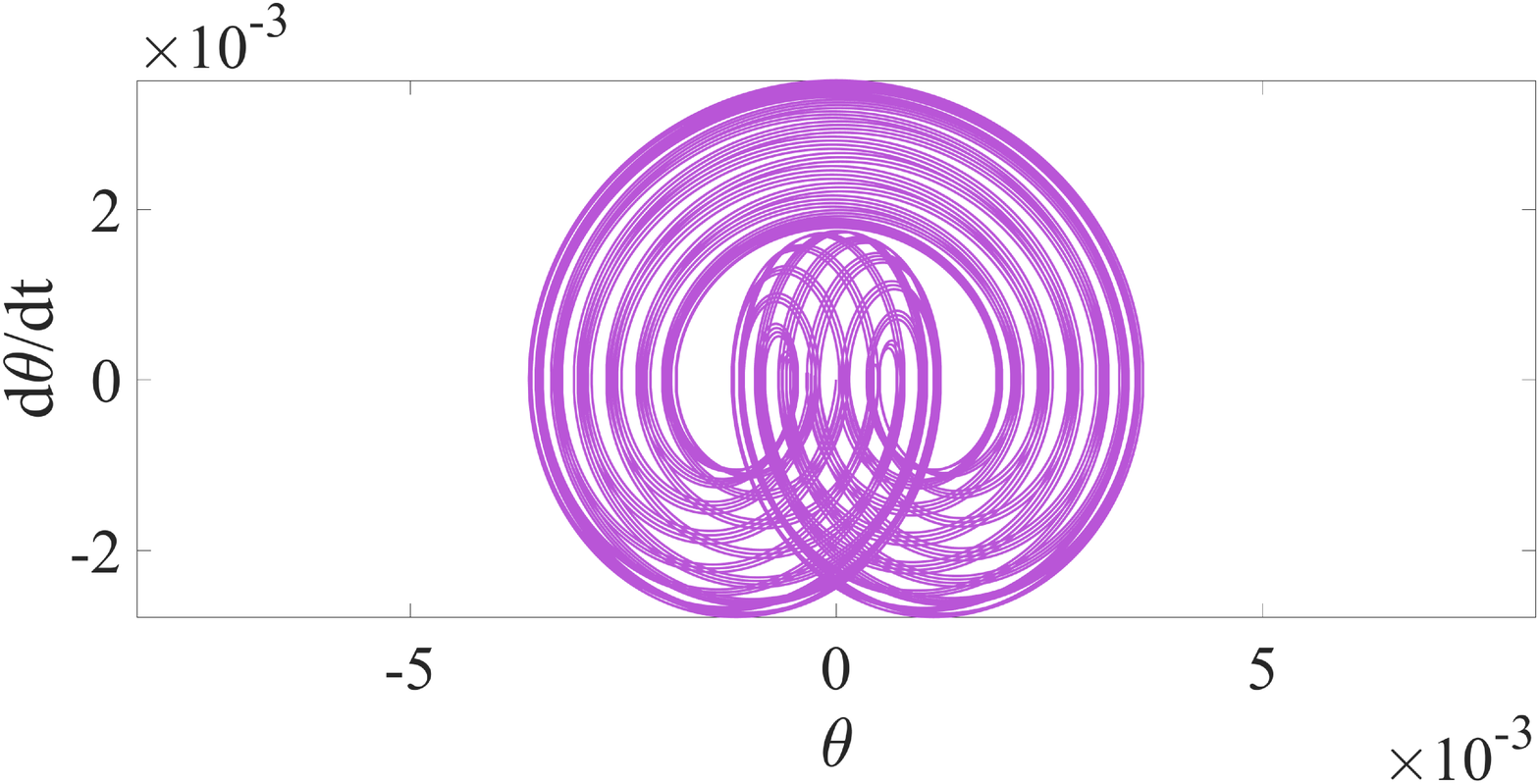} %------v1 = 1.999
}
\subfloat[\footnotesize $\dot{\varphi}(0) = 2$]{
\includegraphics[width = 0.33\textwidth]{fig1e.eps} %------v1 = 2 (crit)
}
\subfloat[\footnotesize $\dot{\varphi}(0) = 2.002$]{
\includegraphics[width = 0.33\textwidth]{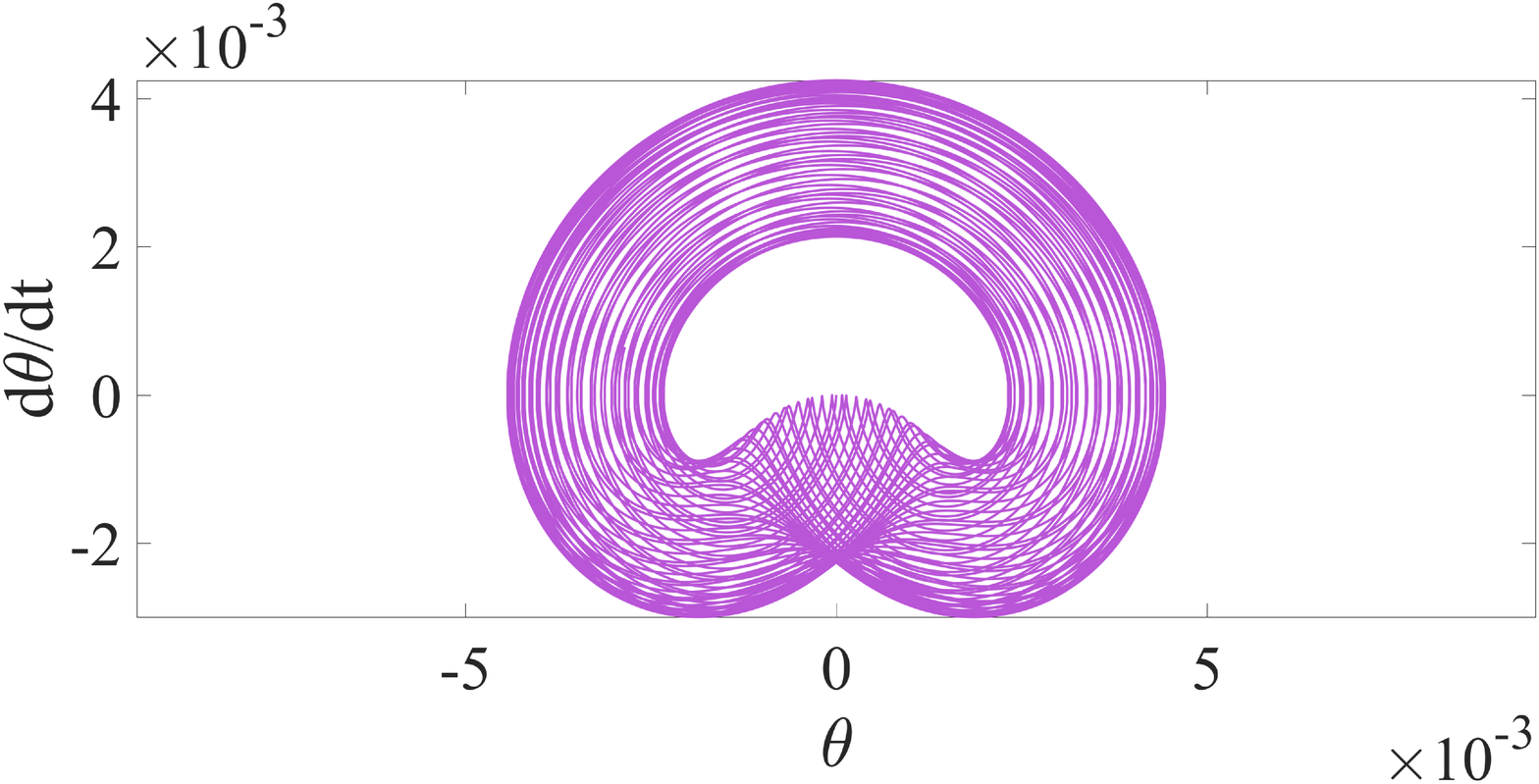} %------v1 = 2.002
}
\caption{Initial angular velocity dependence: first and third rows: axion phase portraits in the $(\theta, \dot{\theta})$-plane; second row: the corresponding $(\dot{\varphi}, \theta, \dot{\theta})$-subspace ($t_{max} = 500$). Again an eversion process is visible, but this time as
a function of the initial angular velocity $\dot{\varphi} (0)$.}
\label{v1}
\end{figure}

Interestingly, for $\dot{\varphi}(0) \approx 1.95$ we see another eversion process (first row in fig.\ref{v1}), which has not been observed in the previous sections. Note: for each of the three cases in the first row in fig.\ref{v1} the phase portrait of the Josephson junction is just a standard `eye-shaped' closed curve, while for the cases in the last row the phase variable of the Josephson junction is monotonically increasing because, in the mechanical analogue, it has enough energy to cross the top (unstable) point of the pendulum, it performs a circular motion and continues to do so. 

Comparing the last row in the above plots with the second row in fig.\ref{c-weak}, increasing the initial angular velocity of the Josephson junction around $2$ is somewhat equivalent to increasing the coupling around $2.0545\times 10^{-3}$ --- both produce to a `cardioid' deformation and an eversion phenomenon.
 
\begin{figure} [H]
\centering
\subfloat[\footnotesize $\dot{\varphi}(0) = 2.05$: time series]{
\includegraphics[width = 0.33\textwidth]{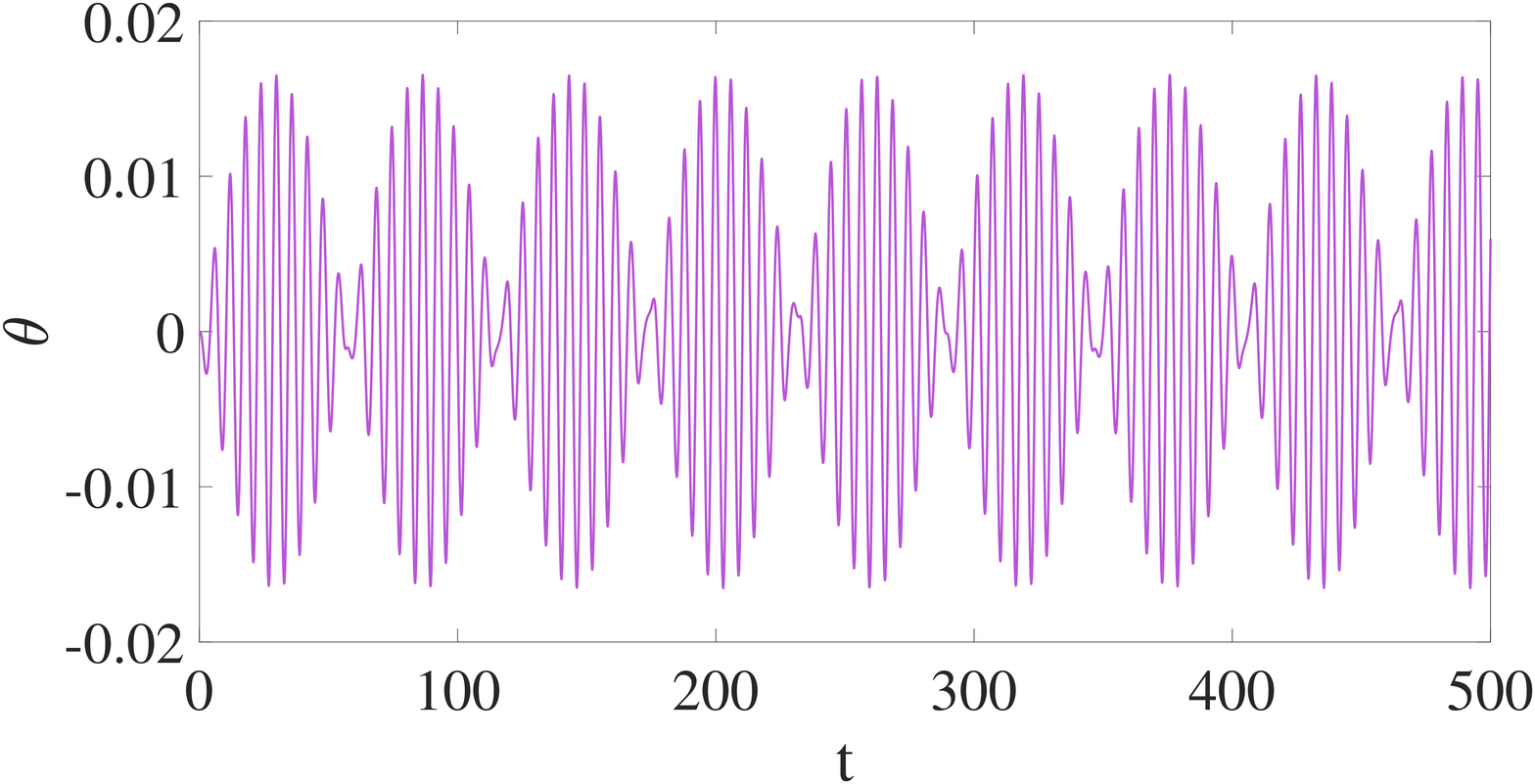} %------sol 2.05
}
\subfloat[\footnotesize $\dot{\varphi}(0) = 2.5$: time series]{
\includegraphics[width = 0.33\textwidth]{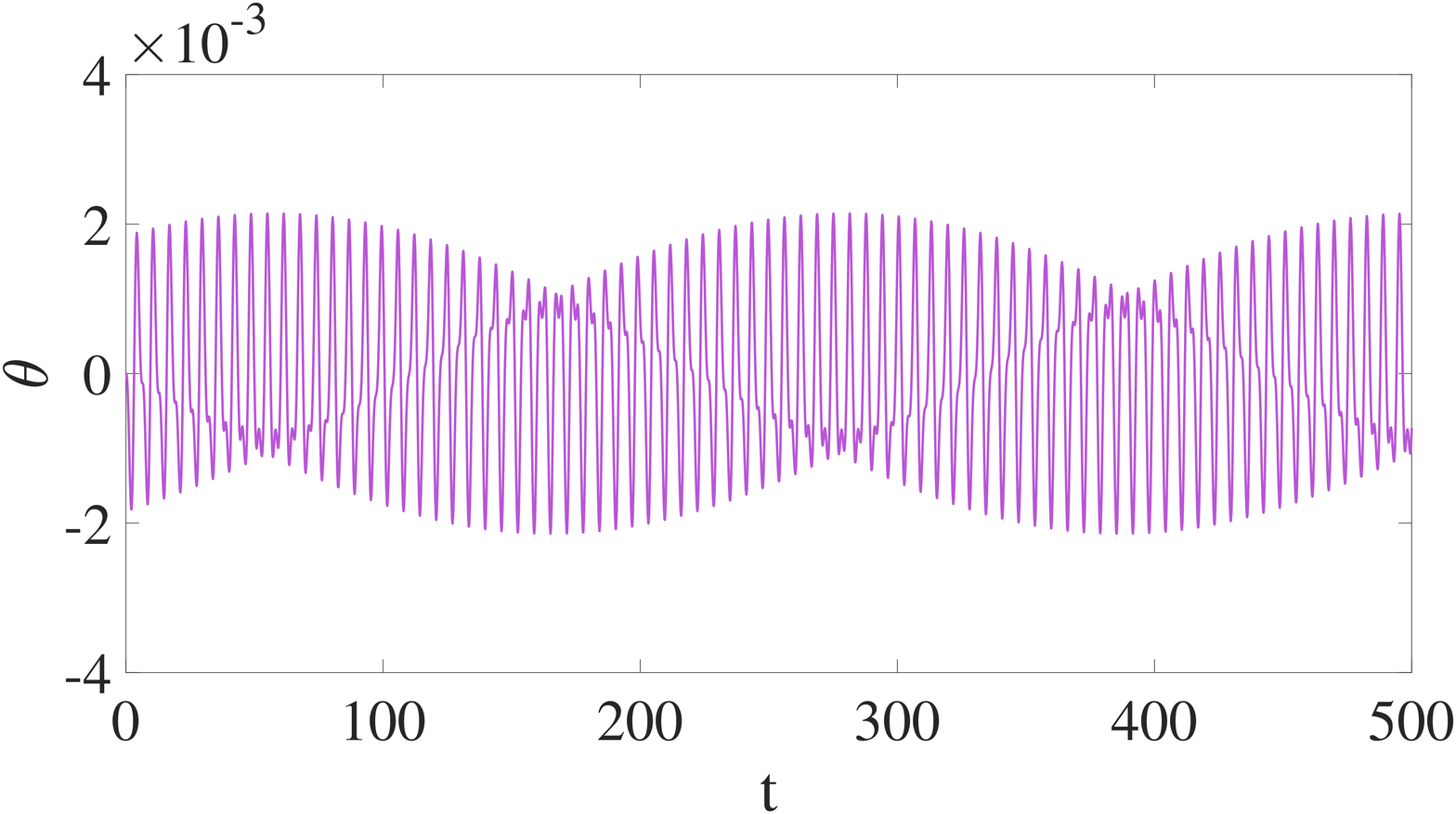} %------sol 2.5
}
\subfloat[\footnotesize $\dot{\varphi}(0) = 10$: time series]{
\includegraphics[width = 0.33\textwidth]{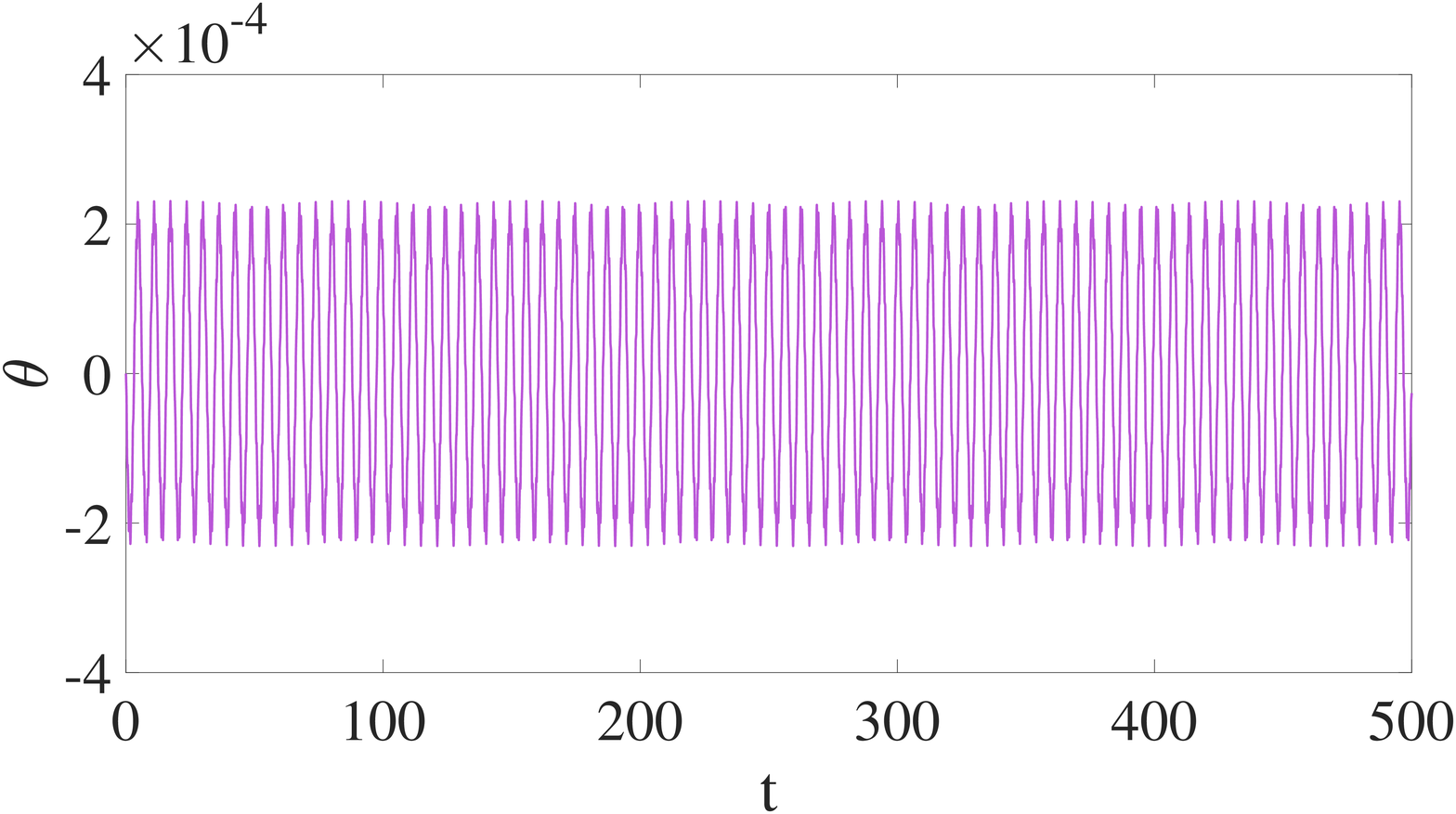} %------sol 10
}
\caption{Initial angular velocity dependence:  the axion time series $(t, \theta (t))$ ($t_{max} = 500$)}
\label{v1ctd}
\end{figure}

Continuously increasing the initial angular velocity of the Josephson junction results in different patterns in the time series and in the phase portrait of the axion, see fig.\ref{v1ctd}. Each time series has a high-frequency component that oscillates within a slowly varying profile, so the solution $\theta (t)$ can be approximated by a periodic (sinusoidal) function with a slowly varying amplitude. 

\subsection{Dissipative system}
For the dissipative case it is physically reasonable to assume that the damping coefficient of the axion, $a_{2}$,  is small, so for simplicity we set $a_{2} = 0$. By varying $a_{1}$ from $10^{-7}$ to $1$ we found that the system approaches various limit cycles depending on the initial value of the axion angular velocity $\dot{\theta} (0)$, and the time for reaching a limit cycle depends on the value of $a_{1}$.

\section{Analytic solutions in the limit of small oscillations}

While many of the phenomena in the previous section are clearly produced by strong nonlinearities
in the system dynamics, we now present some analytic results for small elongations.
If the oscillation amplitudes of both Josephson junction and axion are small, then the original system \eqref{1} can be approximated as a linearised model, i.e. by the first-order approximation
\begin{subequations}
\begin{align}
\ddot{\varphi} &= -a_{1}\dot{\varphi} - b_{1}\varphi + c(\ddot{\theta} - \ddot{\varphi})\\
\ddot{\theta} &= -a_{2}\dot{\theta} - b_{2}\theta + c(\ddot{\varphi} - \ddot{\theta})
\end{align}
\label{2}
\end{subequations}
For simplicity we set $a_{1} = a_{2} = a > 0$, $b_{1} = b_{2} = 1$ and $c > 0$. 

The system becomes uncoupled via the transformation $(\theta_{+}, \theta_{-}) = (\frac{1}{2}(\varphi + \theta), \frac{1}{2}(\varphi - \theta))$ and it reads in the new coordinates
\begin{subequations}
\begin{align}
\ddot{\theta}_{+} &= -a\dot{\theta}_{+} - \theta_{+}\\
\ddot{\theta}_{-} &= \frac{-a}{1 + 2c} \dot{\theta}_{-} - \frac{1}{1 + 2c}\theta_{-}
\end{align}
\end{subequations}
Note that the above coordinate transformation decouples the system for arbitrary coupling strengths,
including very large $c$.
The general solutions in the new coordinates are then given by
\begin{subequations}
\begin{align}
\theta_{+}(t) &= C_{1} e^{\frac{1}{2}(-a - \sqrt{a^{2} - 4})t} + C_{2} e^{\frac{1}{2}(-a + \sqrt{a^{2} - 4})t}\\
\theta_{-}(t) &= C_{3} e^{\frac{1}{2}(-\alpha - \sqrt{\alpha^{2} - 4\beta})t} + C_{4} e^{\frac{1}{2}(-\alpha + \sqrt{\alpha^{2} - 4\beta})t}
\end{align}
\label{theta-pm}
\end{subequations}
where $\alpha = \frac{a}{1 + 2c}$ and $\beta = \frac{1}{1 + 2c}$, provided $a \neq 2$ and $\alpha^{2} \neq 4\beta$. Otherwise, 

1) if $a = 2$, the solution for $\theta_{+}$ is
\begin{equation}
\theta_{+}(t) = C_{1}e^{-t} + C_{2}e^{-t}t
\label{a=2}
\end{equation}

2) if $\alpha^{2} = 4\beta$, or $a = 2\sqrt{1 + 2c}$, the solution for $\theta_{-}$ is
\begin{equation}
\theta_{-}(t) = C_{3}e^{-\frac{1}{\sqrt{1 + 2c}}t} + C_{4}e^{-\frac{1}{\sqrt{1 + 2c}}t}t
\label{a=2sqrt}
\end{equation}
where the constant coefficients $C_{1}$, $C_{2}$, $C_{3}$ and $C_{4}$ are to be determined by the initial conditions $(\varphi, \dot{\varphi}, \theta, \dot{\theta})_{t = 0}$, or equivalently $(\theta_{+}, \dot{\theta}_{+}, \theta_{-}, \dot{\theta}_{-})_{t = 0}$. 

\subsection{Non-dissipative case}
In the case that $a = 0$ the solutions are simply given by
\begin{subequations}
\begin{align}
\theta_{+}(t) &= C_{1}\cos{t} + C_{2}\sin{t}\\
\theta_{-}(t) &= C_{3}\cos{\left(\frac{1}{\sqrt{1 + 2c}}t\right)} + C_{4}\sin{\left(\frac{1}{\sqrt{1 + 2c}}t\right)}
\end{align}
\label{cos-sin}
\end{subequations}
Imposing our example of initial conditions $(\varphi, \dot{\varphi}, \theta, \dot{\theta})_{t = 0} = (0, 2, 0, 0)$, we get $C_{1} = C_{3} = 0$, $C_{2} = 1$ and $C_{4} = \sqrt{1 + 2c}$. Thus, the solutions in the original coordinates are given by 
\begin{subequations}
\begin{align}
\varphi(t) = \theta_{+} + \theta_{-} &= \sin{t} + \sqrt{1 + 2c}\sin{\left(\frac{1}{\sqrt{1 + 2c}}t\right)}\\
\theta(t) = \theta_{+} - \theta_{-} &= \sin{t} - \sqrt{1 + 2c}\sin{\left(\frac{1}{\sqrt{1 + 2c}}t\right)}
\end{align}
\label{a=0}
\end{subequations}
Note that the solutions involve two frequencies, $\omega_{+} = 1$ and $\omega_{-} = \frac{1}{\sqrt{1 + 2c}}$, which are commensurate if $\sqrt{1 + 2c}$ is rational. The phase trajectory is dense if the two frequencies are incommensurate. 

An interesting result is concerned with possible synchronisation behavior of the two coupled oscillators.
We found that, for some values of the coupling $c$ (of measure zero on $\mathbb{R}$), the phase trajectories of $\varphi$ and $\theta$ are identical up to a time translation. A simple proof of this result is provided here: %simple proof of two identical trajectories, with 3 examples in the end

We set the initial conditions as $(\varphi , \dot{\varphi}, \theta , \dot{\theta})_{t = 0} = (0, v_{1}, 0, v_{2})$. Then by eqns.\eqref{cos-sin} the constants $C_{1}$ and $C_{3}$ are always zero; hence we get solutions $\varphi(t)$ and $\theta(t)$ which are combinations of two sine functions: 
\begin{subequations}
\begin{align}
\varphi(t) &= \theta_{+0}\sin{t} + \frac{1}{\omega_{-}}\theta_{-0}\sin{\omega_{-}t}\\
\theta(t) &= \theta_{+0}\sin{t} + \frac{1}{\omega_{-}}\theta_{-0}\sin{\omega_{-}t}
\end{align}
\label{eqn9}
\end{subequations}
where $\omega_{-} = \frac{1}{\sqrt{1 + 2c}}$, with $\theta_{+0} = \frac{\dot{\varphi}(0) + \dot{\theta}(0)}{2} = \frac{v_{1} + v_{2}}{2}$ and $\theta_{-0} = \frac{\dot{\varphi}(0) - \dot{\theta}(0)}{2} = \frac{v_{1} - v_{2}}{2}$.\\
With $y_{i} = \dot{x}_{i}$ ($i = 1, 2$) the parametrised trajectories in each phase plane are written as
\begin{subequations}
\begin{align}
x_{1}(t_{1}) &= \theta_{+0}\sin{t_{1}} + \frac{1}{\omega_{-}}\theta_{-0}\sin{\omega_{-}t_{1}}\\
y_{1}(t_{1}) &= \theta_{+0}\cos{t_{1}} + \theta_{-0}\cos{\omega_{-}t_{1}}
\end{align}
\end{subequations}

\begin{subequations}
\begin{align}
x_{2}(t_{2}) &= \theta_{+0}\sin{t_{2}} - \frac{1}{\omega_{-}}\theta_{-0}\sin{\omega_{-}t_{2}}\\
y_{2}(t_{2}) &= \theta_{+0}\cos{t_{2}} - \theta_{-0}\cos{\omega_{-}t_{2}}
\end{align}
\end{subequations}
Denote the differences by
\begin{subequations}
\begin{align}
\Delta x &= x_{1} - x_{2} = \theta_{+0}(\sin{t_{1}} - \sin{t_{2}}) + \frac{1}{\omega_{-}}\theta_{-0}(\sin{\omega_{-}t_{1}} + \sin{\omega_{-}t_{2}})\\
\Delta y &= y_{1} - y_{2} = \theta_{+0}(\cos{t_{1}} - \cos{t_{2}}) + \theta_{-0}(\cos{\omega_{-}t_{1}} + \cos{\omega_{-}t_{2}})
\end{align}
\end{subequations}
or 
%by the trigonometric identities $\sin{A} \pm \sin{B} = 2\cos{\frac{A \mp B}{2}}\sin{\frac{A \pm B}{2}}$, $\cos{A} + \cos{B} = 2\cos{\frac{A + B}{2}}\cos{\frac{A - B}{2}}$ and $\cos{A} - \cos{B} = -2\sin{\frac{A + B}{2}\sin{\frac{A - B}{2}}}$ they can be written as
\begin{subequations}
\begin{align}
\Delta x &= 2\theta_{+0}\cos{\frac{t_{1} + t_{2}}{2}}\sin{\frac{t_{1} - t_{2}}{2}} + \frac{2}{\omega_{-}}\theta_{-0}\cos{\frac{\omega_{-}(t_{1} - t_{2})}{2}}\sin{\frac{\omega_{-}(t_{1} + t_{2})}{2}}\\
\Delta y &= -2\theta_{+0}\sin{\frac{t_{1} + t_{2}}{2}}\sin{\frac{t_{1} - t_{2}}{2}} + 2\theta_{-0}\cos{\frac{\omega_{-}(t_{1} + t_{2})}{2}}\cos{\frac{\omega_{-}(t_{1} - t_{2})}{2}}
\end{align}
\end{subequations}
By performing a linear transformation
\begin{subequations}
\begin{align}
t_{1} &= t + \tau \\
t_{2} &= t
\end{align}
\end{subequations}
the above differences become
\begin{subequations}
\begin{align}
\Delta x &= 2\theta_{+0}\cos{\frac{2t + \tau}{2}}\sin{\frac{\tau}{2}} + \frac{2}{\omega_{-}}\theta_{-0}\cos{\frac{\omega_{-}\tau}{2}}\sin{\frac{\omega_{-}(2t + \tau)}{2}}\\
\Delta y &= -2\theta_{+0}\sin{\frac{2t + \tau}{2}}\sin{\frac{\tau}{2}} + 2\theta_{-0}\cos{\frac{\omega_{-}(2t + \tau)}{2}}\cos{\frac{\omega_{-}\tau}{2}}
\end{align}
\end{subequations}
The differences are identically zero for all $t$ and all initial velocities if we set 
\begin{subequations}
\begin{align}
\sin{\frac{\tau}{2}} &= 0\\
\cos{\frac{\omega_{-}\tau}{2}} &= 0
\end{align}
\end{subequations}
or equivalently,
\begin{subequations}
\begin{align}
\tau &= 0 \mod 2\pi\\
\omega_{-}\tau &= \pi \mod 2\pi
\end{align}
\end{subequations}
If $\omega_{-} = \frac{p}{q}$ rational with $\gcd (p, q) = 1$, this can be written as $\omega_{-} = \frac{2k_{1} + 1}{2k_{2}}$ ($k_{1}, k_{2} \in \mathbb{N}$), and if we write $\tau = 2k\pi$ ($k \in \mathbb{N}$), then the above conditions are satisfied
\begin{equation*}
\frac{2k_{1}+1}{2k_{2}}2k\pi = \pi \mod 2\pi
\end{equation*}
provided $k = k_{2}$. 

In conclusion, the phase trajectories for the non-dissipative linearised system \eqref{2} are identical\footnote{`Identical' in the strict sense that the two trajectories must be closed, i.e., the two frequencies are commensurate, which is guaranteed by setting $\omega_{-}$ rational. One can also define what `identical' (in a loose way) means for two open trajectories which not necessarily requires $\omega_{-}$ to be rational.} if $\omega_{-} := \frac{1}{\sqrt{1 + 2c}} = \frac{2k_{1} + 1}{2k}$, or equivalently, $c = \frac{2k^{2}}{(2k_{1} + 1)^2} - \frac{1}{2}$ with $k, k_{1} \in \mathbb{N}$, and the time series of the two phase variables are shifted by time $\tau = 2k\pi$. These values of $c$ form a subset of the set of rational numbers in $\mathbb{R}$, which is obviously of measure zero. 

\begin{figure} [H]
\centering
\subfloat[\footnotesize phase trajectory for $\varphi$]{
\includegraphics[width = 0.33\textwidth]{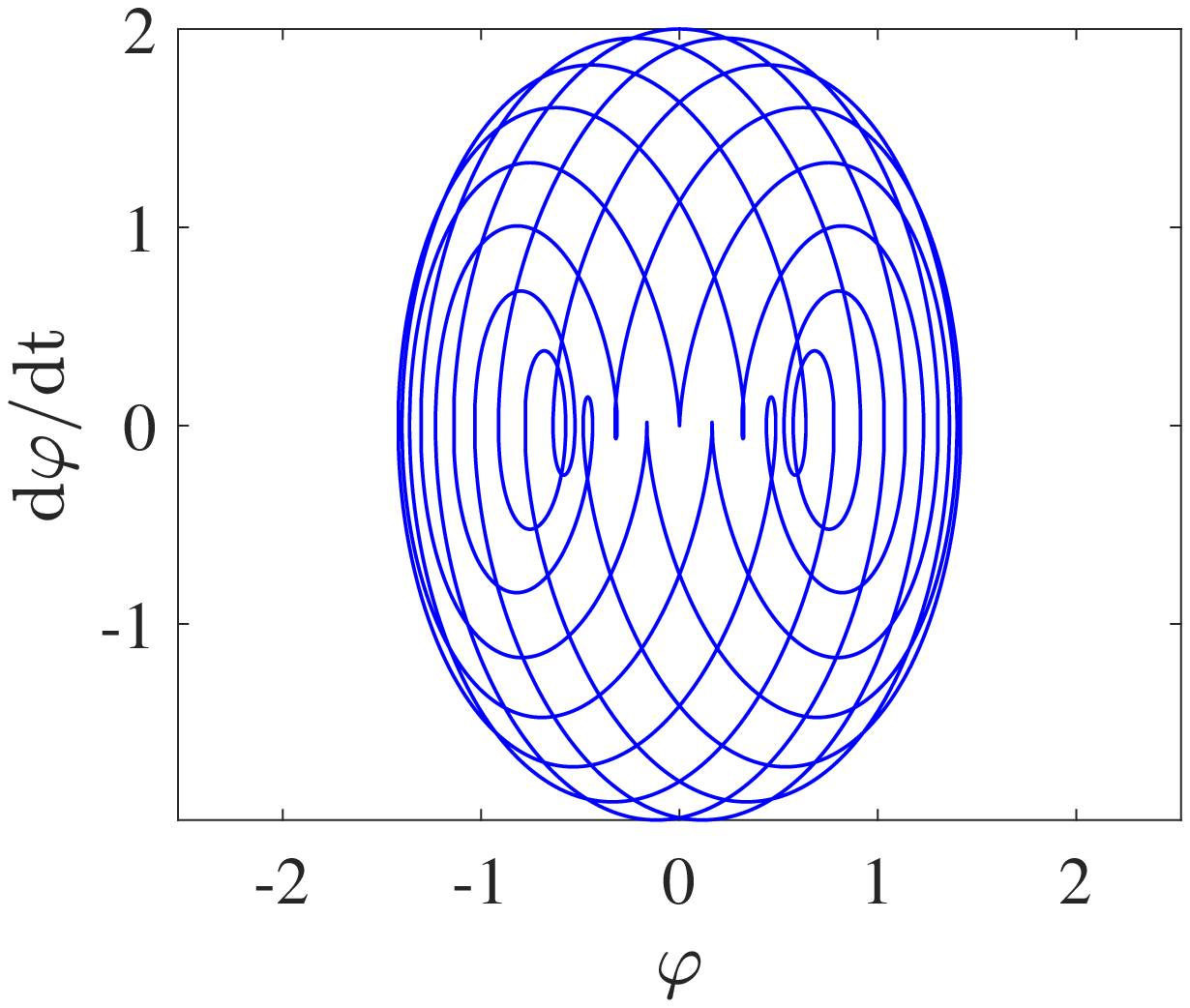} %------c = -297/722 (freq = 19/8) symmetry eg1 (k1 = 9 & k = 4)
}
\subfloat[\footnotesize phase trajectory for $\theta$]{
\includegraphics[width = 0.33\textwidth]{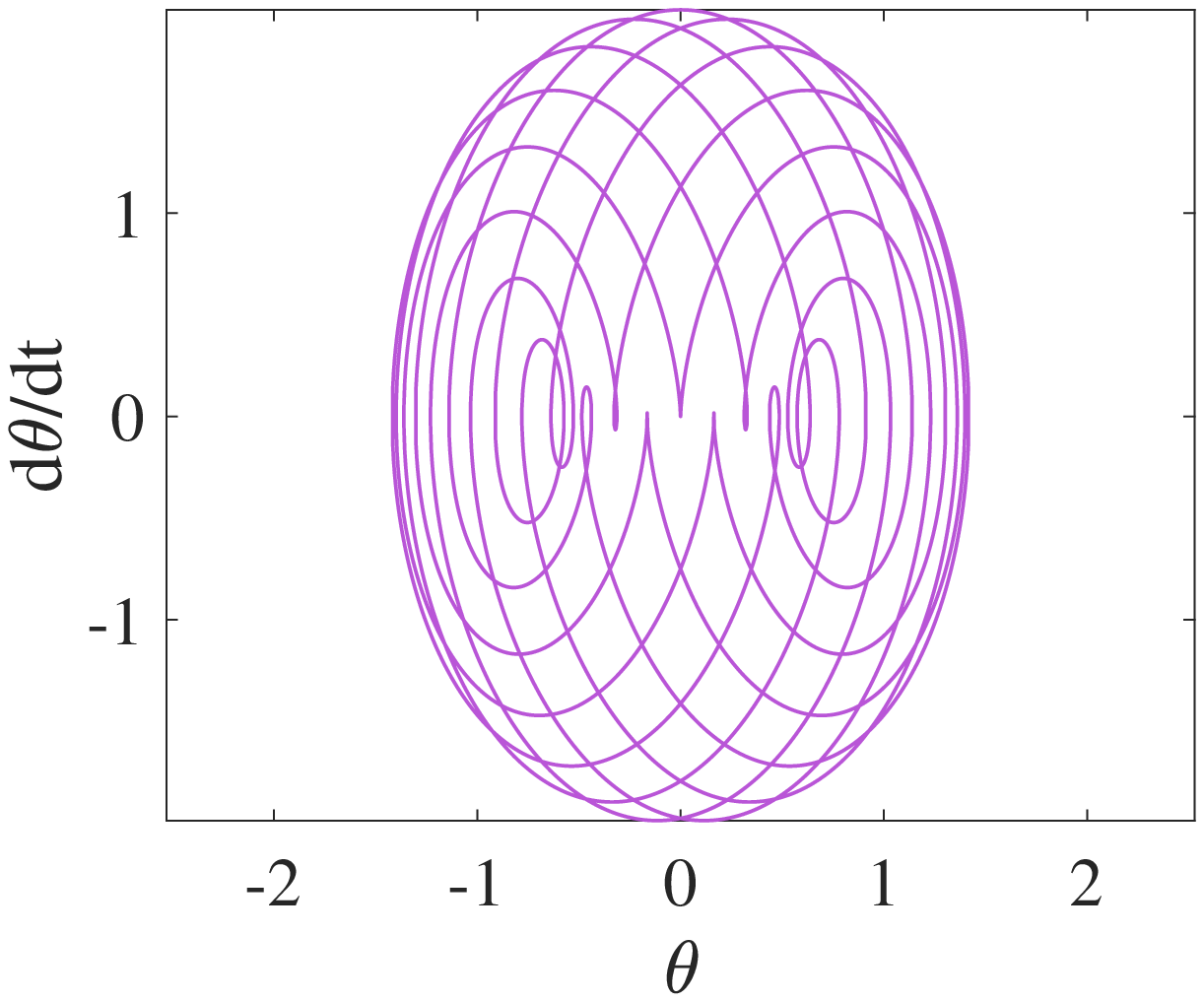} 
}
\subfloat[\footnotesize time series ($\tau = 8\pi$)]{
\includegraphics[width = 0.33\textwidth]{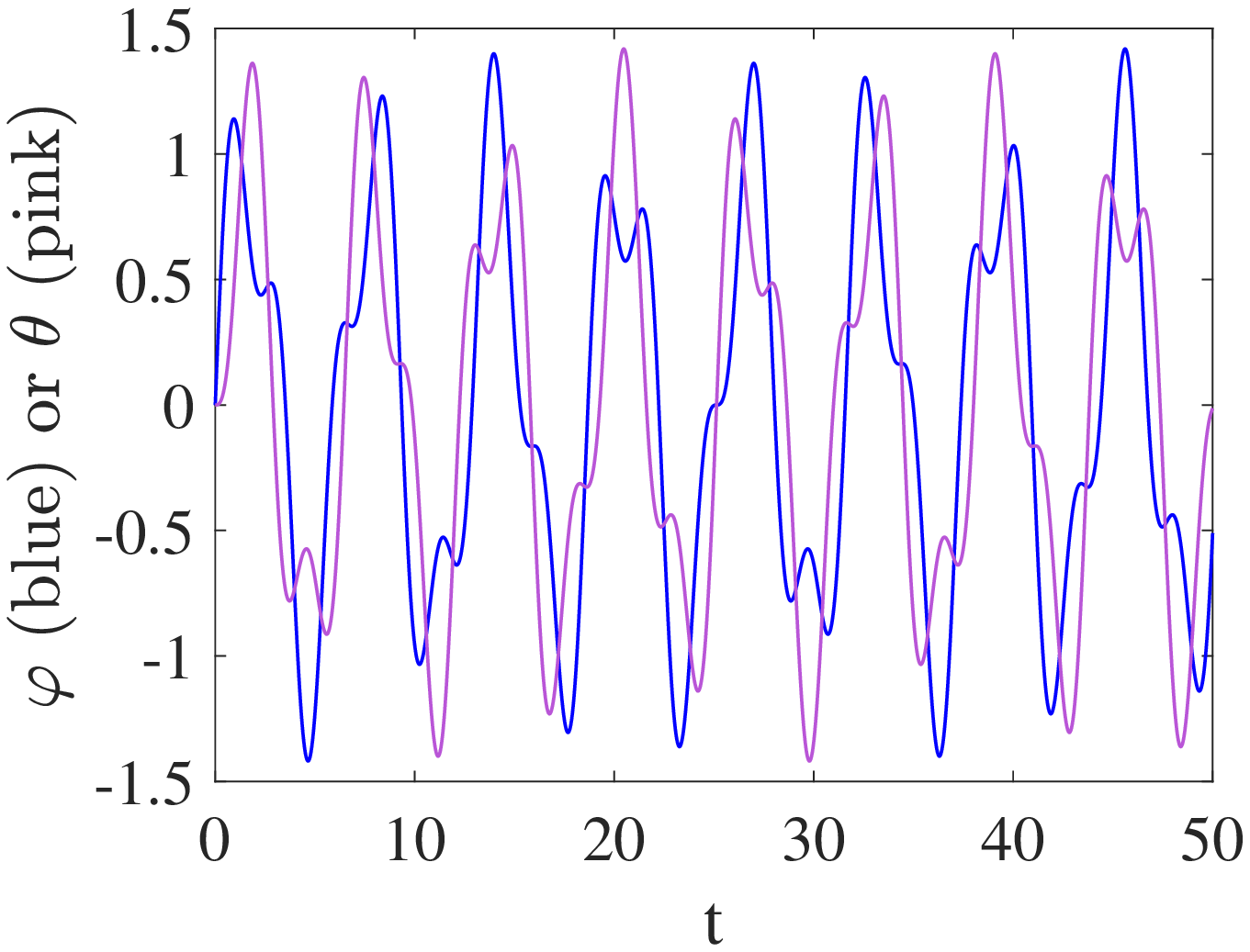} 
}
\qquad
\subfloat[\footnotesize phase trajectory for $\varphi$]{
\includegraphics[width = 0.33\textwidth]{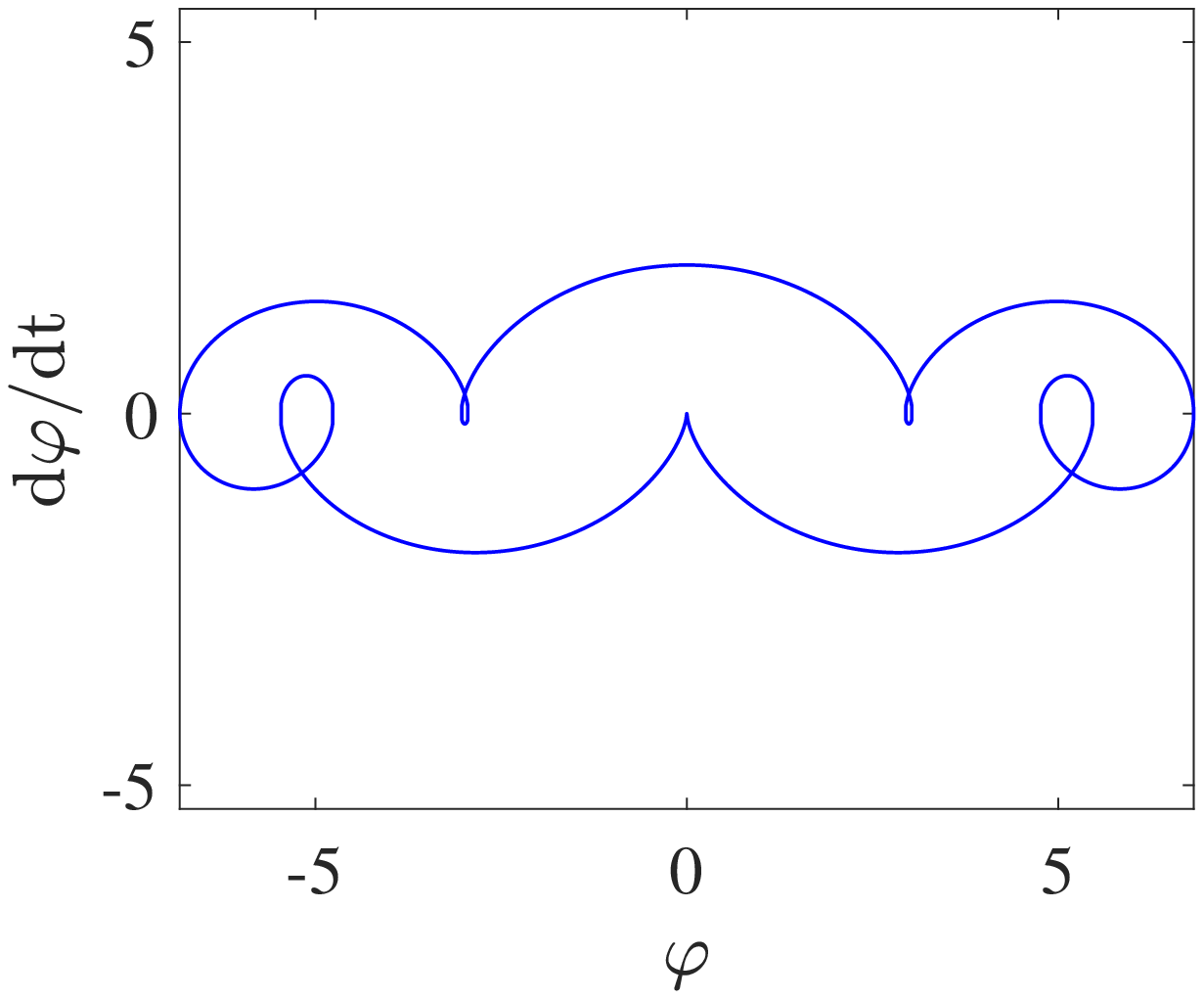} %------c = 35/2 (freq = 1/6) symmetry eg2 (k1 = 0 & k = 3)
}
\subfloat[\footnotesize phase trajectory for $\theta$]{
\includegraphics[width = 0.33\textwidth]{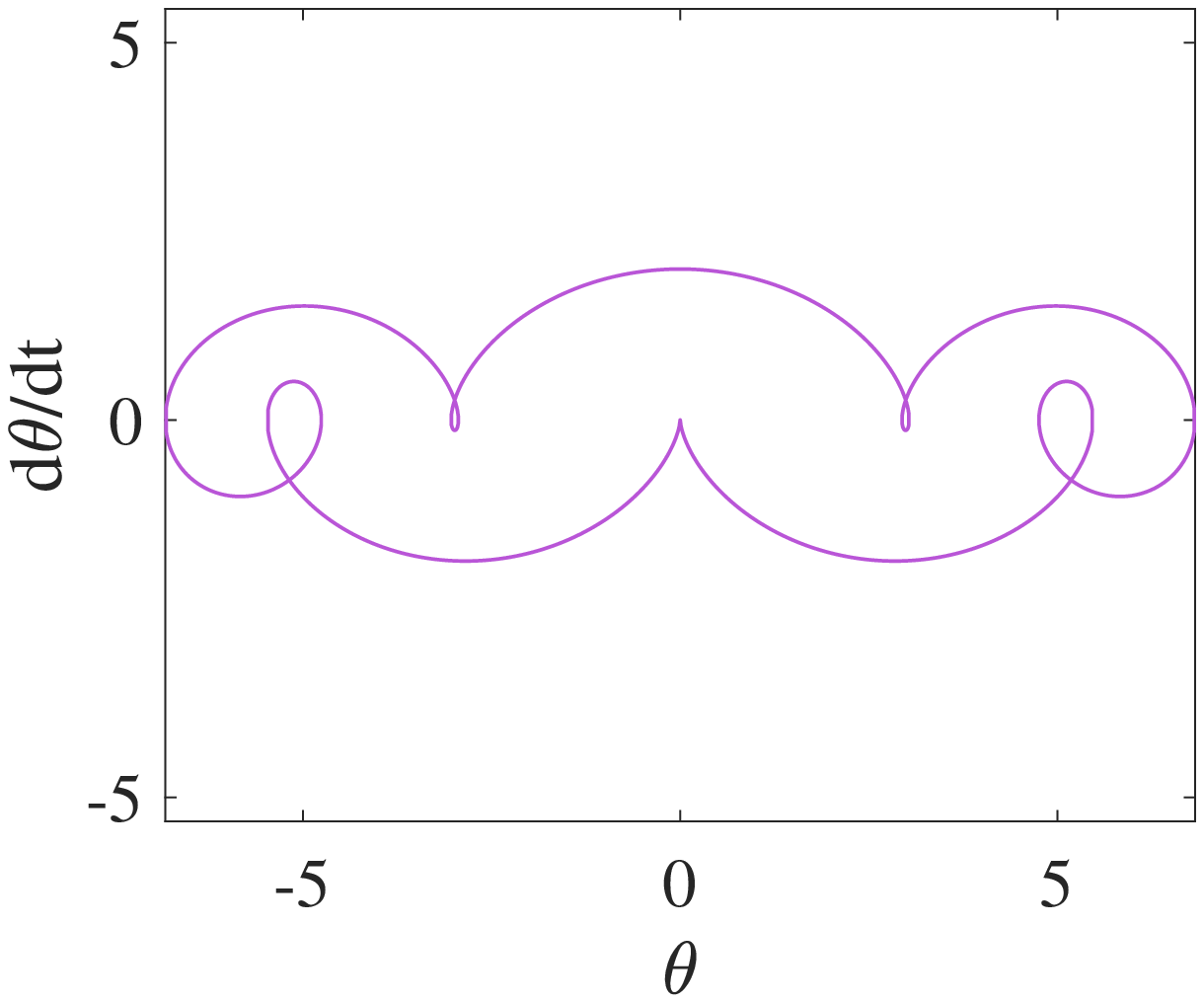} 
}
\subfloat[\footnotesize time series ($\tau = 6\pi$)]{
\includegraphics[width = 0.33\textwidth]{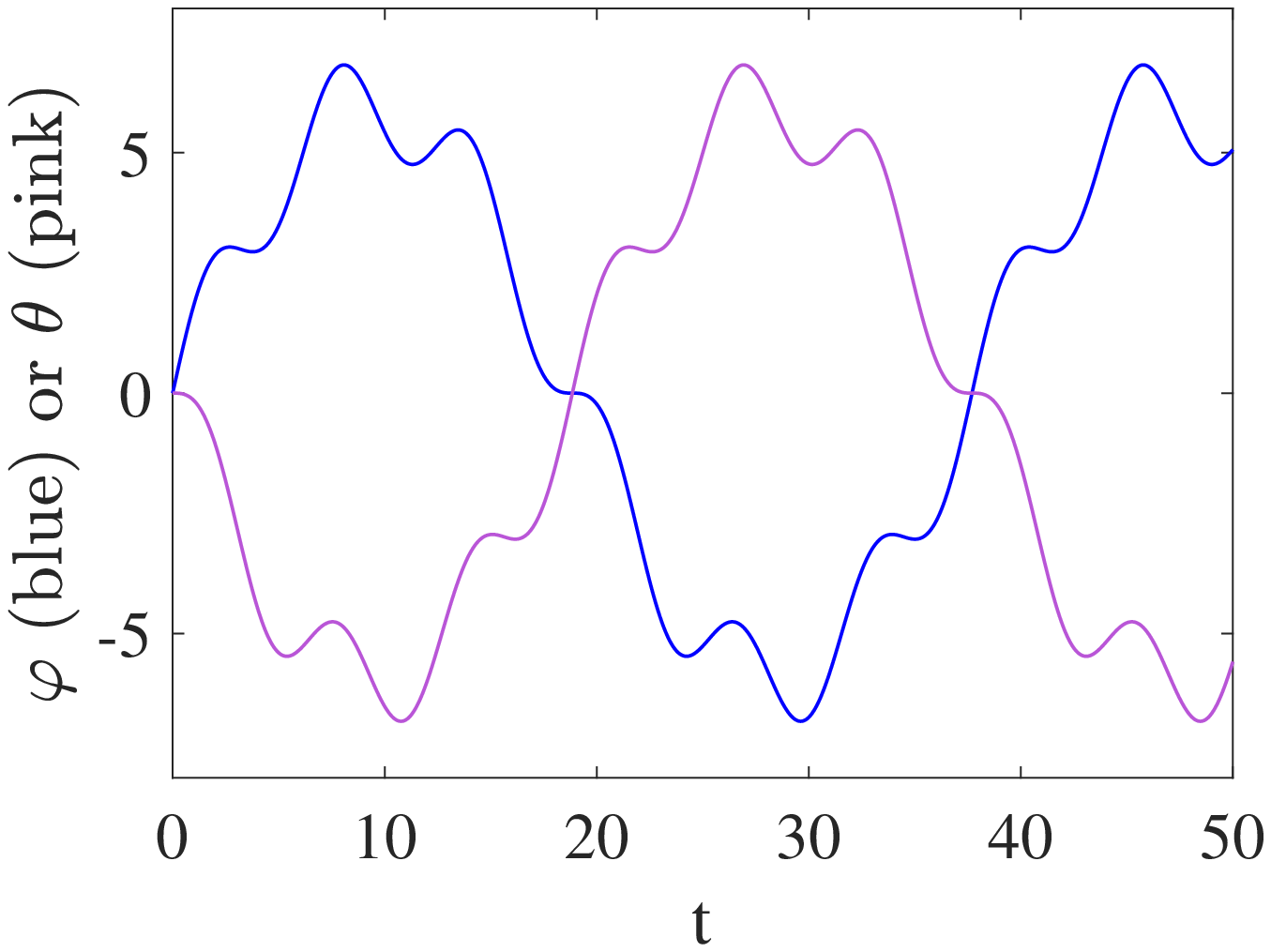} 
}
\qquad
\subfloat[\footnotesize phase trajectory for $\varphi$]{
\includegraphics[width = 0.33\textwidth]{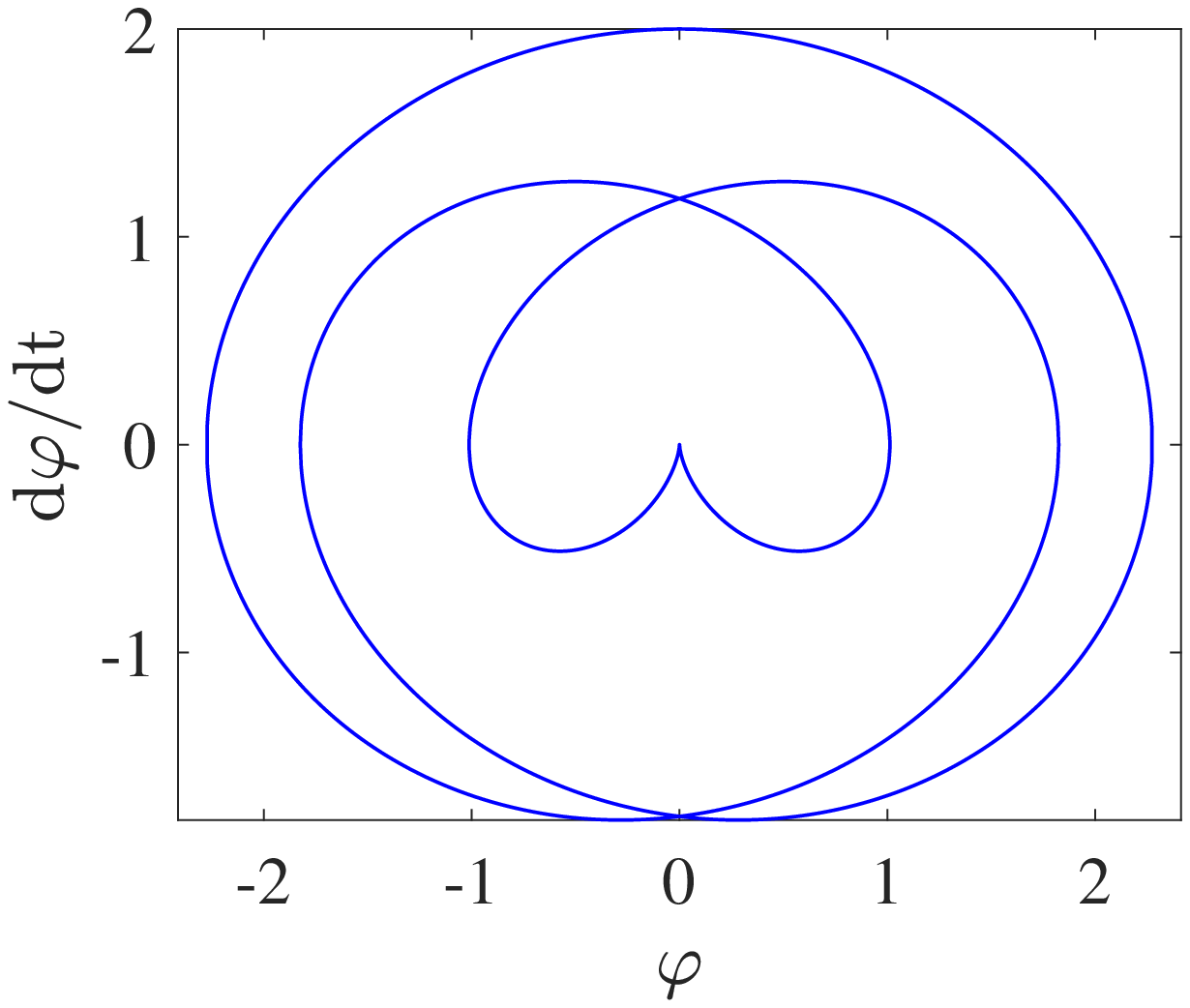} %------c = 7/18 (freq = 3/4) symmetry eg3 (k1 = 1 & k = 2)
}
\subfloat[\footnotesize phase trajectory for $\theta$]{
\includegraphics[width = 0.33\textwidth]{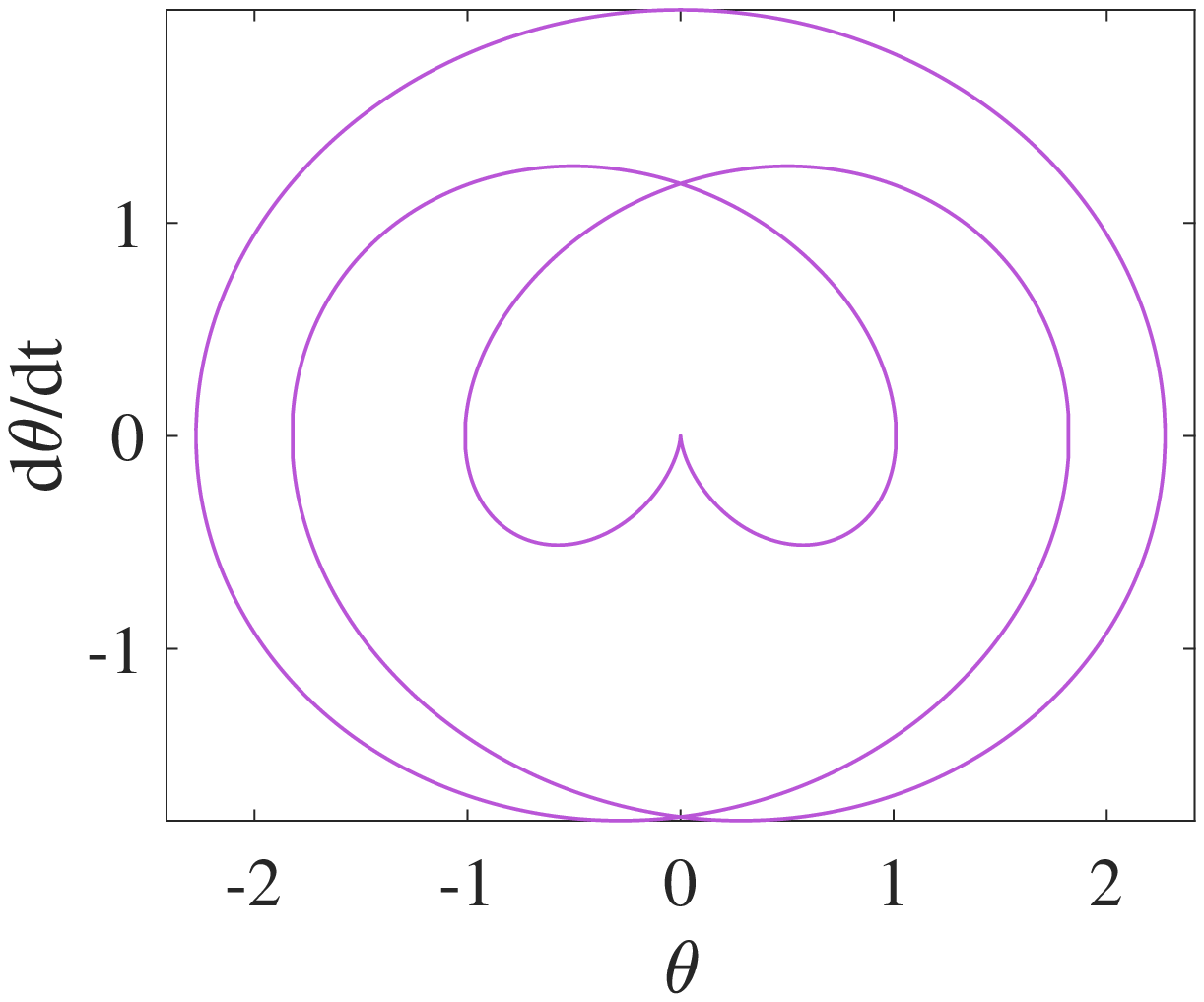} 
}
\subfloat[\footnotesize time series ($\tau = 4\pi$)]{
\includegraphics[width = 0.33\textwidth]{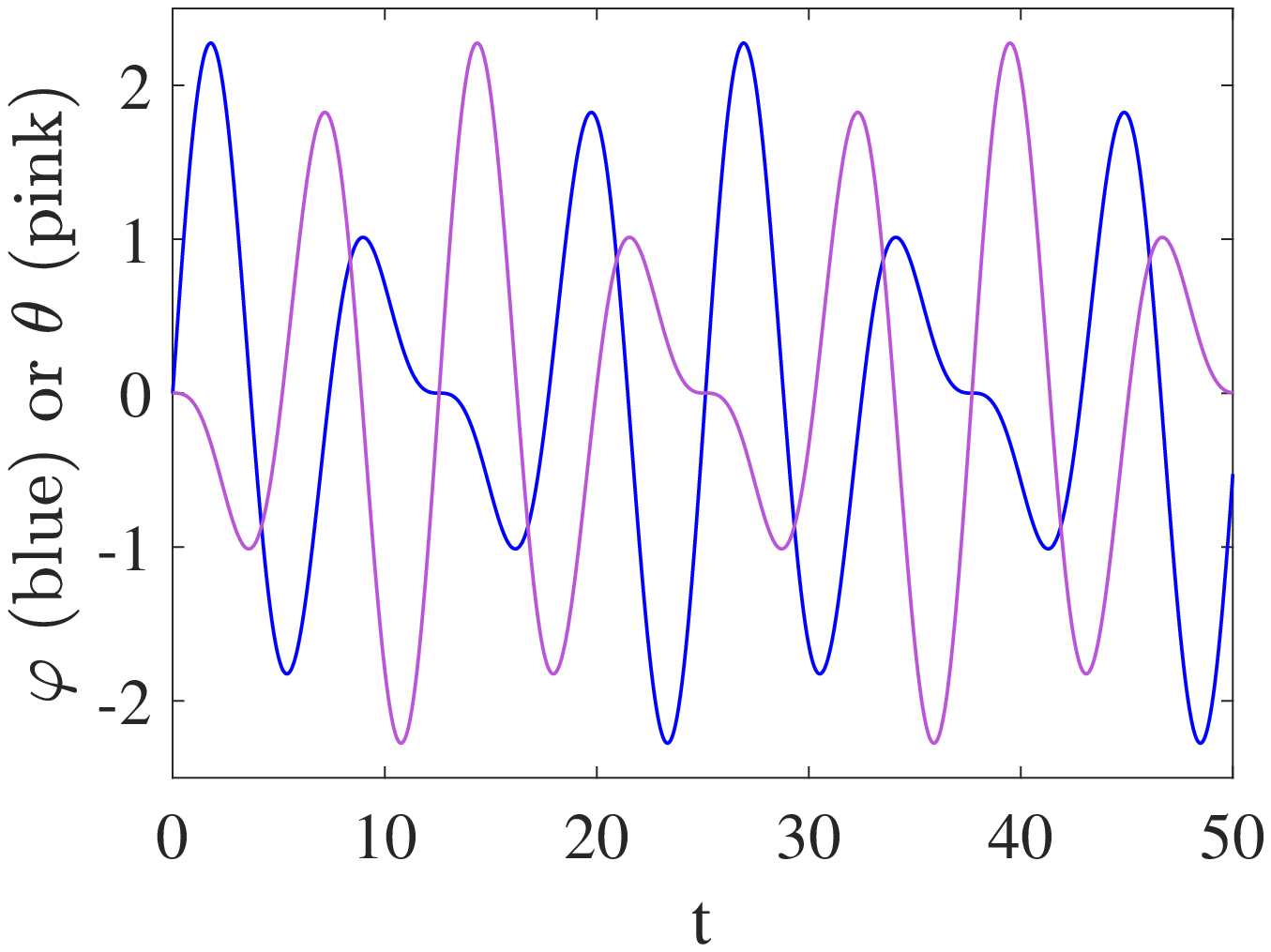} 
}
\caption{Examples of identical trajectories: by time shifting by $\tau$ in each case the two solution curves are identical ($t_{max} = 50$). First row: $\omega_{-} = \frac{19}{8}$ (or $c = -\frac{297}{722}$); second row: $\omega_{-} = \frac{1}{6}$ (or $c = \frac{35}{2}$); third row: $\omega_{-} = \frac{3}{4}$ (or $c = \frac{7}{18}$).
Still a highly nontrivial phase portrait arises in the phase space, which depends on the integers $k, k_1$ that make up the coupling $c$.}
\label{identical}
\end{figure}

Let us illustrate these general results with some examples:
When $\omega_{-} = \frac{19}{8}$ (or $c = -\frac{297}{722}$), $\omega_{-} = \frac{1}{6}$ (or $c = \frac{35}{2}$) and $\omega_{-} = \frac{3}{4}$ (or $c = \frac{7}{18}$), the two identical trajectories are shown in fig.\ref{identical} in row 1, 2, and 3, respectively. 

Notice that if $k \approx k_{1} \rightarrow \infty$, then $c \rightarrow 0$ (very weak coupling), $\omega_{-} \rightarrow 1 = \omega_{+}$, and the two trajectories will be identical although it requires a long time shift, $\tau \rightarrow \infty$, for their solution curves to coincide. On the other hand, for the solutions \eqref{eqn9} the period $p$ given by $p = \gcd (p_{1}, p_{2}) = \gcd (\frac{2\pi}{\omega_{+}}, \frac{2\pi}{\omega_{-}}) = \gcd (2\pi, 2\pi \cdot \frac{2k}{2k_{1} + 1}) = 2\pi \cdot \gcd (1, \frac{2k}{2k_{1} + 1}) = 2\pi (2k_{1} + 1) \rightarrow \infty$. 

In particular, for different relations between  $k_{1}$ and $k$ the asymptotic trajectory follows different patterns. For example, the last row in fig.\ref{identical} has a heart-shaped pattern in the centre, for which $(k_{1}, k) = (k_{1}, k_{1} + 1) = (1, 2)$. As $k_{1}$ increases, the coupling parameter $c$ decreases to zero from above, and the winding number (with respect to the origin of the phase plane) increases, see fig.\ref{heart} below. Inspired by the first and second rows in fig.\ref{identical}, one can also choose some other relations between the two integers $(k_{1}, k)$ to get different sequences of trajectories and eventually obtain a different asymptotic structure. 
\begin{figure} [H]
\centering
\subfloat[\footnotesize $(k_{1}, k) = (4, 5)$]{
\includegraphics[width = 0.33\textwidth]{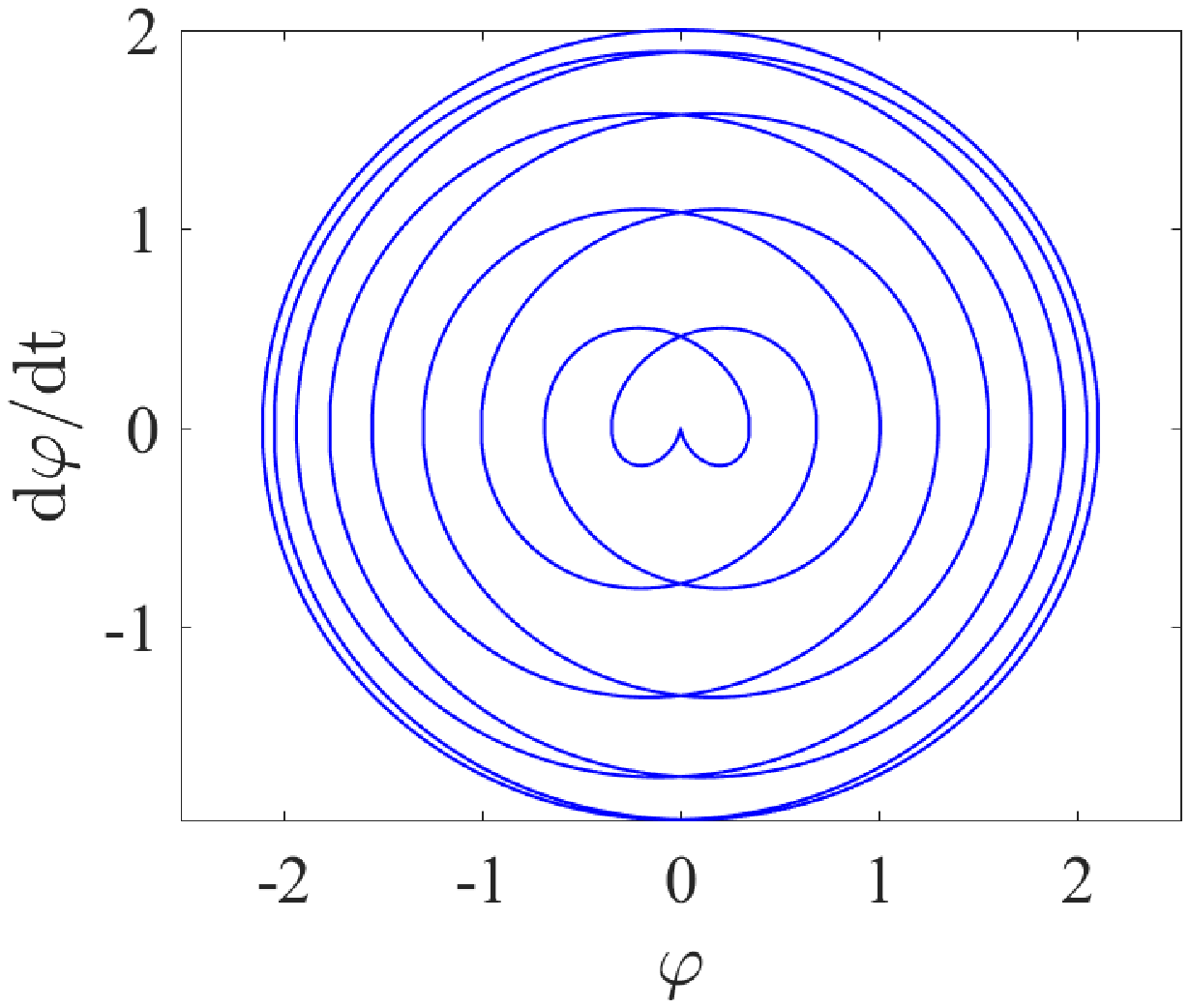} %(k1 = 4 & k = 5)
}
\subfloat[\footnotesize $(k_{1}, k) = (9, 10)$]{
\includegraphics[width = 0.33\textwidth]{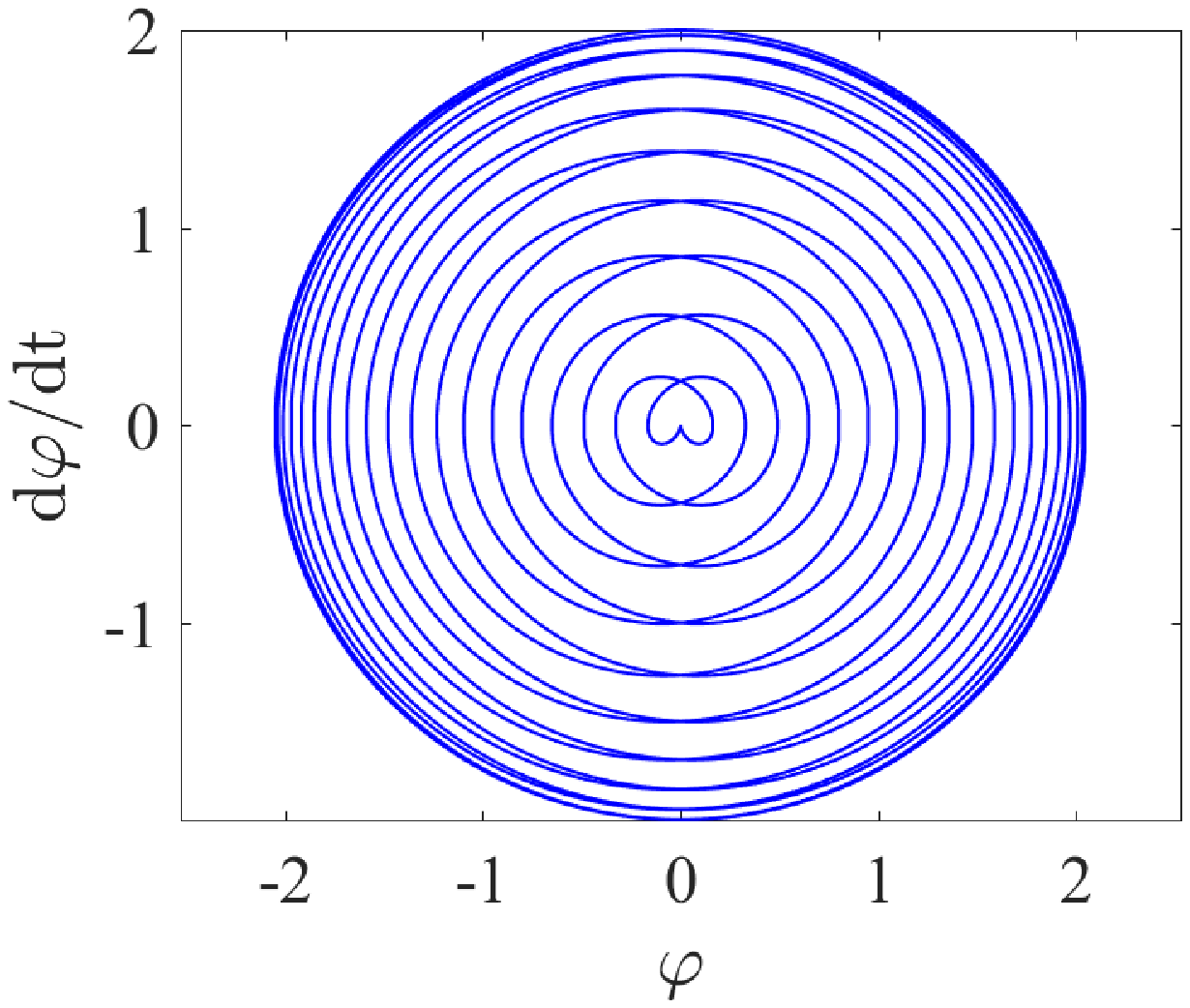} %(k1 = 9 & k = 10)
}
\subfloat[\footnotesize $(k_{1}, k) = (19, 20)$]{
\includegraphics[width = 0.33\textwidth]{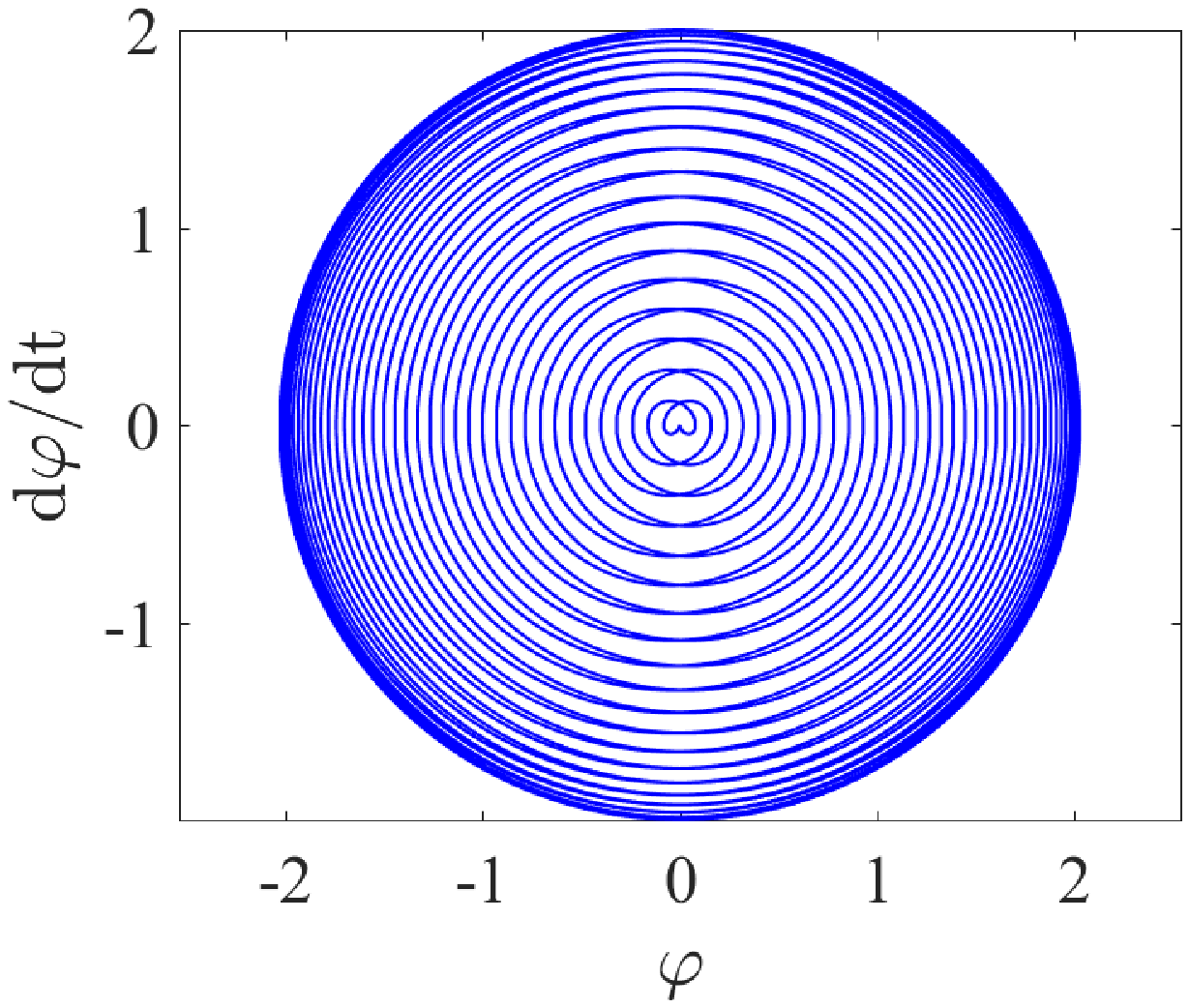} %(k1 = 19 & k = 20)
}
\caption{An example of trajectory patterns for $k = k_{1} + 1$ as $k_{1} \in \mathbb{R}_{+}$ increases. Phase trajectories for $\varphi$ (and also for $\theta$ as they are identical) are shown.}
\label{heart}
\end{figure}

\subsection{Dissipative cases}
If $a \neq 0$, the two critical values of $a$ divide the solutions into the following five categories: ($c > 0$)

(i) $0 < a < 2$ (small dissipation): %'small' comparing to the rest four categories, but in our model it is not small (different scales)
\begin{subequations}
\begin{align}
\varphi (t) &= \frac{2}{\sqrt{4 - a^{2}}}e^{-\frac{a}{2}t} \sin{\left(\frac{\sqrt{4 - a^{2}}}{2}t\right)} + \frac{2}{\sqrt{4\beta - \alpha^{2}}}e^{-\frac{\alpha}{2}t} \sin{\left(\frac{\sqrt{4\beta - \alpha^{2}}}{2}t\right)}\\
\theta(t) &= \frac{2}{\sqrt{4 - a^{2}}}e^{-\frac{a}{2}t} \sin{\left(\frac{\sqrt{4 - a^{2}}}{2}t\right)} - \frac{2}{\sqrt{4\beta - \alpha^{2}}}e^{-\frac{\alpha}{2}t} \sin{\left(\frac{\sqrt{4\beta - \alpha^{2}}}{2}t\right)}
\end{align}
\end{subequations}
where $\alpha = \frac{a}{1 + 2c}$ and $\beta = \frac{1}{1 + 2c}$.
Because of the presence of small dissipation, both phase variables exhibits oscillatory decay, with two frequencies $(\omega_{+}, \omega_{-}) = (\frac{\sqrt{4 - a^{2}}}{2}, \frac{\sqrt{4\beta - \alpha^{2}}}{2}) = (\frac{\sqrt{4 - a^{2}}}{2}, \frac{\sqrt{4(1 + 2c) - a^{2}}}{2(1 + 2c)})$ and two decaying rates $(\nu_{+}, \nu_{-}) = (\frac{a}{2}, \frac{\alpha}{2}) = (\frac{a}{2}, \frac{a}{2(1 + 2c)})$, which are proportional to $a$. It makes sense that when $a$ is very small, $\omega_{+} \approx \omega_{-}$ and $\nu_{+} \approx \nu_{-}$; when $a$ approaches zero, we recover the solutions \eqref{a=0} of the non-dissipative case.

(ii) $a = 2$:
\begin{subequations}
\begin{align}
\varphi(t) &= e^{-t}t + \frac{1 + 2c}{\sqrt{2c}}e^{-\frac{1}{1 + 2c}t} \sin{\left( \frac{\sqrt{2c}}{1 + 2c}t \right)}\\
\theta(t) &= e^{-t}t - \frac{1 + 2c}{\sqrt{2c}}e^{-\frac{1}{1 + 2c}t} \sin{\left( \frac{\sqrt{2c}}{1 + 2c}t \right)}
\end{align}
\end{subequations}

(iii) $2 < a < 2\sqrt{1 + 2c}$ (medium dissipation):
\begin{subequations}
\begin{align}
\varphi(t) &= \frac{2}{\sqrt{a^{2} - 4}}e^{-\frac{a}{2}t} \sinh{\left( \frac{\sqrt{a^{2} - 4}}{2}t \right)} + \frac{2}{\sqrt{4\beta - \alpha^{2}}}e^{-\frac{\alpha}{2}t} \sin{\left(\frac{\sqrt{4\beta - \alpha^{2}}}{2}t\right)}\\
\theta(t) &= \frac{2}{\sqrt{a^{2} - 4}}e^{-\frac{a}{2}t} \sinh{\left( \frac{\sqrt{a^{2} - 4}}{2}t \right)} - \frac{2}{\sqrt{4\beta - \alpha^{2}}}e^{-\frac{\alpha}{2}t} \sin{\left(\frac{\sqrt{4\beta - \alpha^{2}}}{2}t\right)}
\end{align}
\end{subequations}
where $\alpha = \frac{a}{1 + 2c}$ and $\beta = \frac{1}{1 + 2c}$.

(iv) $a = 2\sqrt{1 + 2c}$:
\begin{subequations}
\begin{align}
\varphi(t) &= \frac{1}{\sqrt{2c}} e^{-\sqrt{1 + 2c} t} \sinh{(\sqrt{2c} t)} + e^{-\frac{1}{\sqrt{1 + 2c}}t}t\\
\theta(t) &= \frac{1}{\sqrt{2c}} e^{-\sqrt{1 + 2c} t} \sinh{(\sqrt{2c} t)} - e^{-\frac{1}{\sqrt{1 + 2c}}t}t
\end{align}
\end{subequations}

(v) $a > 2\sqrt{1 + 2c}$ (large dissipation):
\begin{subequations}
\begin{align}
\varphi(t) &= \frac{2}{\sqrt{a^{2} - 4}} e^{-\frac{a}{2}t} \sinh{\left( \frac{\sqrt{a^{2} - 4}}{2}t\right)} + \frac{2}{\sqrt{\alpha^{2} - 4\beta}} e^{-\frac{\alpha}{2}t} \sinh{\left( \frac{\sqrt{\alpha^{2} - 4\beta}}{2}t\right)}\\
\theta(t) &=  \frac{2}{\sqrt{a^{2} - 4}} e^{-\frac{a}{2}t} \sinh{\left( \frac{\sqrt{a^{2} - 4}}{2}t\right)} - \frac{2}{\sqrt{\alpha^{2} - 4\beta}} e^{-\frac{\alpha}{2}t} \sinh{\left( \frac{\sqrt{\alpha^{2} - 4\beta}}{2}t\right)}
\end{align}
\end{subequations}
where $\alpha = \frac{a}{1 + 2c}$ and $\beta = \frac{1}{1 + 2c}$.

\begin{figure} [H]
\centering
\subfloat[\footnotesize $c = 0.1$]{
\includegraphics[width = 0.33\textwidth]{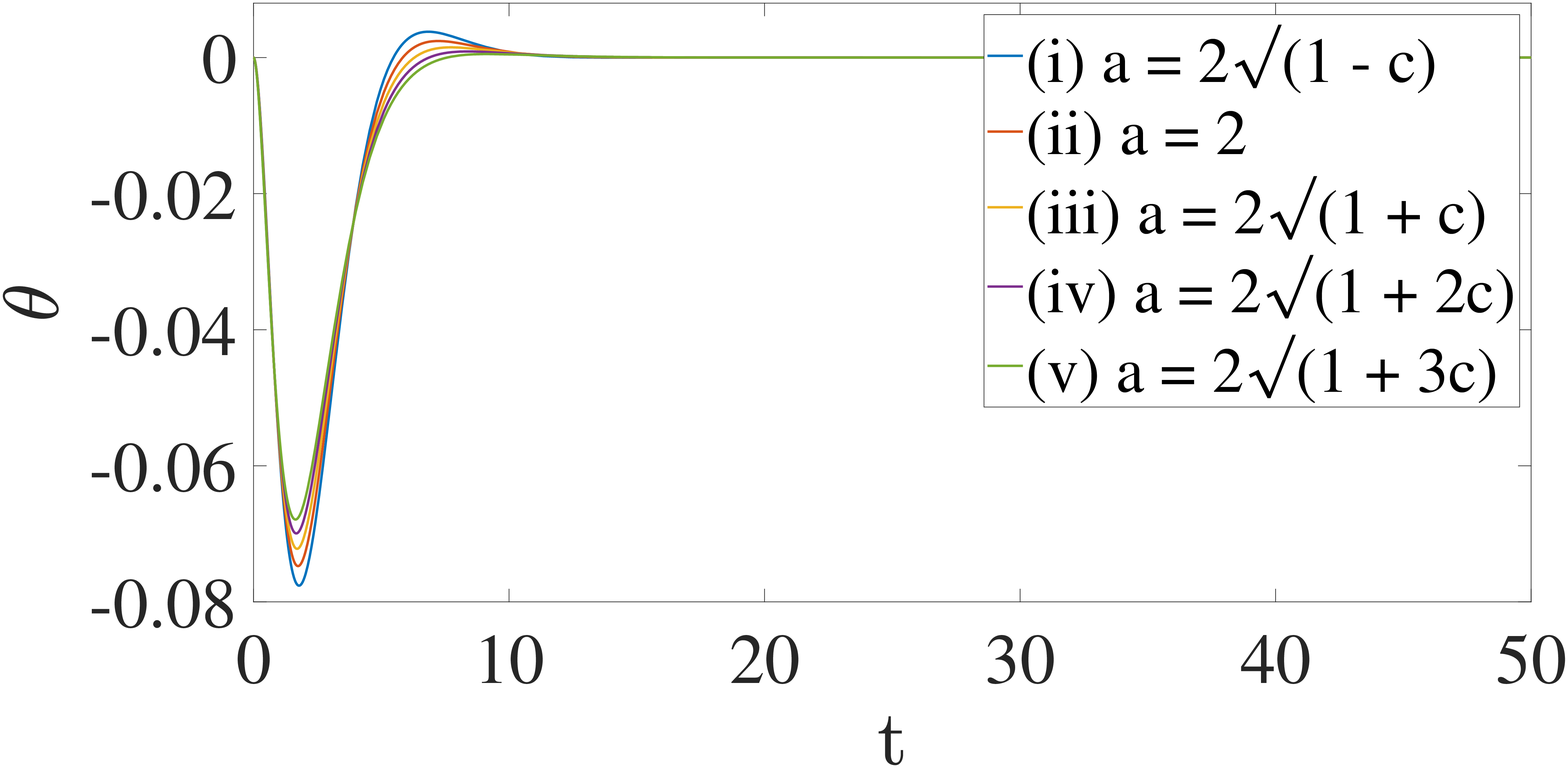} %c = 0.1
}
\subfloat[\footnotesize $c = 0.5$]{
\includegraphics[width = 0.33\textwidth]{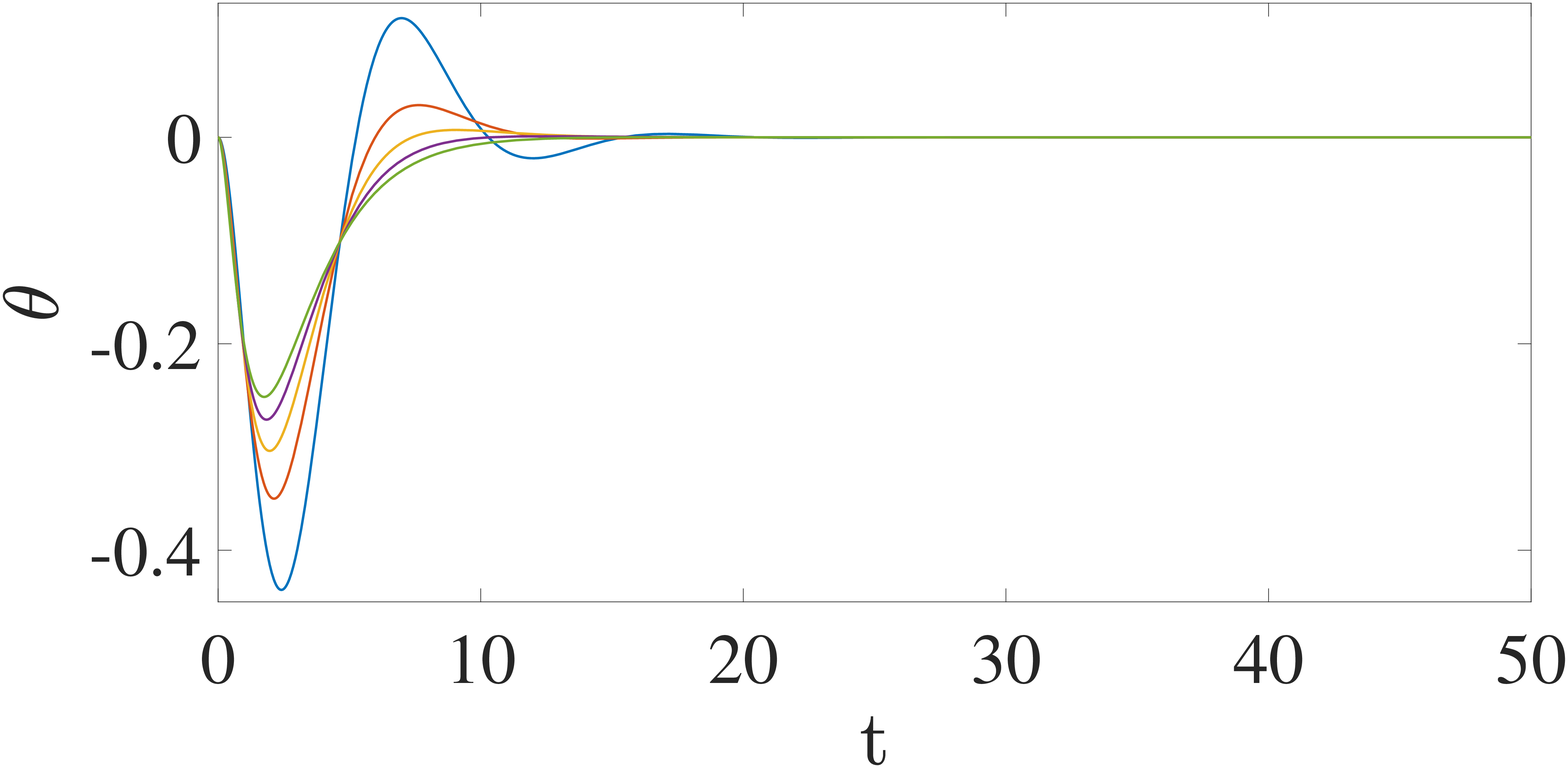} %c = 0.5
}
\subfloat[\footnotesize $c = 0.9$]{
\includegraphics[width = 0.33\textwidth]{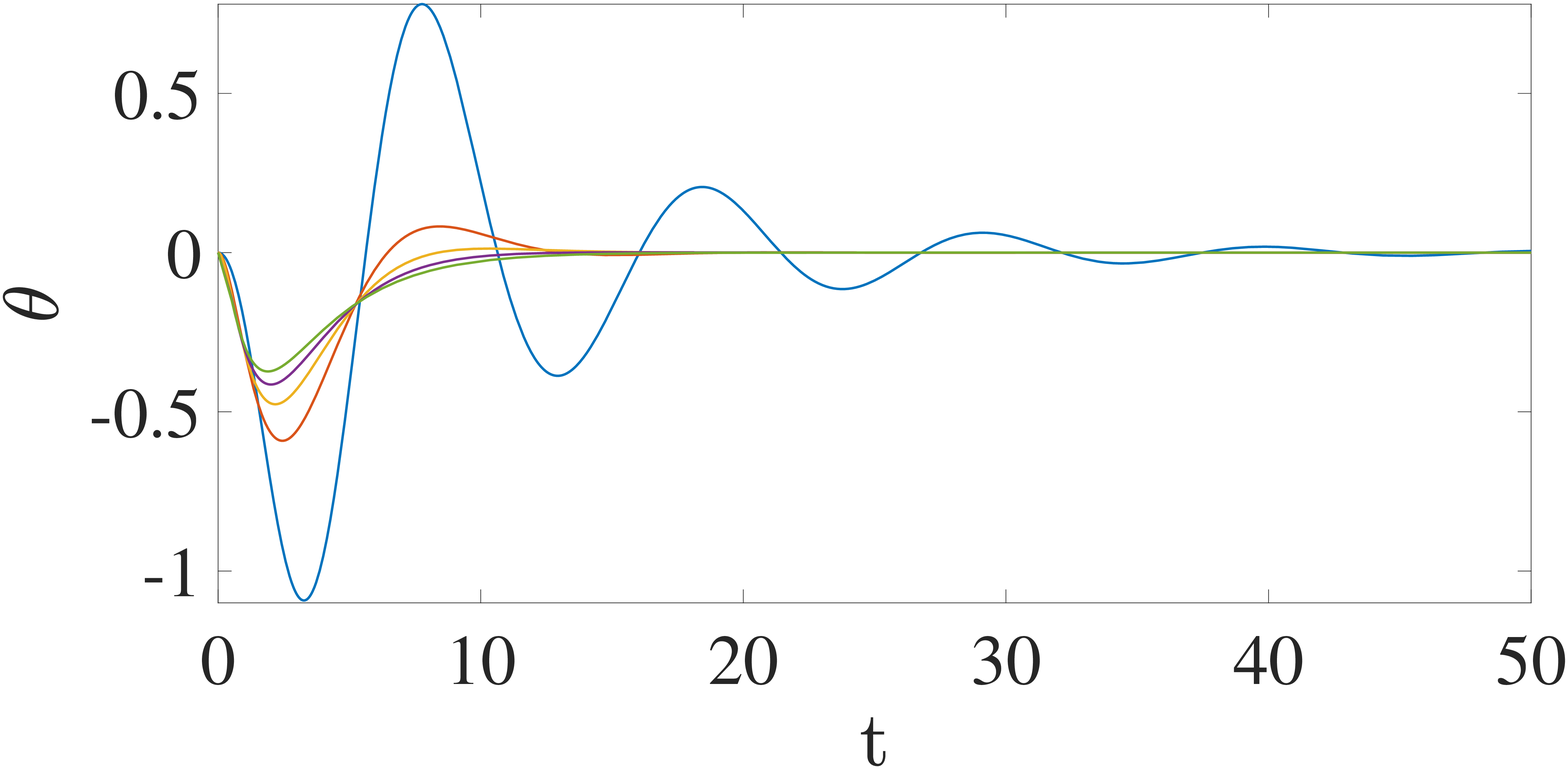} %c = 0.9
}
\caption{Examples of axion solution curves ($t_{max} = 50$) in the case of dissipation.}
\label{5cate}
\end{figure}

As an example, for the above categories we set (i) $a = 2\sqrt{1 - c}$, (ii) $a = 2$, (iii) $a = 2\sqrt{1 + c}$, (iv) $a = 2\sqrt{1 + 2c}$ and (v) $a = 2\sqrt{1 + 3c}$ (for $0 < c < 1$). The solutions are shown in fig.\ref{5cate} with $c = 0.1, 0.5$ and $0.9$, respectively.

\section{The effect of an external magnetic field}

Coupled Josephson junctions, when treated quantum mechanically, can form
SQUIDs (superconducting quantum interference devices), which are very sensitive to small magnetic fields.
Moreover, axions also have non-trivial interactions with magnetic fields, for example they can decay
in strong magnetic fields into two photons. 
It has also been suggested that magnetic flux noise phenomena in coupled $q$-bits and SQUIDs can yield
valuable information about axion physics \cite{fluxnoise}.
It is thus interesting to study
the effects of weak and strong magnetic fields for our coupled (classical) axion-Josephson system.

To take into account magnetic fields, our system dynamics is extended to
\begin{subequations}
\begin{align}
\ddot{\varphi} + a_{1} \dot{\varphi} + b_{1} \sin{\varphi} &= c(\ddot{\theta} - \ddot{\varphi}) + d_{1}(\varphi + e_{1})\\
\ddot{\theta} + a_{2} \dot{\theta} + b_{2} \sin{\theta} &= c(\ddot{\varphi} - \ddot{\theta})
\end{align}
\label{full}
\end{subequations}
where $d_{1} = -\Phi_{0}/2\pi L I_{c}$ is a material constant and $e_{1} = 2\pi \Phi/\Phi_{0}$ is the normalised applied flux (with $\Phi_{0}$ the flux quantum, $L$ the inductance, $I_{c}$ the critical current, and $\Phi = |\vec{B}|\times Area$), the external magnetic field is $\vec{B}$, cf. \cite{beck1}). Referring to experiments described in \cite{coup2}, we choose $a_{1} = a_{2} = 5\times 10^{-5}, b_{1} = b_{2} = 1, c = 2.31\times 10^{-3}, d_{1} = -0.352$ and allow $e_{1}$, which is determined by the strength of the external magnetic field, to vary from $0$ to $100$. Note that unlike the model in \cite{coup2}, the second equation in the above system describes the axion dynamics and there is no \textit{a priori} reason to include a similar flux term as in the first equation. For our numerical simulations in the following,
the initial conditions are chosen in the same way as in section 2.

The phase space structure for this full system is more complicated than in the previous cases: an interesting phenomenon occurs when $e_{1}$ increases from $\frac{\pi}{2}$ via $\pi$ to $\frac{3\pi}{2}$, see fig.\ref{full-e1-pi} below. 
\begin{figure} [H]
\centering
\subfloat[\footnotesize $e_{1} = \pi/2$]{
\includegraphics[width = 0.3\textwidth]{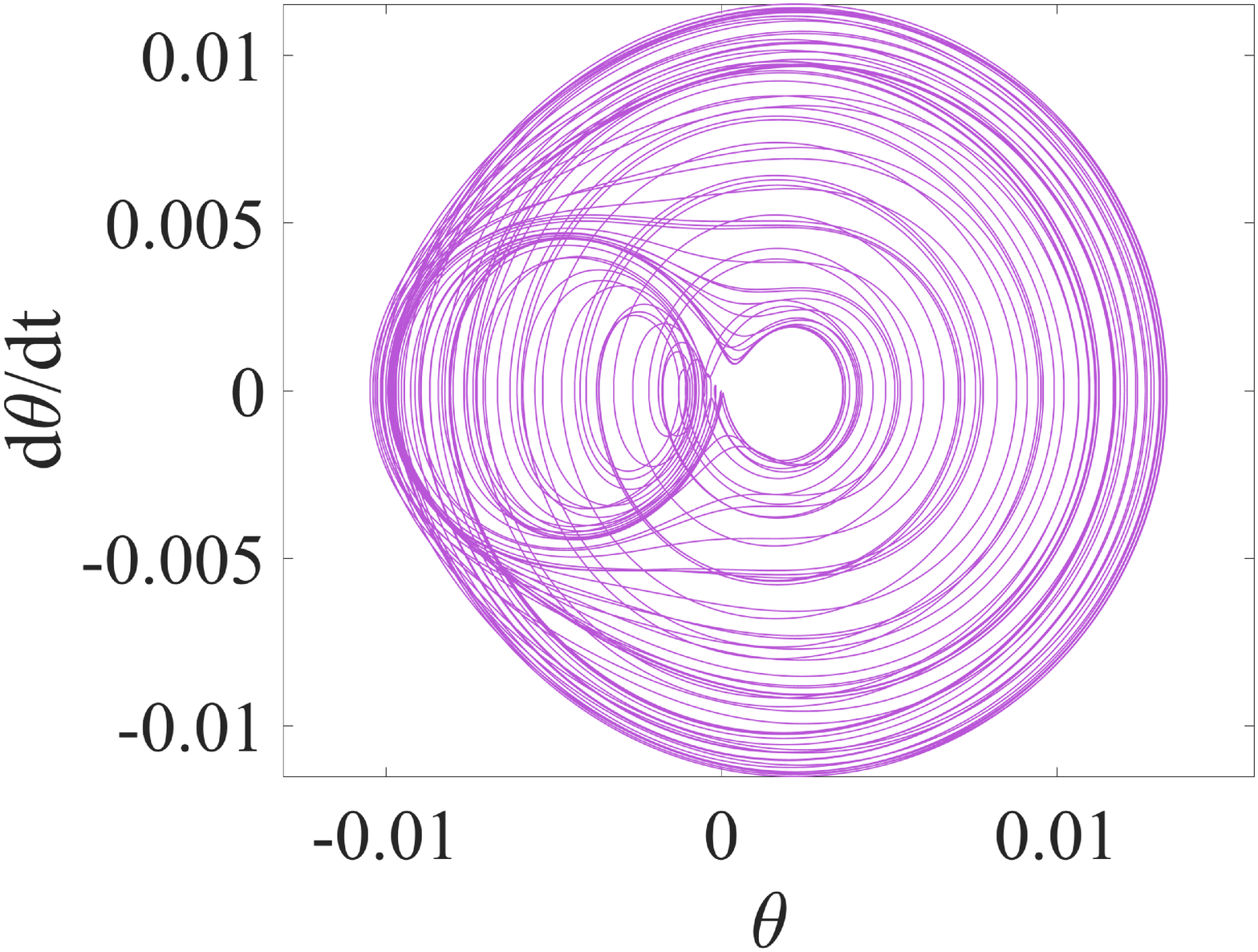} %---full e1 = pi/2, %traj for axion
}
\subfloat[\footnotesize $e_{1} = \pi$]{
\includegraphics[width = 0.3\textwidth]{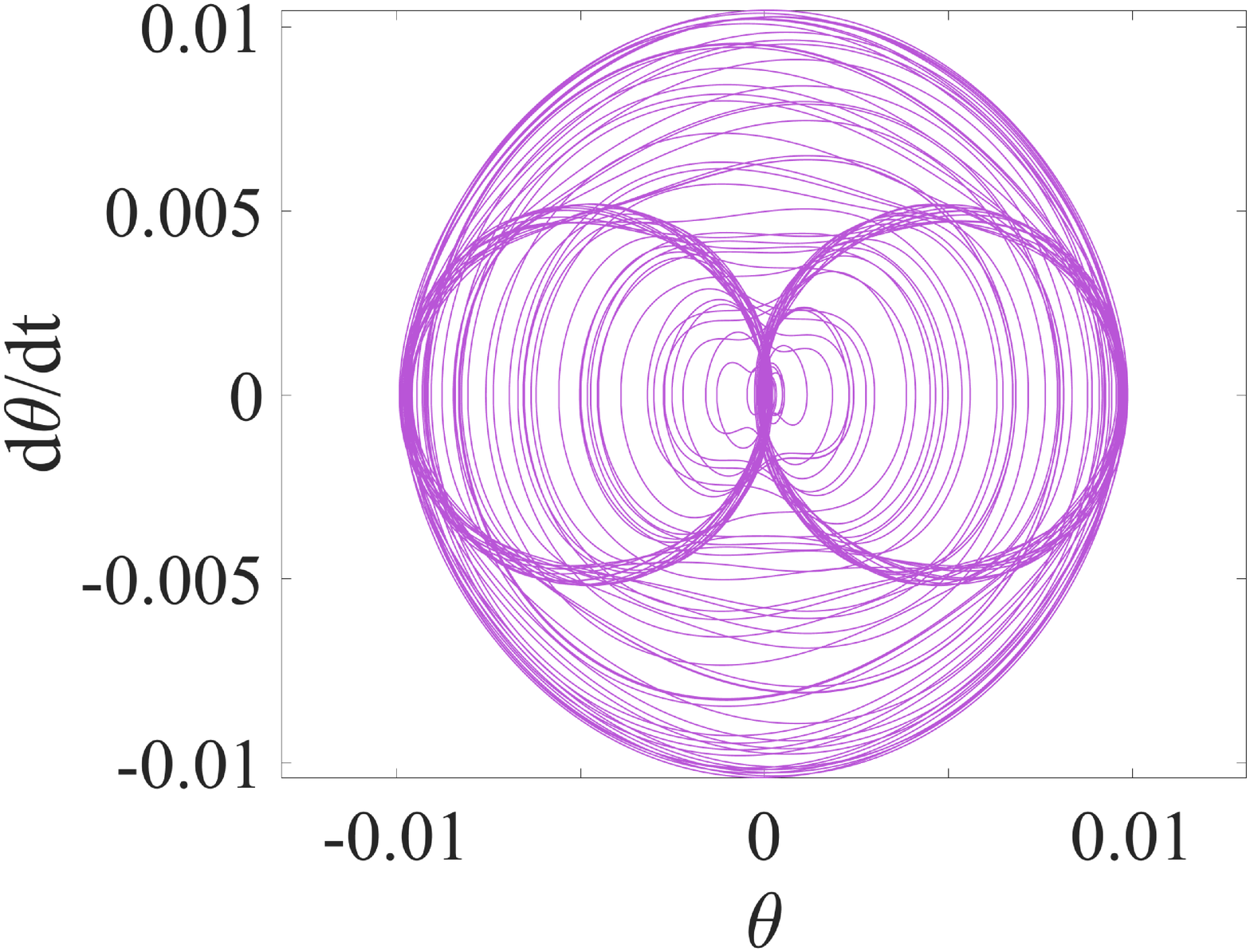} %---full e1 = pi
\label{pi}}
\subfloat[\footnotesize $e_{1} = 3\pi/2$]{
\includegraphics[width = 0.3\textwidth]{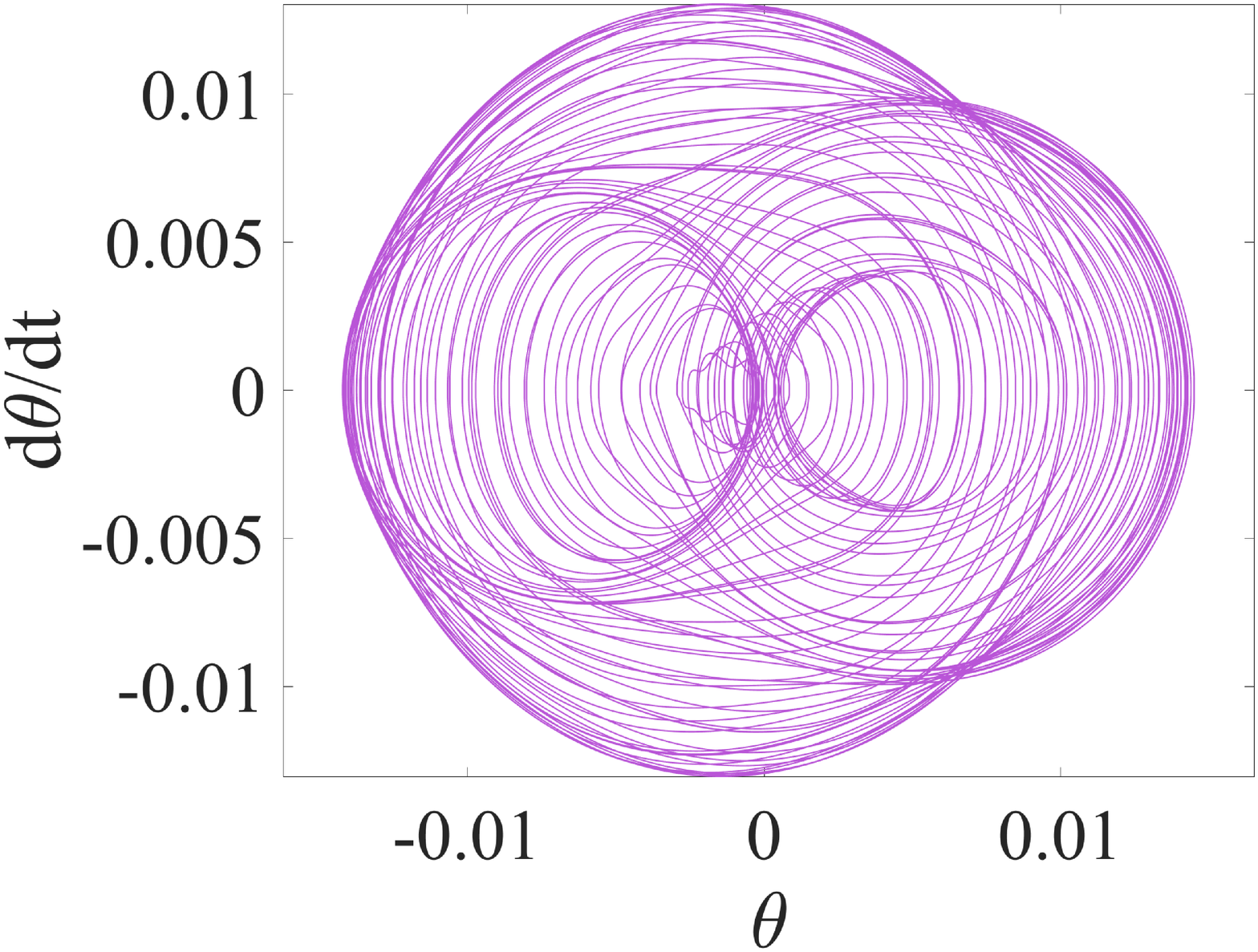} %---full e1 = 3pi/2
}
\qquad
\subfloat[\footnotesize $e_{1} = \pi/2$: JJ in blue]{
\includegraphics[width = 0.3\textwidth]{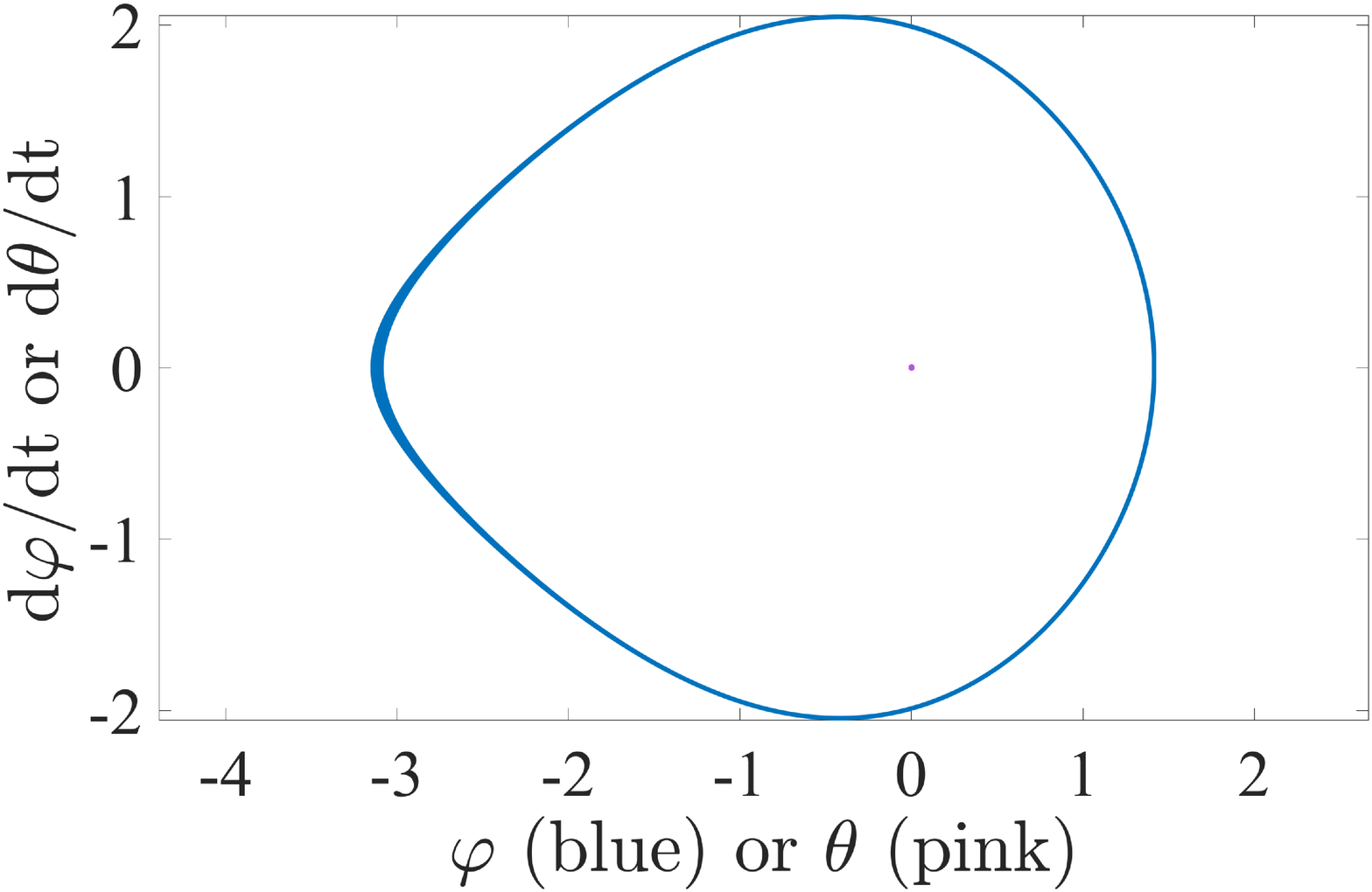} %corresponding phase space for both (axion as a small dot)
}
\subfloat[\footnotesize $e_{1} = \pi$: JJ in blue]{
\includegraphics[width = 0.3\textwidth]{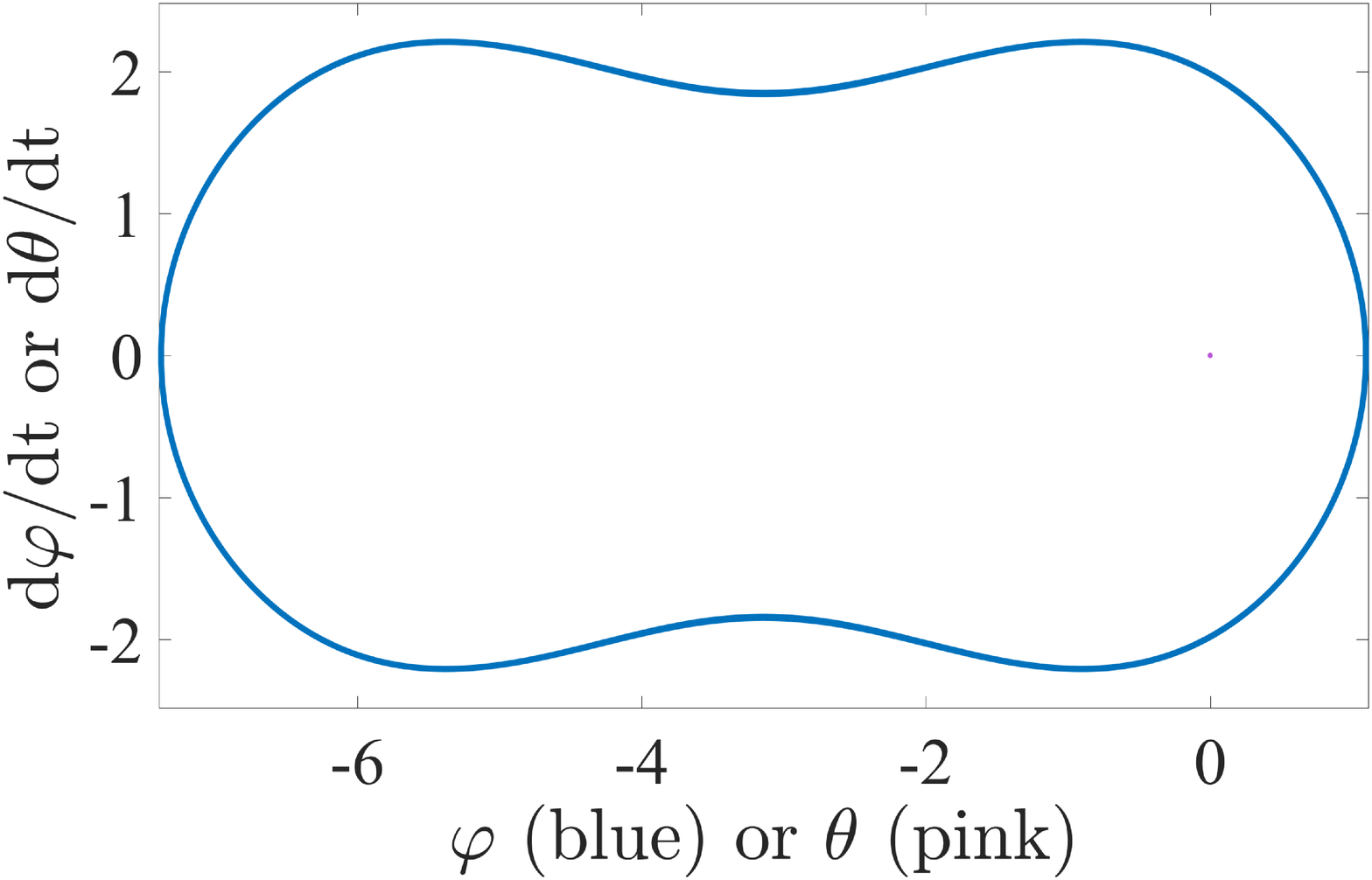} 
}
\subfloat[\footnotesize $e_{1} = 3\pi/2$: JJ in blue]{
\includegraphics[width = 0.3\textwidth]{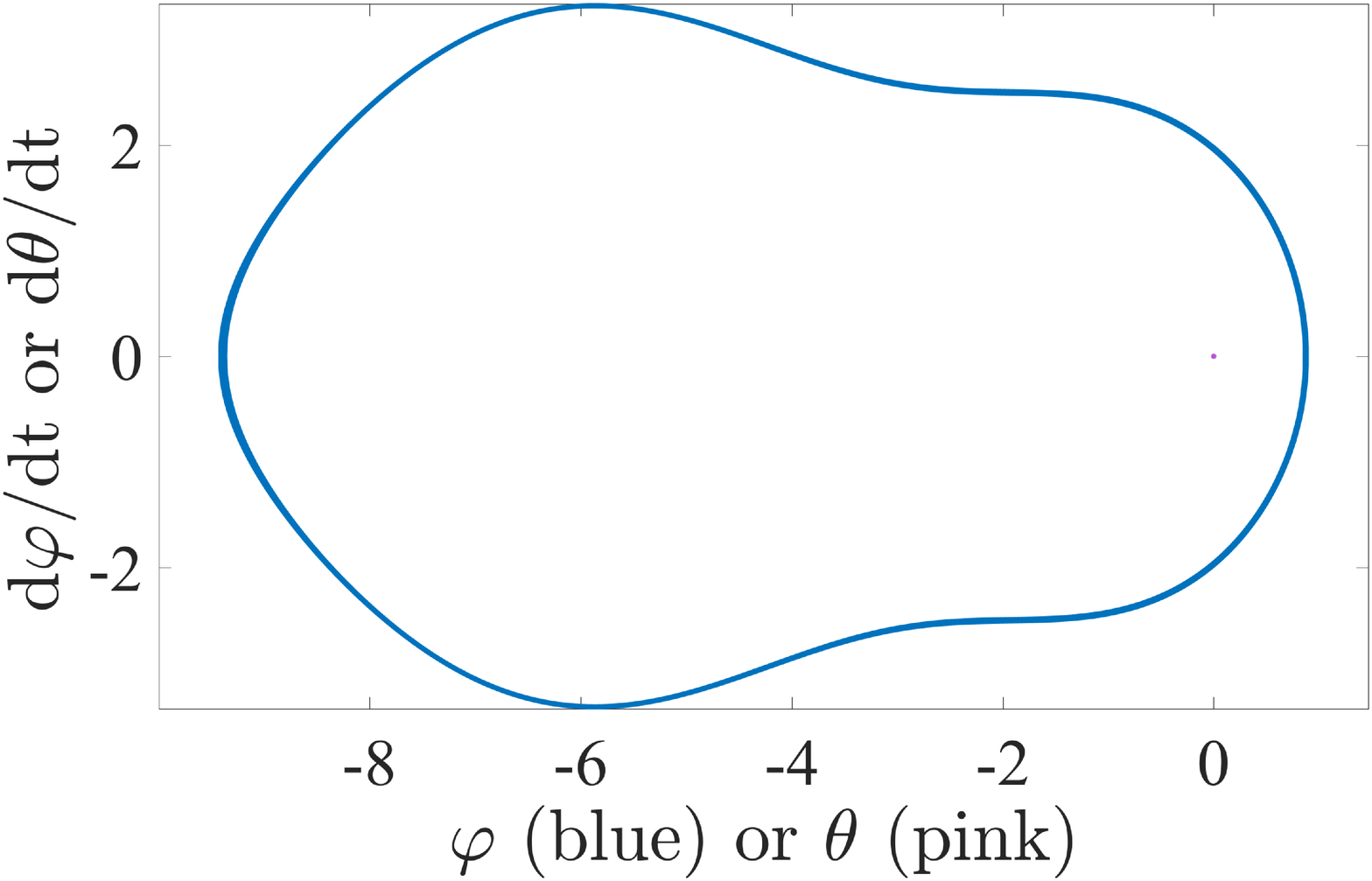} 
}
\qquad
\subfloat[\footnotesize $e_{1} = \pi/2$]{
\includegraphics[width = 0.3\textwidth]{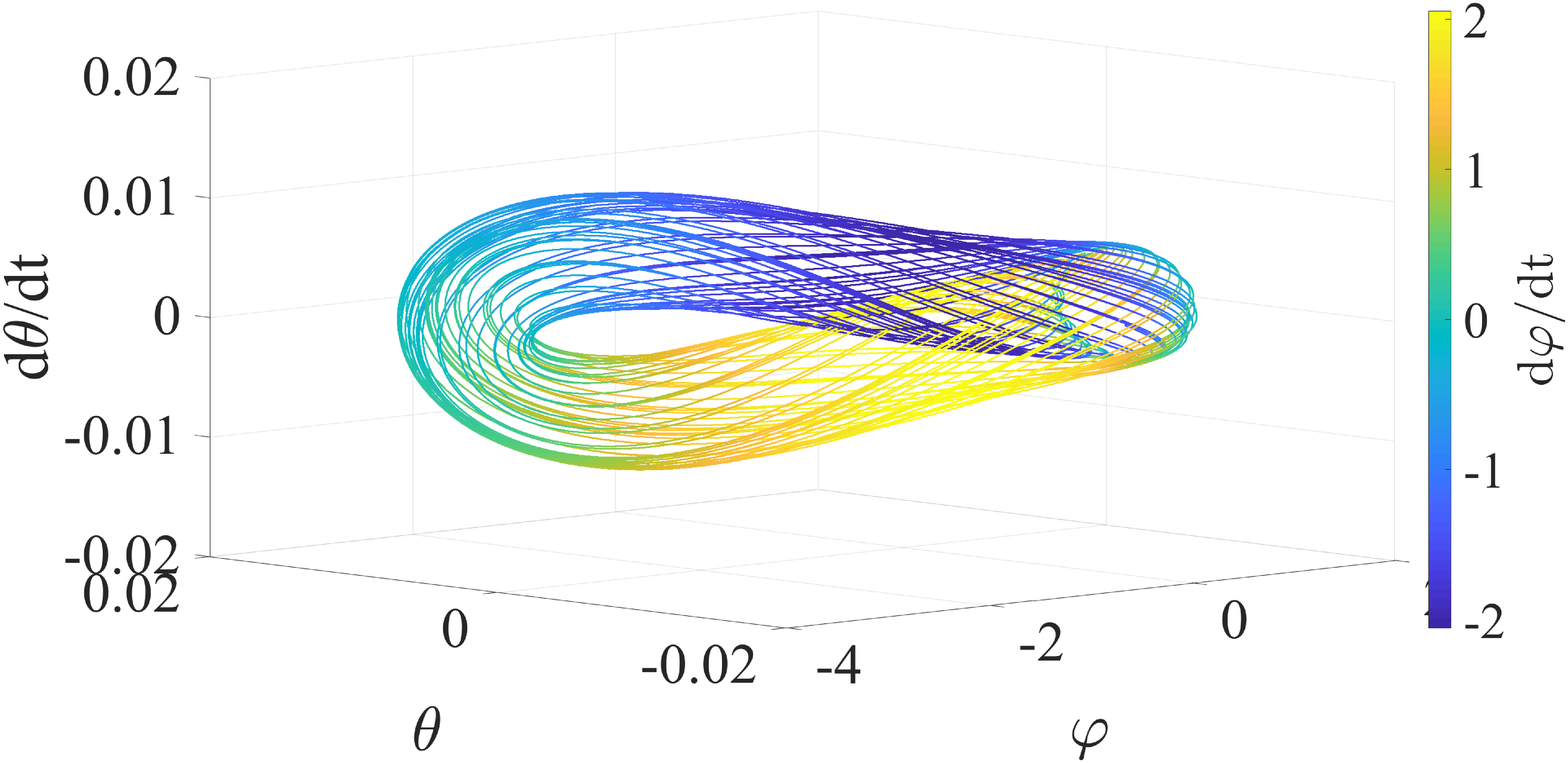} %corresponding 3D (x1, x3, x4)-subspace
\label{klein}}
\subfloat[\footnotesize $e_{1} = \pi$]{
\includegraphics[width = 0.3\textwidth]{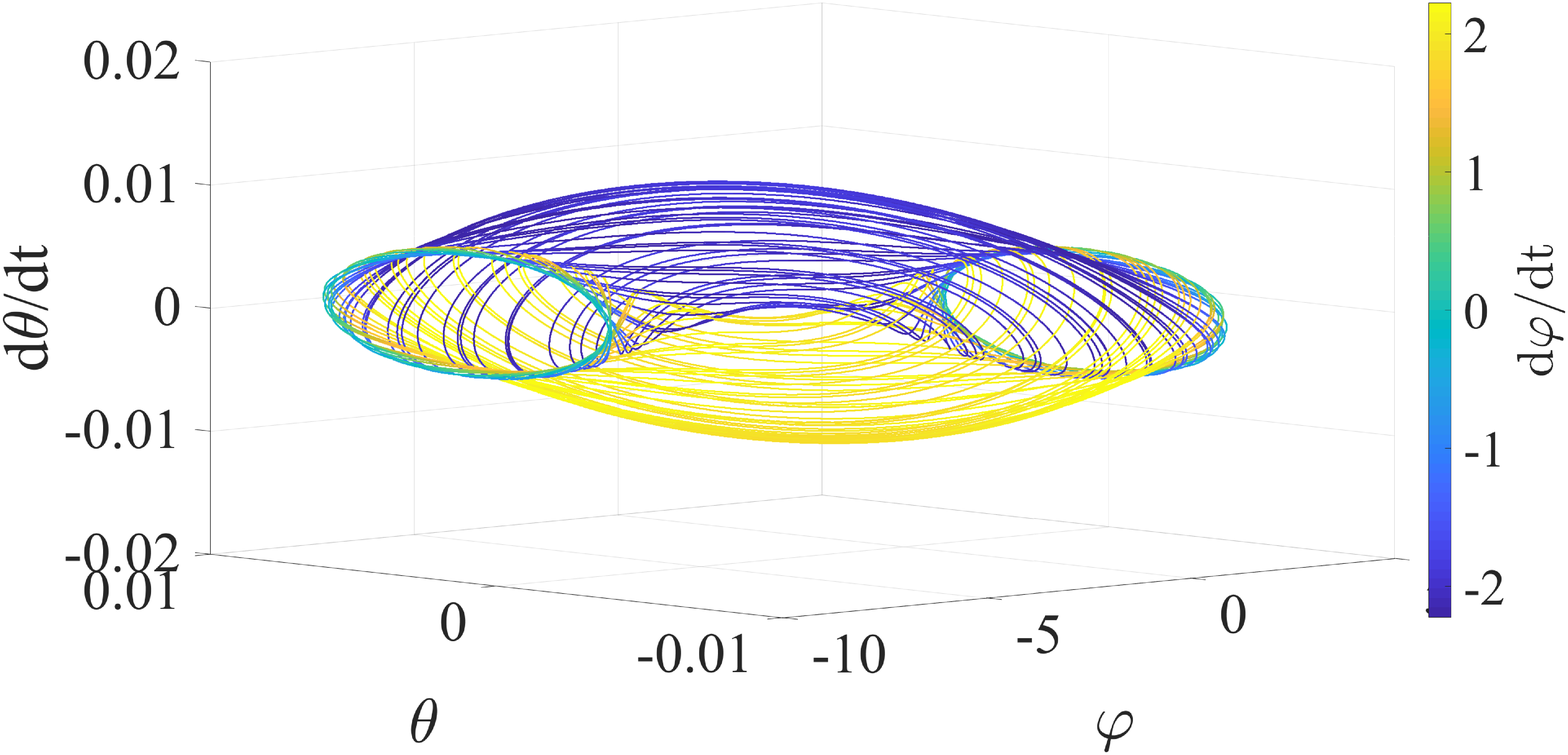} 
}
\subfloat[\footnotesize $e_{1} = 3\pi/2$]{
\includegraphics[width = 0.3\textwidth]{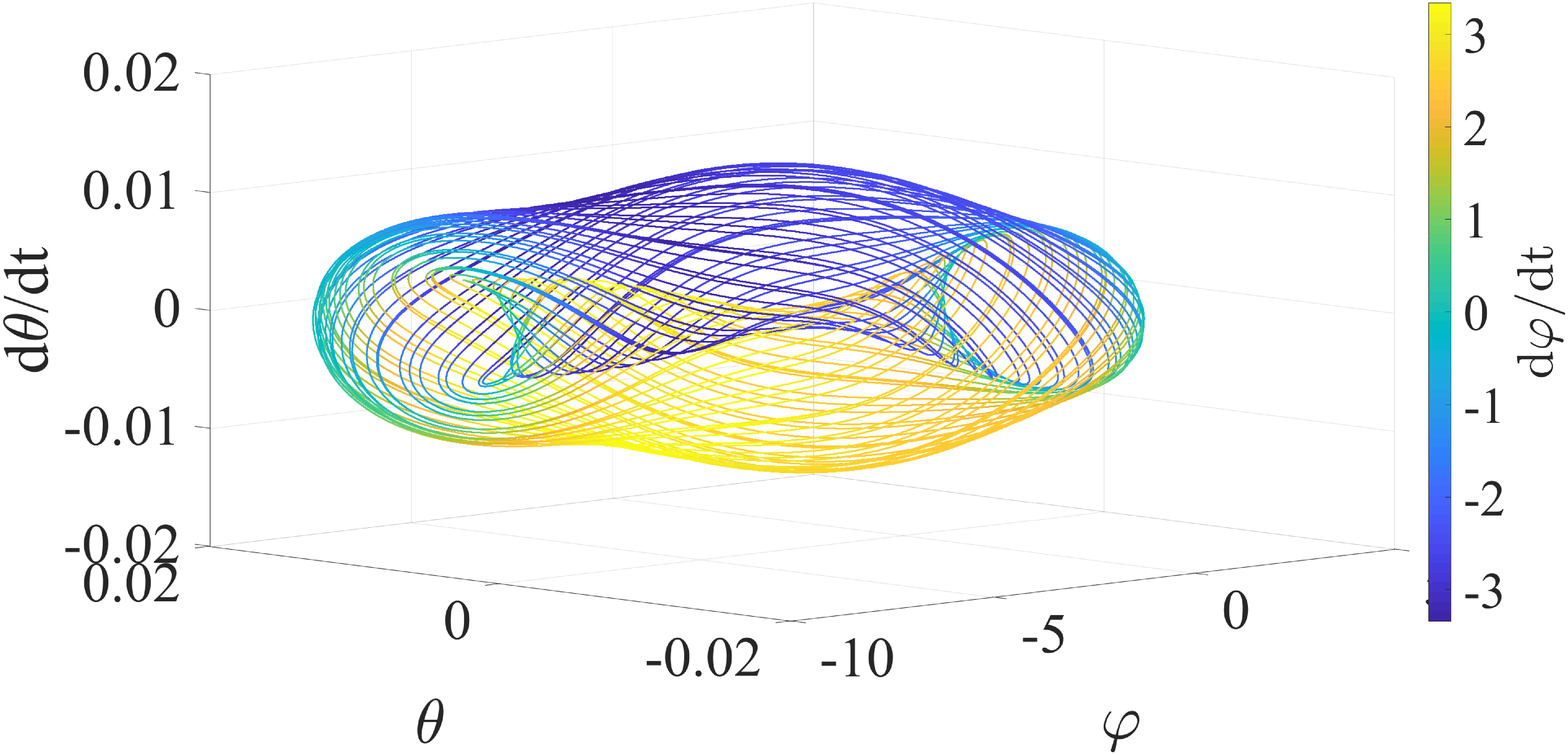} 
}
\caption{Effect of an external magnetic field with $e_{1} = \pi/2, \pi, 3\pi/2$: first row: axion phase trajectories; second row: the corresponding trajectories for both Josephson junction (JJ) and axion  (in each plot the almost invisible pink dot at the origin represents the relative size of the axion oscillation); last row: the corresponding $(\varphi, \theta, \dot{\theta})$-subspace with color coding representing the value of  $\dot{\varphi}$ ($t_{max} = 500$)}
\label{full-e1-pi}
\end{figure}

Comparing the phase portrait structure near the transition point at $e_{1} = \pi$, we observe an
interesting topological phase transition: The cyan side of fig.\ref{klein} is everted to make the whole structure being twisted on both ends --- the dark blue and light yellow bands cross each other twice (like an inside-out torus), and the cyan parts become two thin circles which correspond to the `$\infty$' that appears in the axion phase trajectory (fig.\ref{pi}). The overall topological deformation undergoes a procedure like a torus eversion: from a torus (for $e_{1} \sim 0$) to a one-side everted torus (for $e_{1} \sim \frac{\pi}{2}$) to a two-side everted torus (for $e_{1} \sim \pi$) to finally an inside-out torus (for $e_{1} \sim \frac{3\pi}{2}$). 
This complicated topological phase transition illustrates that the effect of small magnetic fields for coupled
Josephson systems is profound, and occurs already in the classical treatment.
%Also notice that the presence of the external magnetic field `destroys' the left-right symmetry of the phase trajectory but it can be recovered by setting the opposite sign to $e_{1}$. 
For a large enough magnetic field $e_{1} \sim 100$ the phase portrait of the Josephson junction (in blue) approaches an ellipse while the two bands in the subspace are densely interwoven and eventually fill the whole structure in phase space.

\section{Conclusion}

The classical dynamics of coupled axion-Josephson junction dynamical systems exhibits a surprisingly large complexity which we have investigated in detail in this paper.
When changing either the coupling constant or the initial angular velocity, we observed eversion processes where a simple 
cardioid-shaped trajectory splits into multiple copies, with a different topology above and below the eversion point. Close to resonance points, where the 
plasma frequency of the junction coincides with the axion mass ($b_1 = b_2$), there is extreme sensitivity of the structure of the phase portrait depending on the ratio of Josephson to axion frequency. In the limit of small elongations, we proved that for certain distinguished coupling constants the coupled system generates time-shifted identical trajectories for the axion and the Josephson junction, however these oscillations have much richer structure and the phase portrait depends in a complicated way on two integers $k,k_1$ that make up the coupling constant $c=\frac{2k^2}{(2k_1+1)^2}-\frac{1}{2}$. From a physical point of view this result is very interesting, because it shows that even in the limit of extremely small coupling $c \to 0$ the axion and Jospsephson junction trajectory can synchronize in a time-shifted way, meaning that the (measured) Josephson junction phase angle mirrors the behavior of the axion.
The introduction of a magnetic field makes the phase portrait even more complex, with topological transitions and torus eversions at critical values of the magnetic field. Our investigation was motivated by the need to understand the dynamics (and possible signals) in future axion detectors based on Josephson junctions or coupled Josephson junctions, which have been suggested in the recent literature \cite{beck1,beck2,beck3,beck4}. For realistic detector scenarios one would need to proceed from a classsical description to a quantum description, which is out of the scope of the current paper. However, interestingly enough already the classical dynamics is extremely complex, and this will imprint onto the quantum dynamics.

%\newpage
%\appendix

%\subsection*{Acknowledgements}

%C.B.'s research is supported by EPSRC under grant No. EP/N013492/1.

%\subsection*{Author Contributions}

%C.B. did the research and wrote the manuscript.

%\subsection*{Additional Information}

%{\bf Competing financial interests:} The author declares no competing financial interests.

\end{document}